\documentclass[11pt]{article}

\usepackage{cite,graphicx}

\RequirePackage{amsgen}[1995/01/01]
\RequirePackage{amsfonts}[1995/01/01]
\RequirePackage{amssymb}[1995/01/01]
\RequirePackage{amsbsy}[1995/01/01]

\newcommand{\ppbox}[1]{\parbox{6.7cm}{\small #1}}
\newcommand{\pppbox}[1]{\,\parbox{3.5cm}{\small #1}}
\newcommand{\lra}{{}\!\!\!\!\leftrightarrow{~}}

\newcommand{\Pop}{{\bf P}}

\usepackage{amssymb,epsf}
\newcommand{\hepth}[1]{{[arXiv:hep-th/#1]}}
\newcommand{\heplat}[1]{{[arXiv:hep-lat/#1]}}
\newcommand{\hepph}[1]{{[arXiv:hep-ph/#1]}}
\newcommand{\mathph}[1]{{[arXiv:math-ph/#1]}}
\newcommand{\condmat}[1]{{[arXiv:cond-mat/#1]}}
\newcommand{\physics}[1]{{[arXiv:physics/#1]}}
\newcommand{\mathCA}[1]{{[arXiv:math-CA/#1]}}
\newcommand{\mathSP}[1]{{[arXiv:math-SP/#1]}}
\newcommand{\quantph}[1]{{[arXiv:quant-ph/#1]}}
\newcommand{\nlinsys}[1]{{[arXiv:nlin-si/#1]}}

\newcommand{\toline}[1]{}
\newcommand{\phup}{^{\phantom{p}}}

\renewcommand{\bar}{\overline}
\newcommand\blank[1]{}
\newcommand{\xtra}[1]{{.}}
\renewcommand{\xtra}[1]{{, \tt hep-th/#1.}}

\newcommand{\xtrac}[1]{{.}}
\renewcommand{\xtrac}[1]{{, \tt cond-mat/#1.}}

\newcommand{\mathematica}[1]{{}}

\newcommand{\mm}[2]{{\vphantom{\vbox to 6mm{}}}}
\newcommand{\fract}[2]{{\textstyle\frac{#1}{#2}}}

\newcommand{\ri}{\right}
\newcommand{\ep}{\varepsilon}
\newcommand{\lf}{\left}
\newcommand{\te}{\theta}
\newcommand{\CS}{{\cal S}}

\newcommand{\beqcol}{\begin{array}{rcl}}
\newcommand{\eeqcol}{\end{array}}

\newcommand\ZZ{{\mathbb Z}}
\newcommand\CC{{\mathbb C}}
\newcommand\RR{{\mathbb R}}

\newcommand{\CT}{{\cal T}}

\newcommand{\CF}{{\cal F}}
\newcommand{\T}{{\bf T}}
\newcommand{\Q}{{\bf Q}}

\newcommand\eq{\begin{equation}}
\newcommand\en{\end{equation}}
\newcommand\bea{\begin{eqnarray}}
\newcommand\eea{\end{eqnarray}}
\newcommand\nn{\nonumber}
\newcommand\half{{\textstyle\frac{1}{2}}}
\newcommand\hf{\frac{1}{2}}

\newcommand{\teta}{\tilde \eta}

\newcommand{\One}{{\hbox{{\rm 1{\hbox to 1.5pt{\hss\rm1}}}}}}
\renewcommand{\One}{{\mathbb 1}}
\renewcommand{\One}{{\rm 1\!\!1}}

\newcommand{\ket}[1]{|#1\rangle}

\newcommand{\opnup}[1]{\renewcommand{\\}{\\[50 pt]}}
\renewcommand{\bar}{\overline}
\renewcommand{\tilde}{\widetilde}

\newcommand{\Balpha}{\balpha}
\newcommand{\Bbalpha}{\balpha}
\newcommand{\Bbdelta}{\bdelta}

\newcommand{\PT}{{\cal P}{\cal T}}
\newcommand{\CH}{{\cal H}}
\newcommand{\CaC}{{\cal C}}

\newcommand\LL{{\mathbb L}}
\newcommand\TT{{\mathbb T}}
\newcommand\QQ{{\mathbb Q}}
\newlength{\indentedwidth}
\newdimen\mathindent
\indentedwidth=\mathindent
\newcommand{\balpha}{\boldsymbol{\alpha}}
\newcommand{\bdelta}{\boldsymbol{\delta}}
\def\Wp{ \raise.4ex\hbox{\textrm{\Large $\wp$}}}
\def\vt{\vartheta}
\newcommand{\fl}{\hspace*{-\mathindent}}

\newcommand{\resection}[1]{\setcounter{equation}{0}\section{#1}}
\begin{document}

%

\begin{titlepage}
\vskip 0.5cm
\begin{flushright}
{\tt hep-th/0703066}
\end{flushright}
\vskip .7cm
\begin{center}
{\Large{\bf 
The ODE/IM correspondence
}}
\end{center}
\vskip 0.8cm \centerline{Patrick Dorey$^1$, Clare Dunning$^2$ and
  Roberto Tateo$^3$} \vskip 
0.9cm \centerline{${}^1$\sl\small Dept.\ of Mathematical Sciences,
University of Durham,} \centerline{\sl\small  Durham DH1 3LE, United
Kingdom\,}
\vskip 0.3cm \centerline{${}^{2}$\sl\small IMSAS, University of
Kent, Canterbury, UK CT2 7NF, United Kingdom}
\vskip 0.3cm \centerline{${}^{3}$\sl\small Dip.\ di Fisica Teorica
and INFN, Universit\`a di Torino,} \centerline{\sl\small Via P.\
Giuria 1, 10125 Torino, Italy}
\vskip 0.2cm \centerline{E-mails:}
\centerline{p.e.dorey@durham.ac.uk, t.c.dunning@kent.ac.uk,
tateo@to.infn.it}

\vskip 1.25cm
\begin{abstract}
\noindent
\looseness+1 This article reviews a recently-discovered link between
integrable quantum
field theories and certain ordinary differential equations in the
complex domain.
Along the way, aspects of $\PT$-symmetric quantum mechanics are discussed,
and some elementary features of the six-vertex model and the Bethe ansatz are
explained.
\end{abstract}

\end{titlepage}
%


\tableofcontents
%

\resection{Introduction}
Our aim in this article is to describe some links which are starting
to emerge between two previously-separated areas of mathematical
physics -- the theory of integrable models in two dimensions, and the
spectral analysis of ordinary differential equations.

The study of integrable lattice models has been an intriguing
part of mathematical physics since Onsager's solution of
the two-dimensional Ising model. Lieb and Sutherland's work on the
six-vertex model showed that this was not an isolated phenomenon, and
with Baxter's work at the beginning of the 1970s 
the full richness of the field began to be appreciated by a wider
community. Since that
time interest
in the subject has grown steadily, receiving a particular boost of late
from the links which exist with integrable quantum field theories.
Many different methods exist for the solution of
these models, and a technique which will be very important in
the following goes
by the name of the `functional relations' approach. The idea,
initially put forward by Baxter, is to show that quantities of
interest satisfy functional relations. When combined with suitable
analyticity properties, these relations can be highly
restrictive and often lead to exact formulae for quantities of
physical interest.

In a parallel chain of development which also dates back at least to
the early 1970s, Sibuya, Voros and others have shown that
functional relations have an important r\^ole to play
in a rather more classical area of mathematics,
namely the theory of Stokes multipliers and spectral determinants
for ordinary differential equations in the complex domain.
However, it is only recently that the existence of a precise link --
an `ODE/IM correspondence' -- between these
two areas has been realised.  The aim of these notes is to provide an
elementary introduction to this connection and its background.
For most of the time the focus will be on the simplest example,
which connects second-order ordinary differential equations to
integrable models associated with the Lie algebra $SU(2)$.

We begin, in the next section,
with a short introduction to the types of spectral problems which
will be
relevant,  mentioning in the process our third main theme, the
intriguing reality properties of certain non-Hermitian spectral
problems which arise in the study of `$\PT$-symmetric' quantum mechanics.
Integrable lattice models
and their treatment via functional relations are
introduced in section \ref{sect:fnrel}; we also discuss briefly
the recent development of these ideas within quantum field
theory. The differential equations side of the story is explained
in section \ref{funode}, after which the link with
integrable models is made precise in section \ref{dict}.
Some applications and generalisations of the correspondence
are outlined in section~\ref{spepro}, and section \ref{dev} contains
our conclusions. Various pieces of background material have been
collected in the appendices, including an explanation
of the algebraic  Bethe ansatz in appendix \ref{aba_app}.

More on the ODE/IM correspondence can be found in references
\cite{DTa,BLZa,Sa,DTb,DTc,Sb,Sc,DDTa,Srev,DDTrevb,DDTb,DDTc,DDTrev,BLZhigh, 
  dst, DMST, bm,Bazhanov:2006ic,ddmst}. 
All of this work rests on
earlier studies by, among others, Sibuya~\cite{Sha},
Voros~\cite{Voros},
and Bender et al~\cite{BT,BB,BBN} (on the ODE side) and by
Baxter~\cite{Bax,Bax2}, Kl\"umper, Pearce and
collaborators~\cite{KP,KBP}, Fendley et al~\cite{FLS},
and Bazhanov, Lukyanov and Zamolodchikov~\cite{BLZ1,BLZ2,BLZ3}
on the integrable models side.

We have aimed to make this article accessible to readers with
backgrounds in both the integrable models and the differential
equations communities; for this reason, we have tried to keep the
treatment relatively elementary. More
details can be found
in many places~\cite{Sha,Voros,baxterbook,FMS}.

Those readers primarily interested in ODEs may prefer to concentrate
on section~\ref{prelude}, briefly read sections \ref{aba}, \ref{blztq} and
\ref{summary} to get  a flavour of the integrable model picture, then
move on to sections \ref{funode}, \ref{dict} and
\ref{spepro}, while
those primarily interested in integrable models may like to look
initially at
sections~\ref{prelude}, \ref{sect:fnrel}, \ref{funode} and
then at \ref{inhom}, \ref{tqcd} and \ref{full}.

%

\resection{Prelude: three reality conjectures in $\PT$-symmetric
quantum mechanics}
\label{prelude}
Spectral problems, some of a rather unconventional nature, will play a
central r\^ole on the `differential equations' side of our story. Rather
than launch straight into technicalities,
we shall warm up
in this preliminary section by describing an intriguing
class of problems much studied by Bender and collaborators and others
in recent years. It all begins\footnote{While this question initiated the
line of work we want to describe here, similar curiosities had in fact
been observed before -- see, for example,
\cite{Lee:1954iq,Brower:1978sr,Denham:1980gs,CGM1980,BG1993}. For
further historical discussions, see \cite{fk1,B2005}.}
 with a question posed by Bessis and
Zinn-Justin, sometime near 1992:

\noindent {\bf Question 1:} What does the spectrum of the Hamiltonian
\eq
\CH=p^2+ix^3
\en
look like?

\begin{figure}[ht]
\begin{center}
\includegraphics[width=0.67\linewidth]{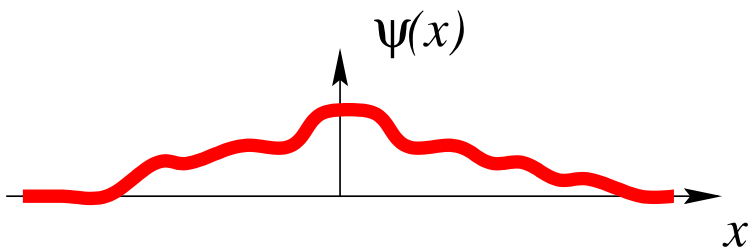}
\end{center}
\caption{A wavefunction decaying at $x=\pm\infty$.\label{wavefun}}
\end{figure}

\noindent
This is a cubic oscillator, with purely imaginary coupling $i$. (Strictly
speaking, Bessis and Zinn-Justin, motivated by considerations of the
Yang-Lee edge
singularity\,\cite{Fisher:1978pf,Cardy:1985yy,Cardy:1989fw},
were initially interested in more general
Hamiltonians of the form $p^2+x^2+igx^3$, from which the above problem
emerged as a strong-coupling limit.)
The
corresponding Schr\"odinger equation is
\eq
-\frac{d^2}{dx^2}\psi(x)+ix^3\psi(x)=E\psi(x)
\label{cseq}
\en
and we shall declare that
the (possibly complex) number $E$ is in the spectrum if and only
if, for that value of $E$, the equation has a solution $\psi(x)$ on the
real axis which decays both at $x\to-\infty$ and at
$x\to+\infty$, as illustrated in figure~\ref{wavefun}.\footnote{To be
  more precise, the decay should be fast 
enough that $\psi(x)$ lies in $L^2(\RR)$, the space of square-integrable
functions on the real axis.
This means that we are actually discussing the so-called {\em
point spectrum} of $H$ -- see, for example, \cite{REL,rich}.}.
Notice  that the differential equation forces the
wavefunction $\psi(x)$ to be complex, even for real values of $x$ and
$E$. And
since the Hamiltonian is not (at least in any obvious way) Hermitian,
the usual arguments to show that all of the eigenvalues
must be real do not apply.
Nevertheless, perturbative and numerical studies led
Bessis and Zinn-Justin to the following claim:

\smallskip

\noindent
{\bf Conjecture 1} \cite{BZJ}:
The spectrum of $\CH$ \underline{is} real, and positive.
\smallskip

What might be behind this strange property? Bender and Boettcher~\cite{BB}
stressed the relevance of $\PT$ symmetry. To be more precise,
`${\cal P}$', or
parity, acts by sending $x$ to $-x$ and $p$ to $-p$ while ${\cal T}$,
time reversal,
sends $x$ to $x$, $p$ to $-p$ and $i$ to $-i$. Note that
both ${\cal P}$ and
${\cal T}$ preserve the canonical commutation relation $[x,p]=i$ of
quantum mechanics even if $x$ and $p$ are complex~\cite{BBN}.

As shown in \cite{BBN}, $\PT$ invariance implies that
eigenvalues are either real, or appear in complex-conjugate pairs,
much like the roots of a real polynomial.
But, just as the typical real polynomial has many complex
roots\footnote{A famous result of M.~Kac shows that the expected
fraction of real zeros of a real polynomial of degree $n$ with random
(normally-distributed) coefficients
tends to zero as $n\to\infty$, as $\frac{2}{\pi}\log(n)/n$.  See
\cite{Kac1,Kac2}, and, for example, \cite{edelmankostlan,ald}.},
on its own $\PT$ invariance of the Hamiltonian does {\em not}\/ guarantee
reality. This is elegantly shown by
the following generalisation of the Bessis-Zinn-Justin problem,
proposed by Bender and Boettcher \cite{BB}:

\smallskip

\noindent {\bf Question 2:} What is the spectrum of
\eq
\CH_M=p^2-(ix)^{2M}\qquad\quad\mbox{($M$ real, $>0$)}\,?
\en
\smallskip

Later, it will turn out that the passage from question 1 to question
2 corresponds to a change in a parameter in a lattice model, or
equivalently to a change of a quantum group deformation parameter in
a Bethe ansatz system. But for now, the generalisation is appealing
because it unites into a single family of eigenvalue problems both
the $M=3/2$ case, for which we have the Bessis-Zinn-Justin
conjecture, and the much more easily-understood $M=1$ case, the
simple harmonic oscillator. Furthermore, for all $M$ the problem is
$\PT$-symmetric. The Schr\"odinger equation is now
\eq
-\frac{d^2}{dx^2}\psi(x)-(ix)^{2M}\psi(x)=E\psi(x)
\label{bbprobs}
\en
and again we ask for those values of $E$ at which there
is a solution along the real $x$-axis which decays at both plus and minus
infinity. Two details need extra care: for non-integer values of $2M$, the
`potential' $-(ix)^{2M}$ is not single-valued; and when $M$ hits
$2$, the naive definition of the eigenvalue problem runs
into difficulties. The first problem is easily cured by adding a
branch cut running up the positive imaginary $x$-axis. The second is
more subtle, and its resolution more interesting; it will be discussed in
greater detail in section~\ref{funode} below.

\begin{figure}[t]
\begin{center}
\includegraphics[width=0.55\linewidth]{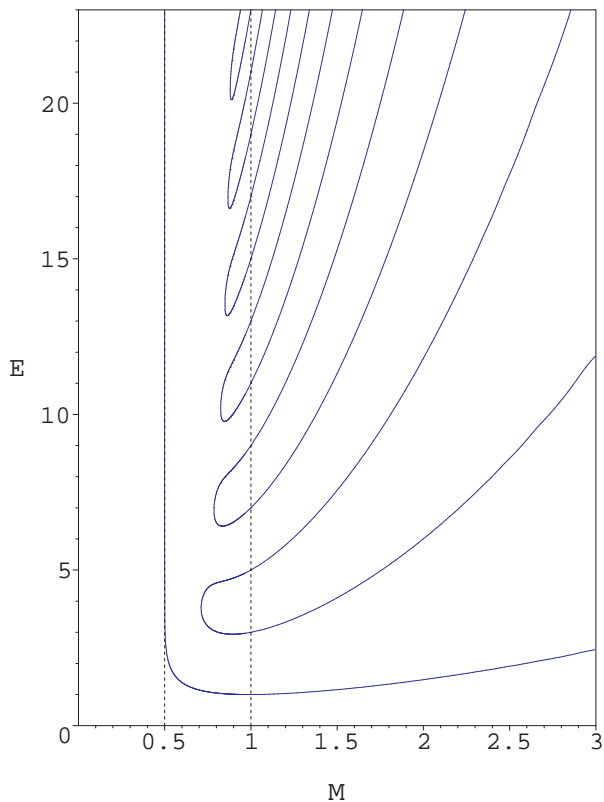}
\end{center}
\caption{$\CH_M=p^2-(ix)^{2M}$\,:
real eigenvalues as a function of $M$.\label{fig1}}
\end{figure}

For the moment, we shall agree to keep $M$
below $2$. Even so, there is an interesting surprise.
Figure~\ref{fig1} is taken from \cite{DTb} --
it reproduces the results of \cite{BB}.
Clearly, something strange occurs as $M$ decreases below $1$.
Infinitely-many real eigenvalues pair off and become complex, and
only finitely-many remain real. By the time $M$ has reached $0.75$, all
but three have become complex, and as $M$ tends to $0.5$ the last real
eigenvalue diverges to infinity. In fact, at $M{=}0.5$ the problem has no
eigenvalues at all, as can be seen by shifting $ix$ to $ix-E$
and solving the resulting equation
using an Airy function.
For $M\ge 1$, numerical results combined with various pieces of
analytical evidence indicated that the spectrum
was entirely real, and
positive, and so Bender and Boettcher generalised
conjecture 1 to

\smallskip
\noindent
{\bf Conjecture 2} \cite{BB}:
The spectrum of $\CH_M$ is real and positive for $M\ge 1$.
\medskip

\noindent
The `phase transition'
to infinitely-many complex
eigenvalues at $M=1$ can be interpreted as a spontaneous
breaking of $\PT$ symmetry \cite{BB}.
\smallskip

One further generalisation of the
Bessis-Zinn-Justin conjecture will be relevant later.
Consider

\smallskip

\noindent
{\bf Question 3:} What is the spectrum of
\eq
\CH_{M,l}=p^2- (ix)^{2M}+ l(l{+}1)/x^2
\qquad\mbox{($M$ and $l$ real, $M>0$)}\,?
\label{q3}
\en
This amounts to studying the effect of an angular-momentum-like
term $l(l{+}1)x^{-2}$ on the Bender-Boettcher problem, and it was
first investigated in~\cite{DTb}.
Note that we continue to impose boundary conditions at $x=\pm\infty$,
in the way stated just after equation (\ref{cseq}) above. With the
angular-momentum term included
we need to specify how the wavefunction should be continued
around the
singularity at $x=0$; given the choice to place a branch cut
on the positive-imaginary $x$ axis this continuation should be
done through the lower half plane. (There will be much more
discussion of boundary conditions later, so we won't go into this
detail any more for now.)
Again, a combination of numerical and analytical
work gave strong evidence for

\nobreak
\smallskip
\noindent
{\bf Conjecture 3} \cite{DTb}:
The spectrum of $\CH_{M,l}$ is real and positive for
$M\ge 1$
and
$|2l{+}1|<M{+}1$.
\medskip

\begin{figure}[t]
\begin{center}
\includegraphics[width=0.55\linewidth]{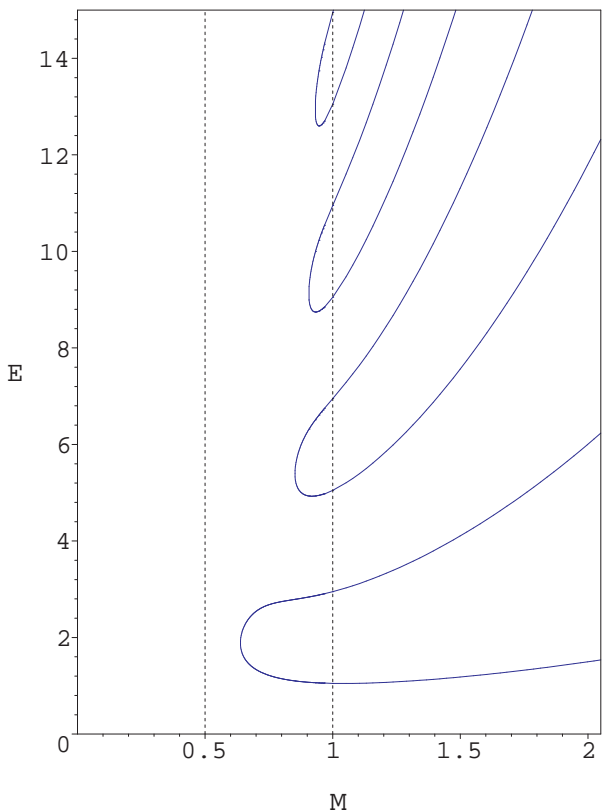}
\end{center}
\caption{$\CH_{M,l}=p^2-(ix)^{2M}+l(l{+}1)\,x^{-2}$\,:
real eigenvalues as a function of $M$ for $l=-0.025$,
$l(l{+}1)=-0.024735$.\label{fig2}}
\end{figure}

\begin{figure}[!t]
\begin{center}
\includegraphics[width=0.28\linewidth]{fm025_review.eps}
{}~~~~~~
\includegraphics[width=0.28\linewidth]{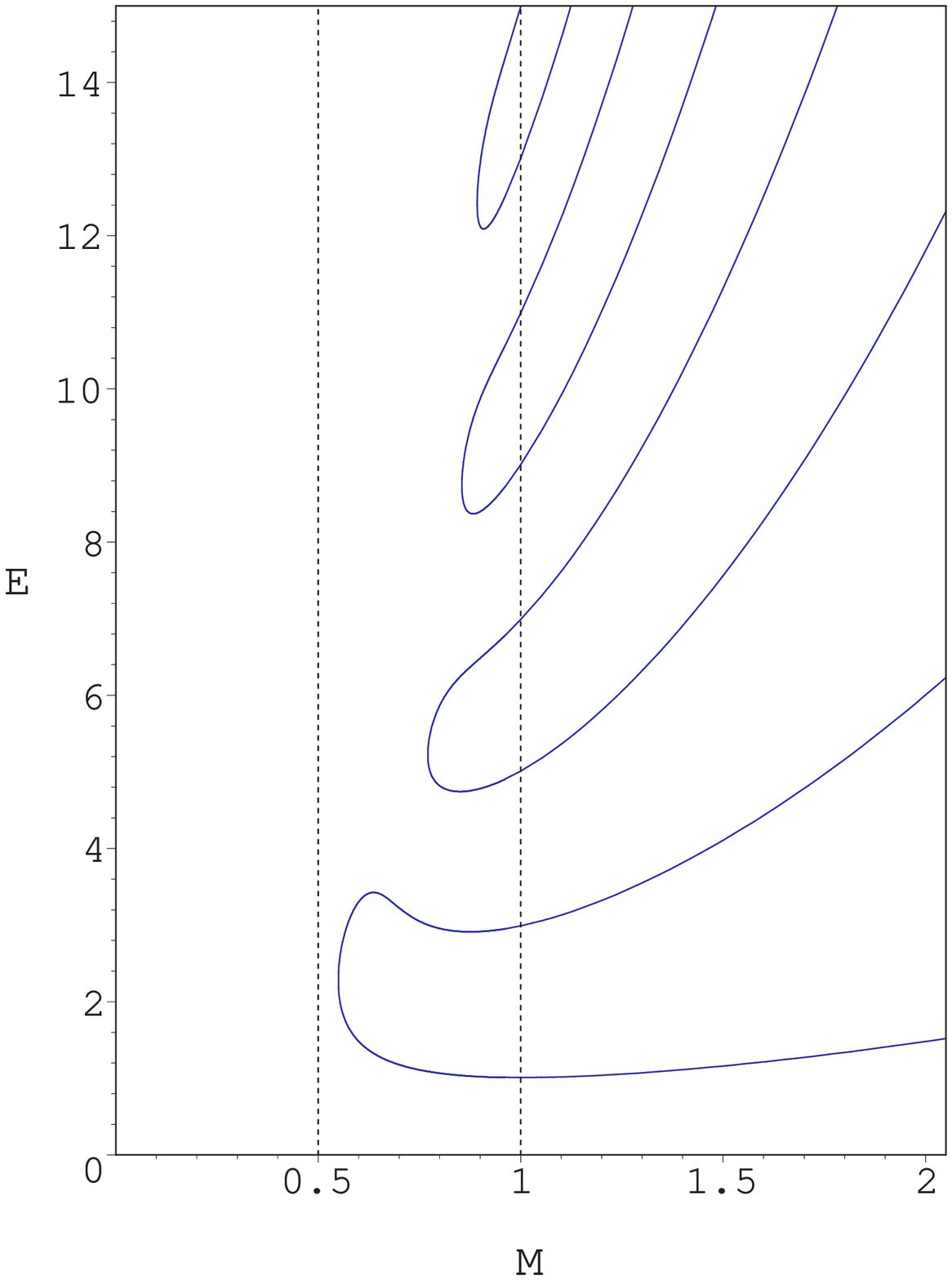}\\[1pt]
\parbox{0.33\linewidth}{~~~~~~\small\protect\ref{numerics}a: $l=-0.025$}~~~~
{}~~~~
\parbox{0.33\linewidth}{~~~~~~\small\protect\ref{numerics}b:
$l=-0.005$}\\[14pt]
\includegraphics[width=0.28\linewidth]{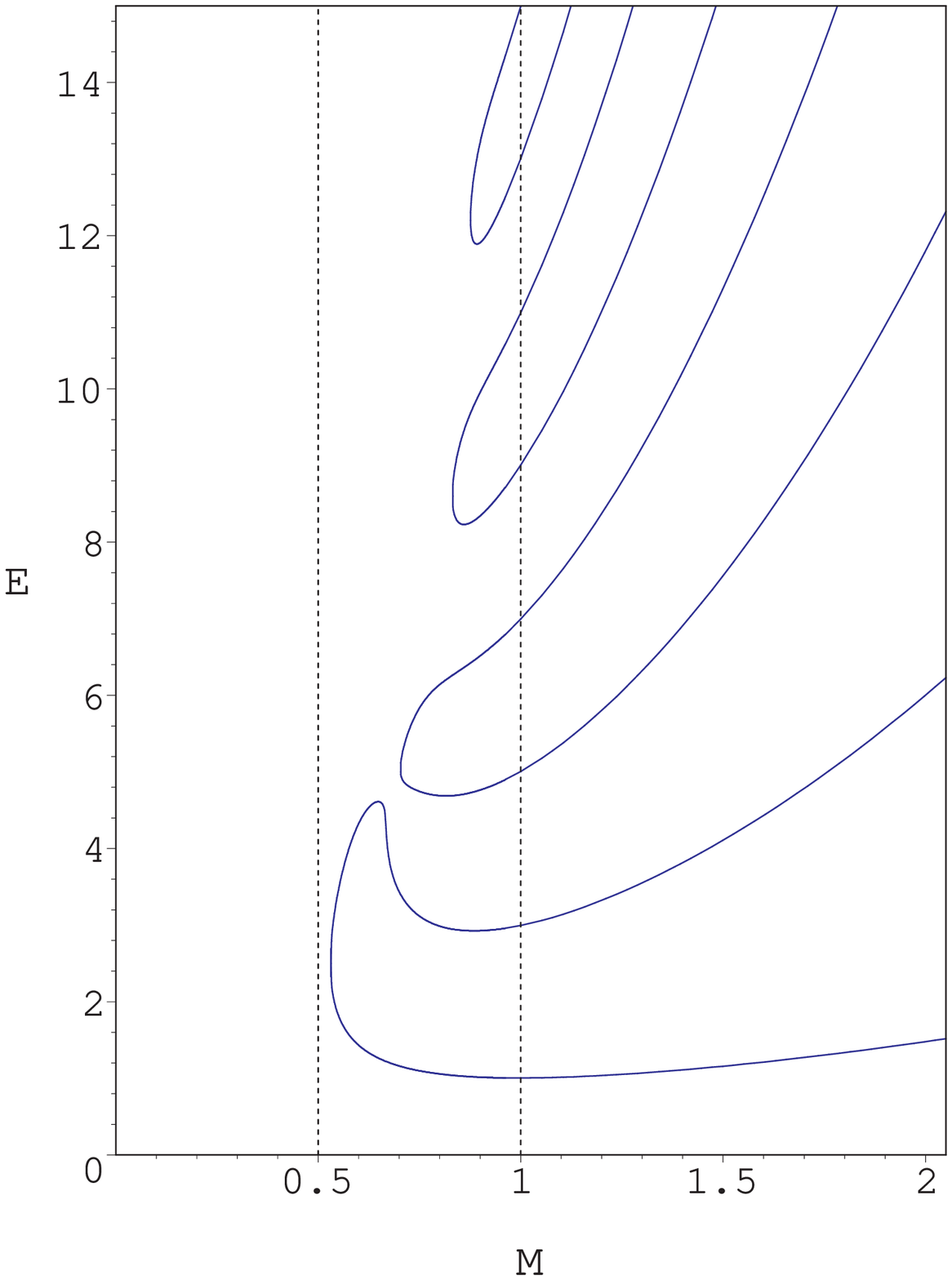}
{}~~~~~~
\includegraphics[width=0.28\linewidth]{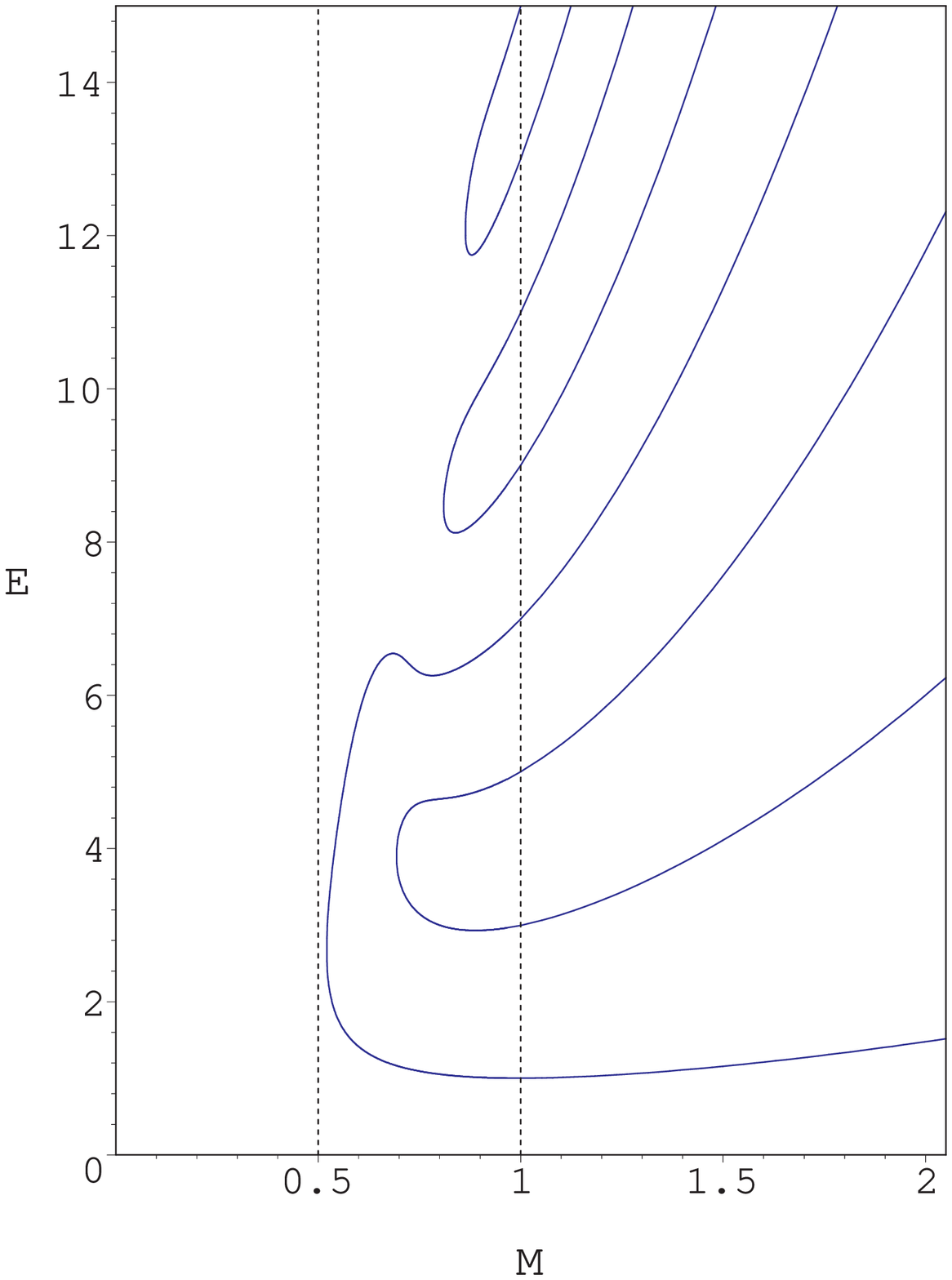}\\[1pt]
\parbox{0.33\linewidth}{~~~~~~\small\protect\ref{numerics}c:
 $l=-0.0025$}~~~~
{}~~~~
\parbox{0.33\linewidth}{~~~~~~\small\protect\ref{numerics}d:
$l=-0.0015$}\\[14pt]
\includegraphics[width=0.28\linewidth]{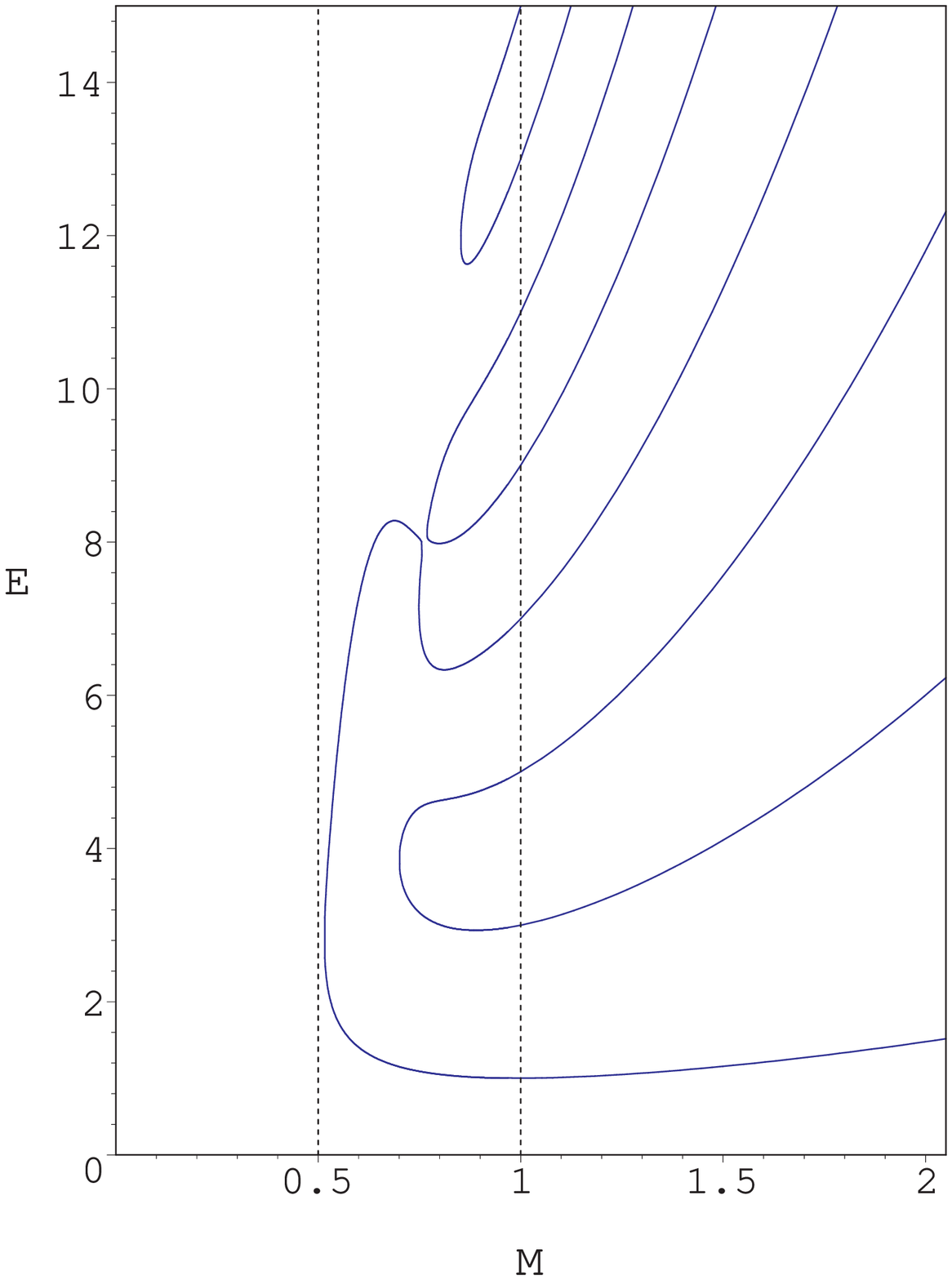} 
{}~~~~~~
\includegraphics[width=0.28\linewidth]{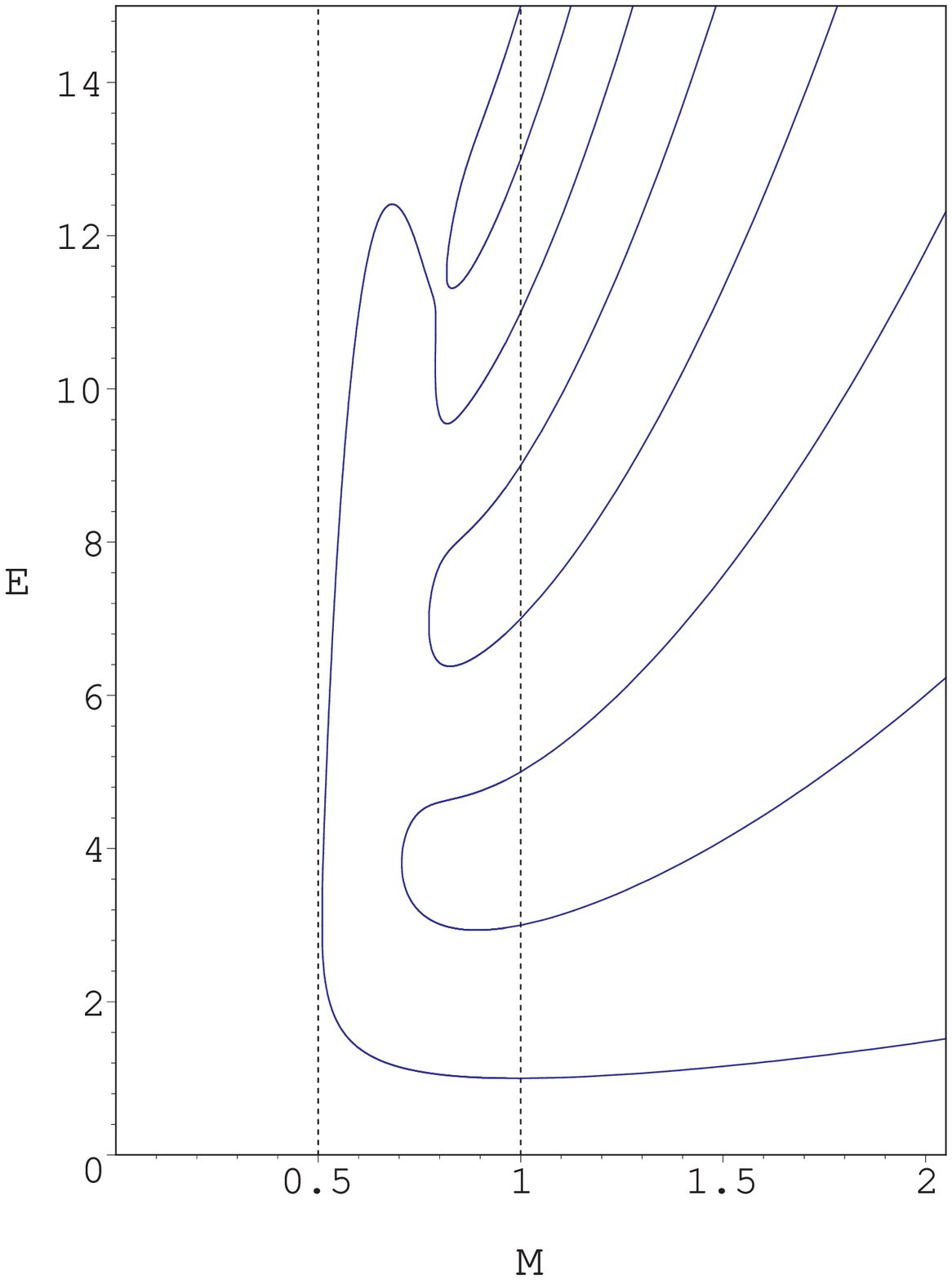}\\[1pt]
\parbox{0.33\linewidth}{~~~~~~\small\protect\ref{numerics}e:
$l=-0.001$} ~~~~
{}~~~~
\parbox{0.33\linewidth}{~~~~~~\small\protect\ref{numerics}f:
$l=-0.0005$}\\[2pt]
\end{center}
\caption{  \protect{ \label{numerics}}Real eigenvalues of
$p^2-(ix)^{2M}+l(l{+}1)/x^2$ as functions of $M$,
for various values of $l$.   }
\end{figure}

Although a small angular-momentum-like term does not have a significant
effect while $M\ge 1$ and the eigenvalues all remain real,
for $M<1$ it can make a remarkable difference to the way in which
they become complex. Figure \ref{fig2} shows the spectral plot for
$l=-0.025$, and reveals a dramatic change from the earlier $l=0$
plot: the connectivity of the real eigenvalues has been reversed, so
that while for $l=0$ the first and second excited states pair off,
for $l=-0.025$ the first excited state is instead paired with the
ground state, and so on up the spectrum.  With this in mind, it may
be hard to see how it is possible to pass between the sets of
spectra depicted in figures \ref{fig2} and \ref{fig1} simply by
varying the continuous parameter $l$ from $-0.025$ to zero.
The  mechanism  should become
clear by looking at the sample of intermediate pictures of
figure~\ref{numerics}.

\begin{figure}[!t]
\begin{center}
\includegraphics[width=0.75\linewidth]{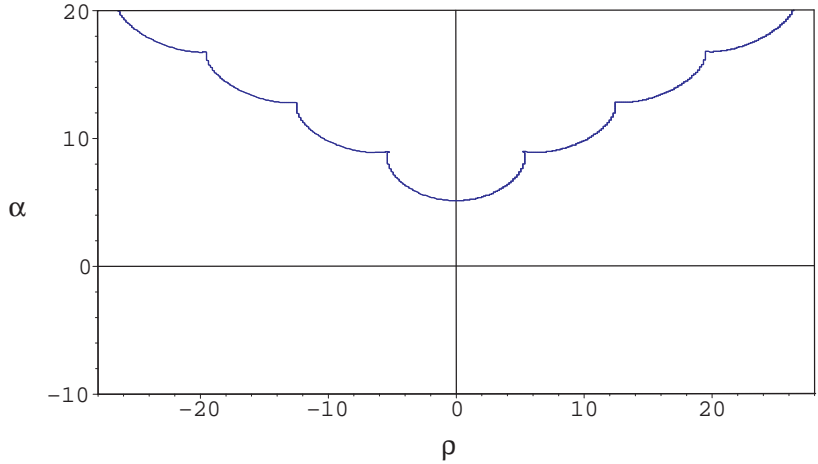}
\end{center}
\caption{$\CH_{3,\alpha,l}$\,:
region of unreality in the $(\alpha,\rho)$ plane, where
$\rho=\sqrt{3}(2l{+}1)$.}
\label{fig3}
\end{figure}

A final generalisation allows for an even richer phenomenology.
Adding an inhomogeneous term $-\alpha (ix)^{M-1}$ to the potential for
$\CH_{M,l}$ gives
a three-parameter family of problems:
\eq
\CH_{M,\alpha,l}=p^2- (ix)^{2M}-\alpha (ix)^{M-1} + l(l{+}1)/x^2.
\label{q4}
\en
Again, the first question to ask
 is whether the spectrum of $\CH_{M,\alpha,l}$ is
entirely real. Some general results
will be described later in this review, but for now we illustrate the
situation by giving
some `experimental' data for the case $M=3$. Special features of
this particular case make it desirable to trade the parameter $l$ for
$\rho:=\sqrt{3}(2l{+}1)$; this being understood, figure \ref{fig3}
shows, below the cusped line,
the region of the $(\alpha,\rho)$ plane where all eigenvalues of
$\CH_{3,\alpha,l}$ are real.
One last comment is worth making here:
the alert reader might protest that substituting $M=3$
into (\ref{q4}) results in a Hamiltonian which is manifestly real,
with a potential bounded from below, which should surely have an
entirely real spectrum for {\em all}\/ values of $\alpha$ and $l$ (or
$\rho$).  The question is a fair one, but it fails to take into
account the fact that we have implicitly imposed boundary conditions
which analytically continue those at $M=1$. This continuation takes
$M$ past the value $2$ at which we already observed that
there would be subtleties in defining the spectrum. We shall return to
this point later.

Bender and Boettcher's  observation \cite{BB} has sparked a
great deal of interest in reality properties in
non-Hermitian quantum mechanics; a (small) sample
of related work on the reality issue is provided by
references~\cite{Cannata:1998bp,Andrianov:1998vy,Fernandez:1998ex,
Bender:1998ry,DP,
 Bender:1999ds,br,Mez,DT,BCQ,Khare:2000ic,Khare:2000hk,
Bagchi:2000yz,Shin, CIRR,BDMS,BBMSS,Bagchi:2001nt, BW,
Mez1,hdy,Handy:2001rf,HX,Yan:2001gp, Bernard:2001wh,
Mostafazadeh:2002hb,Ahmed:2001na,
BBJMS,Znojil:2001ij,Mostafazadeh:2001nr,ZJ1,mon,
Mosta_2,Albeverio:2002ri,Ahmed:2002ht, Bender:2002yp, Fityo:2002wj, 
Shin:2002ay,Mosta:cpt,sw,Shin:2002vu,Jannunnis:2003tj,Yesiltas:2003yv,
Mostafazadeh:2003gz,Ahmed:2003nn,Jia:2003mr,Rosas-Ortiz:2003ix,
Basu-Mallick:2003pt,DDTcz,Ghosh:2003gz,slr,Aktas:2004nx,Shin:2004,Shin:2004ic,
Bender:2005kr, Ghosh:2005br,
Cannata:2005df,sw_det,Jones:2006qs,fariafring,Bender:2006wt}. In this
already-long article, we won't have space for any further discussion
of non-Hermitian quantum mechanics in general; more on current
issues in the field
can be found in, for example, \cite{B2005}, the review article
\cite{benderrev}, the  
conference proceedings \cite{cz}, and references therein.
Instead we just remark that
reality properties in $\PT$-symmetric quantum mechanics of
the sort described above have
turned out to be surprisingly hard to establish by
conventional means. An interesting byproduct of the ODE/IM
correspondence has been a relatively elementary proof of conjectures
1, 2 and 3, and an understanding of many features of
figure~\ref{fig3}. We shall give this proof in section
\ref{realproof} below; it relies heavily on certain functional
relations which had first made their appearance in a very different
context: the theory of integrable lattice models. A rapid
introduction to the background to this material is our next subject.



\resection{Integrable models and functional relations}
\label{sect:fnrel}

In this section we shall introduce
the integrable lattice models and quantum field theories that will be
relevant later. One particular technique for their solution, the
`functional relations' approach, will be highlighted.
We start with the lattice models, and then discuss
what is known as the `continuum limit' in preparation for the link with
quantum-mechanical problems.

\subsection{Generalities}
Lattice models provide a way to understand the
behaviours of magnets, and other substances,
which exhibit a number of distinct
`phases' depending
on the values taken by external parameters, such as the temperature.
The simplest example is the Ising model,
where a macroscopic magnet is modelled by a square lattice of
microscopic `atoms', each of which can exist in one of two states of
magnetisation, up or down. By coupling nearest-neighbour atoms
by an interaction which favours the lining-up of their
magnetisations, a system is produced which captures the tendency of
real magnets to exhibit an overall, macroscopic, magnetisation at
low temperatures, which is then lost as the temperature is increased
past some critical value. At this point, the model is said to
undergo a `phase transition', from an `ordered' to a `disordered' phase.
The qualitative features of this behaviour can often be deduced from
general arguments -- the Peierls argument \cite{peierls}
is a good example -- but for certain two-dimensional
cases it turns out that much more can be said,
and a number of important quantities
can be calculated {\em exactly}\/
as functions of the external parameters. Crudely speaking, these
are the integrable lattice models, and the first example found was
the two-dimensional Ising model, solved by Onsager in
1944~\cite{Onsager:1943jn}.

Much more can be said on this topic, but for the purposes of
this review it is best now to move forward by two decades, and to
describe the model which will be directly relevant to our
subsequent story.

\subsection{The six-vertex model and Bethe ansatz equations}
\label{aba}
We shall be interested in one particular set of integrable lattice models,
called the `ice-type' models, or, a little more prosaically,
the six-vertex models. They were first solved in 1967, by Lieb
\cite{lieb} and Sutherland \cite{sutherland}, and
are among the simplest generalisations of the
Ising model. Good places to look for further details are the book
\cite{baxterbook} by Baxter, and the short review \cite{baxterrev} by
McCoy.

\begin{figure}[ht]
\begin{center}
\includegraphics[width=0.6\linewidth]{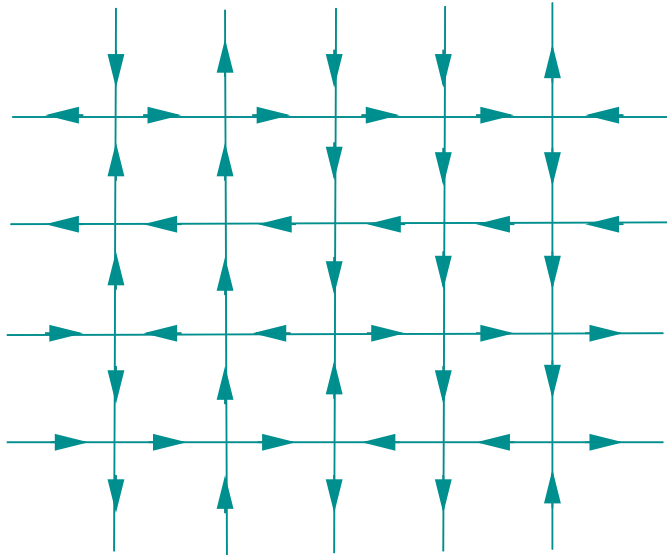}
\end{center}
\caption{A typical configuration of spins for the six-vertex model.
\label{figlattice}}
\end{figure}

The definition of the model begins with an $N\times N'$ lattice, with
periodic boundary conditions in both directions and, in order to
avoid some annoying signs later on, $N/2$ even\footnote{Sometimes
the parity of $N$ can be significant, however: see, for example,
\cite{Stroganov:2000eh}.}.
Ultimately, we
shall take a limit in which $N\to \infty$.
On each horizontal or vertical link of the
lattice, we place a spin $1$ or $2$, conveniently depicted
by an arrow pointing either right or left (for the horizontal links)
or up or down (for the vertical links), as in
figure \ref{figlattice}.

In the `ice' picture, each vertex
represents an oxygen atom, and each link a bond between neighbouring
atoms. Sitting on each bond is a hydrogen ion, lying near to one or
the other end of the bond, according to the direction of the arrow.
The `ice rule' \cite{pauling,slater} states that each oxygen atom
should have exactly two hydrogen ions close to it, and two far away.
Translated into the arrows, this implies that only those
configurations that preserve the flux of arrows through each vertex
are permitted.
This means that there are
six options for the spins around each site, or vertex, of
the lattice (hence the alternative name `six-vertex').
For real ice, all allowed configurations are
equally likely; but if we want to
generalise slightly, then we can allow for differing probabilities
for the different options. These overall probabilities are calculated
in two steps. First,
a number $W$, called a (local) Boltzmann
weight, is assigned to each local possibility.  If we further
restrict ourselves to what is called the zero-field case,
then the Boltzmann weights should be invariant under the
simultaneous reversal of all arrows,
and just three independent quantities need to be specified:\\[-5pt]
\begin{eqnarray}
W\!\left[\mbox{$\rightarrow$\raisebox{1.5ex}{$\uparrow$}%
\raisebox{-1.5ex}{\hspace{-6pt}$\uparrow$}$\rightarrow$} \right]
=
W\!\left[\mbox{$\leftarrow$\raisebox{1.5ex}{$\downarrow$}%
\raisebox{-1.5ex}{\hspace{-6pt}$\downarrow$}$\leftarrow$}
\right]
&=&~a~;\\[7pt]
W\!\left[\mbox{$\rightarrow$\raisebox{1.5ex}{$\downarrow$}%
\raisebox{-1.5ex}{\hspace{-6pt}$\downarrow$}$\rightarrow$} \right]
=
W\!\left[\mbox{$\leftarrow$\raisebox{1.5ex}{$\uparrow$}%
\raisebox{-1.5ex}{\hspace{-6pt}$\uparrow$}$\leftarrow$}\right]
&=&~b~;\\[7pt]
W\!\left[\mbox{$\rightarrow$\raisebox{1.5ex}{$\uparrow$}%
\raisebox{-1.5ex}{\hspace{-6pt}$\downarrow$}$\leftarrow$} \right]
=
W\!\left[\mbox{$\leftarrow$\raisebox{1.5ex}{$\downarrow$}%
\raisebox{-1.5ex}{\hspace{-6pt}$\uparrow$}$\rightarrow$}\right]
&=&~c~.
\end{eqnarray}

\medskip

The relative probability of
finding any given configuration is simply the product
of the Boltzmann weights at the individual vertices.
A first quantity to be calculated is the sum of these numbers over all
possible configurations -- the partition function, $Z$: \\[-7pt]
\eq
Z=\sum_{\{\sigma\}}\,\prod_{\rm sites}
W\!\left[\mbox{$\,\cdot\,$\raisebox{1.5ex}{$~\cdot\,$}%
\raisebox{-1.5ex}{\hspace{-6pt}$\cdot\,$}$~\cdot\,$}\right]~.
\en

\smallskip

\begin{figure}[ht]
\begin{center}
\includegraphics[width=0.6\linewidth]{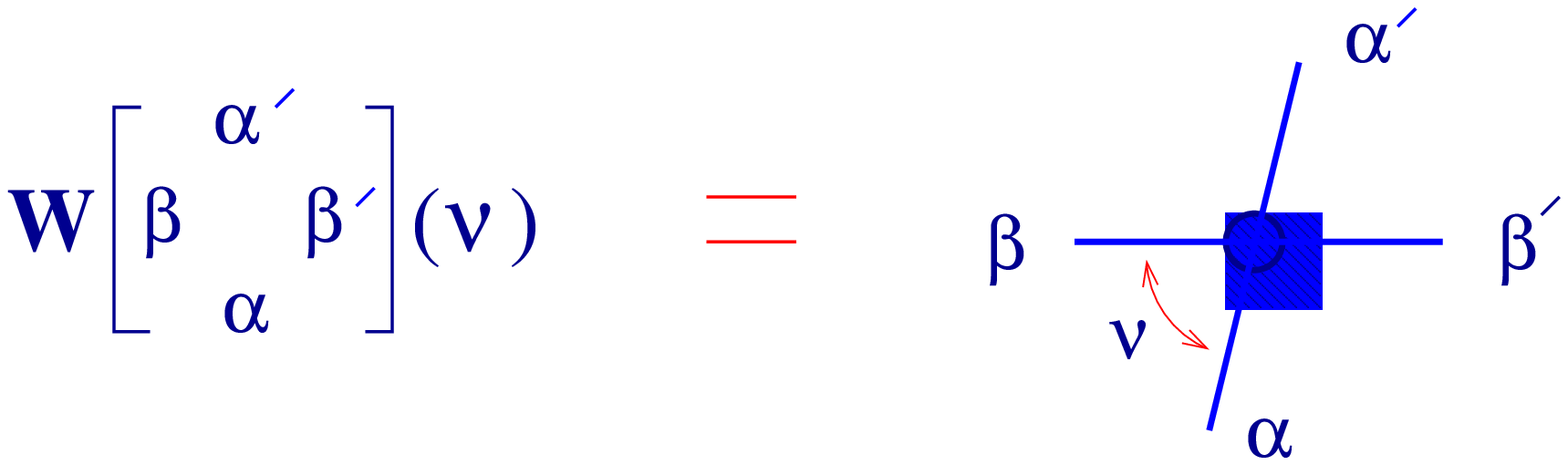}
\end{center}
\caption{The local Boltzmann weights.\label{boltz}}
\end{figure}

One of the special features of integrable models is that
quantities such as the partition function (or even better, the
free energy per site, defined in equation (\ref{freeen}) below)
can be evaluated exactly, at least in the `thermodynamic
limit', where $N$ and $N'$ both tend to infinity. The model under
discussion turns out to be integrable for all values
of $a$, $b$ and $c$. Their overall normalisation factors out
trivially from all quantities, and it is convenient to parameterise
the remaining two degrees of freedom using a pair of variables $\nu$
and $\eta$, called the spectral parameter and the anisotropy:
\eq \fl \quad
a(\nu,\eta)= \sin(\eta+i\nu)~,~~b(\nu,\eta)=
\sin(\eta-i\nu)~, {}~~c(\nu,\eta)=\sin(2\eta)~. \quad~~~
\label{abcdef}
\en
In calculations the anisotropy is usually held fixed,
but it is useful to treat
models with different values of the spectral parameter
$\nu$ simultaneously.
The weights can then be drawn as in figure \ref{boltz}.

For those familiar with integrable quantum field theories, this picture
might suggest
a relationship between Boltzmann weights for integrable
lattice models and S-matrix elements for integrable quantum field theories,
with the spectral parameter of the Boltzmann weight proportional to the
relative rapidity of two particles in the quantum field
theory\footnote{
In fact, in defining the local weights $a$, $b$ and $c$
we have shifted the spectral parameter by $i\eta$ from the value
that would be appropriate for an S-matrix, so as to give later
equations a more standard form.}.  We won't
explore this aspect much further here, but a nice discussion can be found
in~\cite{Zamrev}.

\begin{figure}[ht]
\begin{center}
\includegraphics[height=0.18\linewidth]{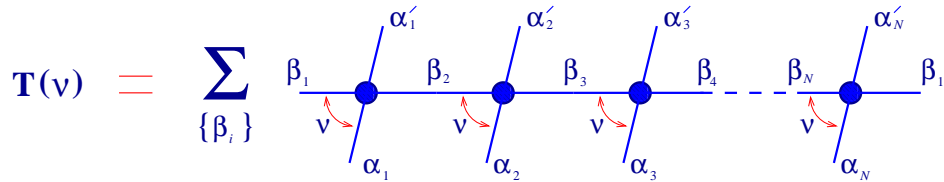}
\end{center}
\caption{The transfer matrix.\label{transf}}
\end{figure}

One popular tactic for the computation of
the partition function
employs the so-called {\em
transfer matrix}, $\TT$.
Introduce multi-indices
$\Balpha=(\alpha_1,\alpha_2\dots\alpha_N)$ and
$\Balpha'=(\alpha'_1,\alpha'_2\dots\alpha'_N)$ and set
\eq
\fl
{}~~\TT^{\Bbalpha'}_{\Bbalpha}(\nu)
{}~{}={}~{}
\sum_{\{\beta_i\}}
W \!\!\left[\mbox{$\beta_1$\raisebox{1.5ex}{$\alpha'_1$}%
\raisebox{-1.5ex}{\hspace{-14pt}
$\alpha_1$}$\beta_2$} \right]\!\!(\nu)
W \!\!\left[\mbox{$\beta_2$\raisebox{1.5ex}{$\alpha'_2$}%
\raisebox{-1.5ex}{\hspace{-14pt}
$\alpha_2$}$\beta_3$} \right]\!\!(\nu)
W \!\!\left[\mbox{$\beta_3$\raisebox{1.5ex}{$\alpha'_3$}%
\raisebox{-1.5ex}{\hspace{-14pt}
$\alpha_3$}$\beta_4$} \right]\!\!(\nu)
\dots
W \!\!\left[\mbox{$\beta_N$\raisebox{1.5ex}{$\alpha'_N$}%
\raisebox{-1.5ex}{\hspace{-17pt}
$\alpha_N$}$\beta_1$} \right]\!\!(\nu)\,.
\label{transfdefa}
\en
A pictorial representation of $\TT(\nu)$ is given in figure
\ref{transf}.
%
The definition involves a sum over one set of
horizontal links, and from the picture it is clear that the matrix
indices of
$\TT$ correspond to the spin variables sitting on the vertical
links. These can now be summed by matrix multiplication, with a
final trace implementing the periodic boundary conditions in the
vertical direction. Thus: \eq Z=\mbox{Trace}\left[\TT^{N'}\right]~. \en
Calculations can now continue via a diagonalisation of $\TT$.
Suppose that the first few eigenvalues, $t_0>t_1>\dots$ are known,
with eigenvectors $\Psi^{(0)}$, $\Psi^{(1)}$, $\dots$ :
\eq
\sum_{\Bbalpha'}
\TT^{\Bbalpha'}_{\Bbalpha}\Psi^{(j)}_{\Bbalpha'}=t_j\Psi^{(j)}_{\Bbalpha}~.
\en
Then, for example, the free energy per site in the limit
$N'\to\infty$ can be obtained as \eq f=-\fract{1}{N\!N'}\log Z=
  -\fract{1}{N\!N'}\log \mbox{Trace}\left[\TT^{N'}\right]\sim -\fract{1}{N}\log
t_0~.
\label{freeen}
\en
The eigenvalues $t_0$, $t_1$\,\dots are functions of $\nu$ and
$\eta$, and the remaining task is to find them. This appears to be a
very tough problem -- $\TT$ is a $2^N\times 2^N$ matrix, and quickly
becomes too large for even the most powerful computers to handle. It
is necessary to exploit some of the special features of the model,
and a popular technique for doing this goes by the name of the Bethe
ansatz. There are two steps:

\noindent
{\bf (i)} Make a (well-informed) \underline{guess} for
a form for an eigenvector of $\TT$, depending on a
finite number $n$ of parameters $\nu_1,\dots \nu_n$ (the
`{\em roots}').

\noindent
{\bf (ii)} Discover that this guess only works if the $\{\nu_i\}$
together solve a certain set of coupled equations (the {\em `Bethe
ansatz equations'}).

Letting $n$ vary over a finite range, and for each $n$ taking the
finite set of solutions to the corresponding Bethe ansatz equations,
should then give totality of the eigenvectors of $\TT$, or at least
all those needed to capture the $N\to\infty$ limit of the
system\footnote{We won't go into the interesting question of the
completeness of the set of BAE solutions here;
see~\cite{Barry,Baxter:2001sx} for recent discussions.}.

The justification of this procedure is an interesting story, or
collection of stories, in its own right; in appendix A we outline
one of the more elegant approaches, a technique called the algebraic
Bethe ansatz.

\begin{figure}[ht]
\begin{center}
\includegraphics[width=0.6\linewidth]{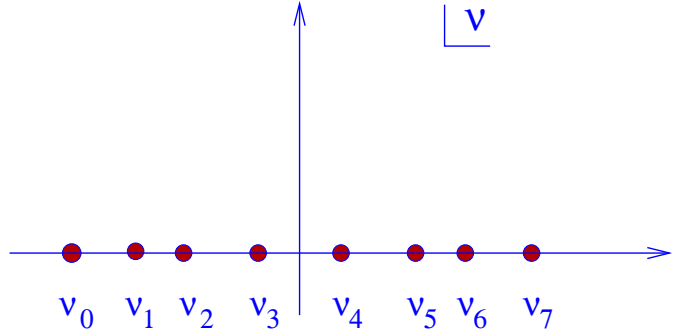}
\end{center}
\caption{The $\nu_i$'s are real for the ground state.\label{nieqfig1}}
\end{figure}

For the six-vertex model, when the dust has settled the Bethe ansatz
equations for the roots $\{\nu_1\dots \nu_n\}$ are
\eq
(-1)^n\prod_{j=1}^n\frac{\sinh(2i\eta-\nu_k+\nu_j)}{\sinh(2i\eta-\nu_j+\nu_k
)}=
-\frac{a^N\!(\nu_k,\eta)}{b^N\!(\nu_k,\eta)}~~,\qquad k=1\dots n\,.
\en
This is a set of $n$ equations for $n$ unknowns.
There is no unique solution, but rather a discrete set.
For each solution, an eigenvector $\ket{\Psi}$
of $\TT$ can be constructed, with eigenvalue
\eq
t(\nu)
=
a^N\!(\nu,\eta)\prod_{j=1}^{n}g(\nu_j-\nu)+
b^N\!(\nu,\eta)\prod_{j=1}^{n}g(\nu-\nu_j)~,
\label{tevalue}
\en
where $g(\nu):=a(\nu-i \eta,\eta)/b(\nu-i \eta,\eta)
=-\sin(2\eta+i\nu)/\sin(i\nu)$\,. To single out a given
eigenvector and eigenvalue, supplementary conditions  must be
imposed on the roots. In  particular, and this will be important
later, the  ground state eigenvalue $t_0(\nu)$ turns out to
correspond, in the parameter region $0< \eta <\pi/2$, $\Re
e(\nu)=0$, $-\eta< \Im m(\nu) <\eta$, to the Bethe ansatz  solution
with $n=N/2$ distinct real roots, packed as closely as possible and
symmetrically placed about the origin~(see for example
\cite{yang,baxterbook, abarefdv}).
 This is depicted in figure~\ref{nieqfig1}.

\subsection{Adding a twist}
\label{twist}
The periodic boundary conditions used above to
define the transfer matrix can be modified in such a way that
integrability is not spoilt (see for example
\cite{KBP,Alcaraz:1988zr,Korff:2004ev}). This does not change the free energy per
site in the thermodynamic limit, but it does modify some subleading
effects.

\begin{figure}[ht]
\begin{center}
\includegraphics[width=0.6\linewidth]{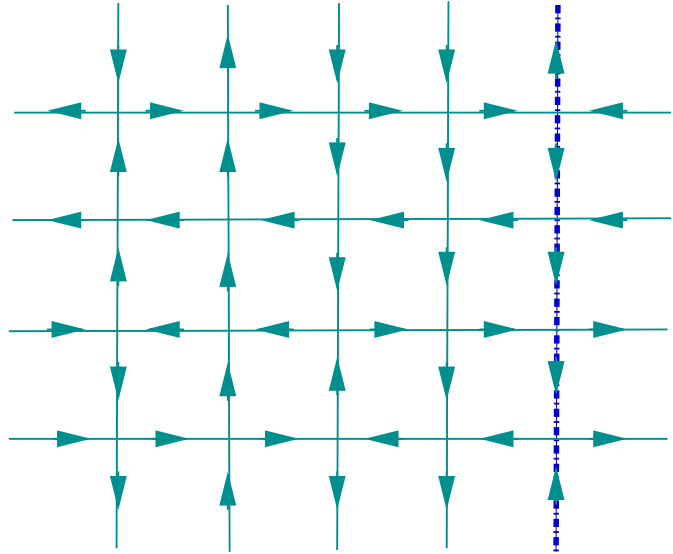}
\end{center}
\caption{The twist is introduced by modifying the local Boltzmann weights
on one column.\label{figlattice2}}
\end{figure}

The twist is introduced by modifying the local Boltzmann weights on
one column, or seam, of the lattice, say the $N^{\rm th}$ (see
figure~\ref{figlattice2}).
In the transfer matrix formulation the modification amounts to making the
substitutions
\eq
W \!\left[\mbox{$\beta_N$\raisebox{1.5ex}{$\alpha'_N$}%
\raisebox{-1.5ex}{\hspace{-17pt}
$\alpha_N$}$\rightarrow$} \right]\!(\nu)
\Longrightarrow
~e^{-i \phi}~W \!\left[\mbox{$\beta_N$\raisebox{1.5ex}{$\alpha'_N$}%
\raisebox{-1.5ex}{\hspace{-17pt}
$\alpha_N$}$\rightarrow$} \right]\!(\nu)
\en
and
\eq
W \!\left[\mbox{$\beta_N$\raisebox{1.5ex}{$\alpha'_N$}%
\raisebox{-1.5ex}{\hspace{-17pt}
$\alpha_N$}$\leftarrow$} \right]\!(\nu)
\Longrightarrow
e^{i \phi}~W \!\left[\mbox{$\beta_N$\raisebox{1.5ex}{$\alpha'_N$}%
\raisebox{-1.5ex}{\hspace{-17pt}
$\alpha_N$}$\leftarrow$} \right]\!(\nu)
\en
in the initial definition (\ref{transfdefa}) of $\TT$. The algebraic
Bethe ansatz works almost unchanged, with the result that the more
general transfer matrix $\TT(\nu,\phi)$ has  eigenvalues given by
\eq
t(\nu,\phi)= e^{-i \phi} a^N\!(\nu,\eta)\prod_{j=1}^{n}g(\nu_j-\nu)+
e^{i \phi} b^N\!(\nu,\eta)\prod_{j=1}^{n}g(\nu-\nu_j)~,
\label{Ntqa}
\en
where the  set of roots $\{\nu_1\dots \nu_n\}$ satisfy the modified
Bethe ansatz equations
\eq
(-1)^n\prod_{j=1}^n
\frac{\sinh(2i\eta-\nu_k+\nu_j)}{\sinh(2i\eta-\nu_j+\nu_k)}=
-e^{-2i\phi} \frac{a^N\!(\nu_k,\eta)}{b^N\!(\nu_k,\eta)}~~,\qquad
k=1\dots n\,.
\label{twistedBAE}
\en

\subsection{The $XXZ$ model}
There is a well-known connection between classical two dimensional
lattice models and quantum spin chains.
The six-vertex model is related to the
(spin $1/2$) $XXZ$ spin chain, a one-dimensional system of
$N$ lattice sites with a  spin variable taking
the value $1$ or $2$ at each site, with each
spin interacting  only with its neighbours.
The Hamiltonian is
\eq
H_{XXZ}
=-\frac 1 2  \sum_{j=1}^N \left(\sigma_j^x \sigma_{j+1}^x +
\sigma_j^y \sigma_{j+1}^y -\cos 2\eta \,\sigma_j^z \sigma_{j+1}^z
\right)~,
\label{hxxz}
\en
where $\sigma_{j}^{\alpha}$ represents a Pauli matrix,
\eq
\sigma^x_j=
\Big(\!\begin{array}{rr}
0&1\\[-1pt]
1&0
\end{array}\!\Big)\,,\quad
\sigma^y_j=
\Big(\!\begin{array}{rr}
0&-i\\[-1pt]
i&0
\end{array}\!\Big)\,,\quad
\sigma^z_j=
\Big(\!\begin{array}{rr}
1&0\\[-1pt]
0&-1
\end{array}\!\Big)\,,
\en
acting on the spin on the $j^{\rm th}$ lattice site.  This model
is sometimes also referred to as the Heisenberg-Ising chain, or the
spin $1/2$ anisotropic Heisenberg chain.
The six-vertex twist can be implemented by
imposing twisted boundary conditions of the form
\eq
\sigma_{N+1}^z =\sigma_{1}^z \quad,\quad   \sigma_{N+1}^x\pm
i\sigma_{N+1}^y   = e^{\pm i2\phi} \bigl(  \sigma_{1}^x\pm
i\sigma_{1}^y\bigr) \,.
\en

{}From these definitions it is not obvious that the six-vertex and $XXZ$
models should be related. However it was noticed in the first work
on the six-vertex model
\cite{lieb,sutherland} that its transfer matrix
eigenvectors coincided with those of $H_{XXZ}$\,, previously studied
in great detail by Yang and Yang \cite{yangyang}. The initial
identification rested on a coincidence of Bethe ansatz equations,
and was given a more direct explanation when Baxter \cite{Baxchain}
showed that the six-vertex transfer matrix $\TT$ and the Hamiltonian
of the spin chain are directly connected through the relation
\eq
H_{XXZ}=-i\sin 2\eta\,   \frac{d}{d\nu} \ln
\TT(\nu)\Bigr|_{\nu=-i\eta}-\frac 1 2  \cos 2\eta\
{\mathbb I}^{\otimes N}\,.
\label{Hdef}
\en
Consequently, if we are able to determine the eigenvalues $t_0,t_1
\dots$ of the transfer matrix we
 gain, for free, information on the spectrum of the $XXZ$ model.

We shall return to a description of the six-vertex spectrum
motivated by this connection with the $XXZ$ model shortly, but for now
we note that for all values of the twist parameter, the Hamiltonian
(\ref{hxxz}) commutes
with the total spin operator
$S^z = \sum_{i=1}^N \sigma_i^z/2$. Therefore the spectrum  splits
into  disjoint sectors labelled by the spin $m=0, 1, 2\dots~$; and
the true ground state lies in the $m=0$ sector.
In the six-vertex model the $XXZ$ spin sectors correspond to taking
the number of Bethe roots $n$ different from the ground state value
of $n=N/2$. The relation between the number of Bethe roots and the
$XXZ$ spin is $n=N/2-m$.

\subsection{Baxter's TQ relation}
\label{baxter}
So much, for now, for the Bethe ansatz. There is a particularly neat
reformulation of the final result, discovered by Baxter, that leads
to an alternative way to solve the model. The first ingredient is
the fact that the transfer matrices at different values of the
spectral parameter $\nu$ {\em commute}\/:
\eq
[\TT(\nu),\TT(\nu')]=0\,.
\en
(The (standard) proof of this fact is given in appendix~A.)
Therefore, the transfer matrices $\TT(\nu)$ can be simultaneously
diagonalised, with eigenvectors which are independent of $\nu$. This
allows us to focus on the individual eigenvalues $t_0(\nu)$,
$t_1(\nu)$,\,\dots~as functions of $\nu$. From the explicit form of
the Boltzmann weights and the claim that the eigenvectors are
$\nu$-independent, these functions are entire, and $i\pi$-periodic.

The second ingredient is the  claim that,
for each eigenvalue function $t(\nu)$,
there exists an auxiliary function $q(\nu)$, also entire and
(at least for the ground state)
$i\pi$-periodic, such that
\eq
t(\nu)q(\nu)=e^{-i\phi}a^N\!(\nu,\eta)q(\nu+2i\eta)+
e^{i\phi}b^N\!(\nu,\eta)q(\nu-2i\eta)\,.
\label{btq}
\en
We shall call this the TQ relation, though this phrase should really
be reserved for the corresponding matricial equation, involving
$\TT(\nu)$ and another matrix $\QQ(\nu)$, from which the above can
be extracted when acting on eigenvectors. At first sight, it is not
clear why this should encode the whole elaborate structure of the
Bethe ansatz equations -- instead of one unknown function $t(\nu)$,
we now have two, and all we know about them is that they enjoy the
curious relationship given by (\ref{btq}). But in fact this
equation, combined with the simultaneous entirety of $t(\nu)$ and
$q(\nu)$, imposes constraints on $t(\nu)$ which are so strong that
there is no need to impose the BAE as a supplementary set of
conditions. (Going further, Baxter
 was able to establish the TQ relation by an independent argument,
thereby finding
an alternative treatment of the six-vertex model which
avoided the explicit construction of eigenvectors.
He then generalised this approach
to the previously unsolved eight-vertex model, but we
shall not
elaborate this aspect any further here.)

The BAE are extracted from (\ref{btq}) as follows. Suppose,
anticipating the final result in the notation, that the zeros of
$q(\nu)$ are at $\nu_1$, \dots $\nu_n$\,.
Given that
 $q(\nu)$ is
$i\pi$-periodic, it can -- up to an irrelevant overall normalisation
-- be written as a product over these zeros as
\eq
q(\nu)=\prod^n_{l=1}\sinh(\nu-\nu_l)~.
\label{Q}
\en
{}From (\ref{btq}), $t(\nu)$
is fixed by $q(\nu)$, and from (\ref{Q}), $q(\nu)$ is fixed by the set
$\{\nu_i\}$. To determine the $\{\nu_i\}$,
set $\nu=\nu_i$ in (\ref{btq}). On the left-hand side we then have
$t(\nu_i)$, which is nonsingular since $t(\nu)$ is entire, multiplied by
$q(\nu_i)$ which is zero by (\ref{Q}). Thus the left-hand side vanishes, and
rearranging we have
\eq
\frac{q(\nu_i-2i\eta)}{q(\nu_i+2i\eta)}=
-e^{-2i\phi}\frac{a^N\!(\nu_i,\eta)}{b^N\!(\nu_i,\eta)}~~
\en
or, using (\ref{Q}) one more time,
\eq
(-1)^n\prod^n_{l=1}\frac{\sinh(2i\eta-\nu_i+\nu_l)}%
{\sinh(2i\eta-\nu_l+\nu_i)}=
-e^{-2i\phi}\frac{a^N\!(\nu_i,\eta)}{b^N\!(\nu_i,\eta)}~~,\qquad
i=1\dots n~.
\en
These are exactly the Bethe ansatz equations (\ref{twistedBAE}) for
the problem, with the $\nu_i$ the roots. The expression for $t(\nu)$
implied by (\ref{btq}) then matches the formula (\ref{Ntqa}), as
would have been found from a direct application of the Bethe ansatz.
\subsection{The quantum Wronskian}
\label{qwronk}
In this review, the TQ relation will be our main tool in making the
link with the theory of ordinary differential equations. However,
there are other sets of functional equations associated with the
six-vertex model which can be equally important. In principle these
can all be obtained first as operator equations~(see for example
\cite{Bax2, baxterbook, BLZ2,Korff:2005AMS}), and then turned into functional relations by specialising
to individual eigenvectors. However a full discussion of this would
take us too far afield, so instead we shall profit from the fact
that we have already obtained the TQ relation from the lattice
model, and give some indications as to why these further properties
should hold.

We continue to consider the model with general
(non-zero) twist, and discuss an important consequence of the
identity~(\ref{inv}):
\eq
t_0(\nu,\phi)=t_0(\nu, -\phi) \equiv t_0(\nu,|\phi|)~,
\label{sym}
\en
where $t_0$ is the  ground state eigenvalue of $\TT(\nu,\phi)$
(with $N/2$ even) and  $\phi$ is the twist parameter. In view of
(\ref{sym}), the following two TQ relations hold simultaneously:
\bea
 \!\!\!\!\!\!\! t_0(\nu,|\phi|)\,q_0(\nu,\phi) &=& a^N\!(\nu, \eta)  e^{-i
\phi} q_0(\nu+ 2 i \eta, \phi)+
 b^N\!(\nu, \eta)  e^{i \phi} q_0(\nu- 2 i \eta, \phi)\,;
\label{tq111} \\
 \!\!\!\!\!\!\!  t_0(\nu,|\phi|)\,q_0(\nu,-\phi) &=& a^N\!(\nu, \eta)  e^{i
\phi} q_0(\nu+ 2 i \eta, -\phi)+
 b^N\!(\nu, \eta)  e^{-i \phi} q_0(\nu- 2 i \eta, -\phi)
\label{tq222}
\eea
where $q_0$ is the corresponding ground state eigenvalue of the
corresponding matrix $\QQ(\nu, \phi)$. 
As equations for $q$, these only differ in the way that the twist
factors $e^{\pm i\phi}$ appear, and even this difference can be
eliminated by defining
\eq
\tilde q_0(\nu,\phi):= e^{-\nu\phi/(2\eta)}q_0(\nu,\phi)
\label{qtdef}
\en
so that\footnote{In the field theory context, $\tilde q_0$ will
correspond to the vacuum eigenvalue of the operator Bazhanov, Lukyanov
and Zamolodchikov denote ${\bf Q}$, while $q_0$ is the vacuum eigenvalue
of their $\bf A$.}
\bea
\fl
\quad t_0(\nu,|\phi|)\,\tilde q_0(\nu,\phi) &=& a^N\!(\nu, \eta)
\tilde  q_0(\nu+ 2 i \eta, \phi)+
 b^N\!(\nu, \eta) \tilde  q_0(\nu- 2 i \eta, \phi)\,;
\label{tq333} \\
\fl \quad
t_0(\nu,|\phi|)\,\tilde q_0(\nu,-\phi) &=& a^N\!(\nu, \eta)
\tilde q_0(\nu+ 2 i \eta, -\phi)+
 b^N\!(\nu, \eta)  \tilde  q_0(\nu- 2 i \eta, -\phi)\,.
\label{tq444}
\eea
Thus $\tilde q_0(\nu,\phi)$ and
$\tilde q_0(\nu,-\phi)$ both solve the single
functional equation
\eq
t_0(\nu,|\phi|)\,\tilde q(\nu) = a^N\!(\nu, \eta)
\tilde  q(\nu+ 2 i \eta)+
 b^N\!(\nu, \eta)  \tilde  q(\nu- 2 i \eta)\,.
\label{tq555}
\en
This is a finite-difference analogue of a second-order ordinary
differential equation, and so it should have two linearly
independent solutions; equations (\ref{tq333}) and (\ref{tq444})
confirm that this is indeed the case. The quasi-periodicity in $\nu$
induced by the definition (\ref{qtdef}), combined with the
periodicity of the `potential' $t_0$, means that $\tilde
q_0(\nu,\phi)$ and $\tilde q_0(\nu,-\phi)$ can be interpreted as the
two Bloch-wave solutions to (\ref{tq555})
\cite{BLZ2,Krichever:1996qd}. Just as in the continuum case, given
two solutions to a single second-order equation it is natural to
construct their Wronskian. To this end, we can multiply
(\ref{tq333}) by $\tilde q_0(\nu,-\phi)$ and (\ref{tq444}) by
$\tilde q_0(\nu,\phi)$, subtract and regroup terms by defining
\eq
\Delta(\nu):=
\tilde q_0(\nu{+}i\eta,-\phi)\tilde q_0(\nu{-}i\eta,\phi)-
\tilde q_0(\nu{+}i\eta,\phi)\tilde q_0(\nu{-}i\eta,-\phi)
\en
to find
\eq
0=a^N\!(\nu,\eta)\Delta(\nu{+}i\eta)-
b^N\!(\nu,\eta)\Delta(\nu{-}i\eta)\,.
\label{pereq}
\en
Recalling from (\ref{abcdef}) the definitions
$a(\nu,\eta)=\sin(\eta+i\nu)$, $b(\nu,\eta)= \sin(\eta-i\nu)$,
and the fact that $N$ is even,
(\ref{pereq}) implies that the function ${\cal W}(\nu):=
\Delta(\nu)/\sinh^N(\nu)$ is
periodic with period $P=2i\eta$. However, from
(\ref{Q}) and (\ref{qtdef}),
we see that  ${\cal W}(\nu)$
also has the period $P'=2i\pi$.
For $P'/P=\eta/\pi$ irrational, ${\cal W}(\nu)$
must therefore be constant; by
continuity in $\eta$, ${\cal W}(\nu)$ is constant for all values of
 $\eta$.

Evaluating ${\cal W}$ at $\nu \rightarrow \infty$ gives an
identity, the finite-lattice version of the quantum Wronskian
relation discussed by Bazhanov, Lukyanov and Zamolodchikov
in~\cite{BLZ2}. Re-expressed in terms of $q_0(\nu,\phi)$ it reads
\eq
\fl
\ e^{-i \phi} q_0(\nu{+} i \eta, \phi) q_0(\nu{-} i \eta, -\phi) -e^{
i \phi} q_0(\nu{+}  i \eta, -\phi) q_0(\nu{-} i \eta, \phi)=  -2i
\sin(\phi) \, \sinh^N(\nu).
\label{qw}
\en

In deriving this result we have only treated the ground state
eigenvalues of $\TT$ and $\QQ$, and we have also assumed that the
twist $\phi$ is non-zero. If these restrictions are dropped, a
number of important subtleties arise, particularly in the so-called
`root of unity' cases when $P'/P$ is rational. For more extensive
discussions which address some of these issues, see, for example,
\cite{Pronko:1998xa} and~\cite{Korff:2004mv, Korff:2005vj}.

\subsection{The fusion hierarchy and its truncation}
\label{fusion}
A further functional equation results if we
multiply the simultaneous TQ relations (\ref{tq333}) and
(\ref{tq444}) by $\tilde q_0(\nu{-}2i\eta,-\phi)$ and $\tilde
q_0(\nu{-}2i\eta,\phi)$, respectively, and subtract. We then obtain
not the quantum Wronskian, but instead an alternative expression for
$t_0$ in terms of ${\tilde q}_0$\,:
\bea
\!\!\!\!\!\!\!\!\!\!\!t_0(\nu,|\phi|)
&=&
 a^N\!(\nu, \eta)\,
{\tilde q_0(\nu{+} 2i\eta,-\phi)\tilde q_0(\nu{-}2i\eta,\phi) -
\tilde q_0(\nu{+} 2 i\eta,\phi)\tilde q_0(\nu{-}2i\eta,-\phi) \over
\tilde q_0(\nu,\phi)\tilde q_0(\nu{-} 2 i \eta,-\phi) -
\tilde q_0(\nu,-\phi)\tilde q_0(\nu{-} 2i\eta,\phi) }\nn\\[3pt]
&=& \frac{-1}{2i\sin\phi}\,\Bigl( \tilde q_0(\nu{+} 2 i
\eta,-\phi)\tilde q_0(\nu{-}2i\eta,\phi)-\tilde q_0(\nu{+}
2i\eta, \phi)\tilde q_0(\nu{-}2i\eta,-\phi) \Bigr)
\label{tt2}
\eea
using the previously-obtained formula for the quantum Wronskian for
the last equality. This is a sign that the quantum Wronskian fits
into a hierarchy of relations which we now describe. It is
convenient to change the normalisations slightly. Defining
\eq
\vec{q}^{\,(k)}:= {1\over\sqrt{-2i \sin\phi}} \Big( e^{-ik\phi/2}
q_0(\nu{-}i k\,\teta,\phi),e^{i k\phi/2} q_0(\nu{-}i
k\,\teta,-\phi)\Big)^T
\en
with  $\teta= -\eta{+}\pi/2$
and
\eq
{\cal W}[k,-k](\nu):=\det(\vec{q}^{\,(k)}, \vec{q}^{\,(-k)})=\left|
\left|\begin{array}{cc}
(q^{(k)})_1& (q^{(-k)})_1\\
(q^{(k)})_2& (q^{(-k)})_2
\end{array}\right| \right| ~,
\en
we set
\eq
t^{(k/2)}(\nu):={\cal W}[k+1,-k-1](\nu)~,~~~k=-1,0,1,2,3\dots~.
\label{tm}
\en
Then $t^{(-1/2)}(\nu)=0$, and (\ref{qw}) and (\ref{tt2}) imply
\eq
t^{(0)}(\nu)=[i\cosh(\nu)]^N~,\quad t^{(1/2)}(\nu)=t_0(\nu)\,.
\en
We can then use the following  Pl\"ucker-type relation
\eq
\fl \quad
\det(\vec{a}_0, \vec{a}_1) \det(\vec{b}_0,\vec{b}_1)=
\det(\vec{b}_0,\vec{a}_1) \det(\vec{a}_0, \vec{b}_1)+
\det(\vec{b}_1,\vec{a}_1) \det(\vec{b}_0,\vec{a}_0) \label{pl}
\en
and the property
\eq
{\cal W}[k+a,-k+a](\nu)={\cal W}[k,-k](\nu-i a\teta)
\en
to show that
\bea
&&
t^{(m)}(\nu-i \teta)t^{(m)}(\nu+i \teta)=\nn\\
&&\qquad t^{(0)}(\nu-i (2m{+}1) \teta) t^{(0)}(\nu+i (2m{+}1)
\teta)+ t^{(m{-}1/2)}(\nu)t^{(m{+}1/2)}(\nu)~,
\label{fus1}
\eea
where the index $m$ takes the half-integer values $1/2$, $1$, $3/2$,
$\dots$~. Another set of  relations among the functions
$t^{(m)}(\nu)$ is also a simple consequence of the identity
(\ref{pl}):
\bea
  && \hspace{-1cm}
t^{(1/2)}(\nu)t^{(m)}(\nu-i(2m{+}1) \teta)=\nn\\
\fl
&&  t^{(0)}(\nu-i\teta)t^{(m{+}1/2)}(\nu-i2m
\teta)+t^{(0)}(\nu+i \teta)t^{(m{-}1/2)}(\nu-i(2m{+}2) \teta)~.
\label{fus2}
\eea
The sets of
functional relations~(\ref{fus1}) and (\ref{fus2}) are called
fusion hierarchies \cite{
Kulish:1981bi, kiresh,KPfu,KNSa,BLZ2}. The name comes from the
fact that they can also be
obtained by a process known as `fusion' of the basic transfer matrix
$\TT$,  without introducing the auxiliary function $q(\nu)$ \cite{KSS}.

An important phenomenon occurs at rational values of $\eta/\pi$,
known as {\em truncation}\/ of the fusion hierarchy.
Here we shall just mention the case $\eta= { \pi M \over 2M+2}$
($\teta={ \pi \over 2M+2}$) with $2M \in \ZZ^+$ and $\phi= {\pi \over
2M+2} $. Due to the $i\pi$ periodicity of $q(\nu)$ we have
\eq
t^{(M+1/2)}(\nu)=0
\label{tr1}
\en
and also
\eq
\fl   \begin{array}{ccc}
\vec{q}^{(2M+1)}&=&{-i\over\sqrt{-2i \sin\phi}} \Big( e^{
\fract{i\pi}{4M+4}} q_0(\nu{+} \fract{i\pi}{2M+2} ,\phi),-e^{-
\fract{i\pi}{4M+4}} q_0(\nu{+}\fract{i\pi}{2M+2} ,-\phi)\Big)^T~; \\
 \ \ \vec{q}^{(-2M-1)}&=&{i\over\sqrt{-2i \sin\phi}} \Big(  e^{-
\fract{i\pi}{4M+4}} q_0(\nu{-} \fract{i\pi}{2M+2} ,\phi),-e^{
\fract{i\pi}{4M+4}} q_0(\nu{-} \fract{i\pi}{2M+2} ,-\phi)\Big)^T~,
 \label{qq1}
\end{array}
\en
which on comparing with
\eq
\fl  \begin{array}{ccc}
 \vec{q}^{(1)} &=&{1 \over\sqrt{-2i \sin\phi}} \Big(  e^{-
\fract{i\pi}{4M+4}} q_0(\nu{-} \fract{i\pi}{2M+2} ,\phi),e^{
\fract{i\pi}{4M+4}} q_0(\nu{-} \fract{i\pi}{2M+2} ,-\phi)\Big)^T~;  \\
 \ \ \vec{q}^{(-1)}&=&{1\over\sqrt{-2i \sin\phi}} \Big(  e^{
\fract{i\pi}{4M+4}} q_0(\nu{+} \fract{i\pi}{2M+2} ,\phi),e^{-
\fract{i\pi}{4M+4}} q_0(\nu{+} \fract{i\pi}{2M+2} ,-\phi)\Big)^T~,
\end{array}
\en
shows that
\eq
t^{(M)}(\nu)=\det(\vec{q}^{\,(2M+1)}, \vec{q}^{\,(-2M-1)})
=\det(\vec{q}^{\,(1)}, \vec{q}^{\,(-1)})=t^{(0)}(\nu)~.
\label{tr22}
\en
Thus the infinite fusion hierarchy has been reduced, or {\em
truncated}, to
a finite set of functional equations (a T-system)
constraining the $t$-functions. It is also easy to check that the
relations  (\ref{fus1}), (\ref{tr1}) and (\ref{tr22}) together
imply the symmetry
\eq
t^{(m)}(\nu)=t^{(M-m)}(\nu) \quad , \quad m=0,1/2,\dots M/2~.
\label{td}
\en

Truncation is important because it leads to closed sets of
functional relations. Subject to suitable analyticity properties,
these can be converted into sets of integral equations which
allow the model to be solved -- an example of this procedure will be
given in appendix \ref{TBAsection}, in the slightly-simpler
context of the large-$N$ limit.
At generic values of $\phi$ and rational
$\eta/\pi$, truncation is also possible, though it has to be implemented
in a more complicated
way to ensure that the analyticity properties
required for the derivation of the integral equations continue to
hold~\cite{BLZ2,KSS,Tateo:1995sz}.

\subsection{Continuum limit of lattice models}
\label{clim}
Very often, physicists are particularly interested in the behaviour
of lattice models in the so-called `thermodynamic limit', when the
size of the system tends to infinity. If this limit is taken in a
suitable way near to a
phase transition, short-distance details of the model get washed
away. The resulting behaviour is then said to be
`universal', and since it does not depend on the precise formulation
of the model, it also
tells us about more realistic systems near to their phase transitions,
beyond the idealised models discussed so far.
If we continue to measure distances by the number of lattice sites,
all of the universal properties will be found in the long-distance
asymptotics of quantities such as the transfer matrix eigenvalues
$t_0$. To focus on these features, it is
common to introduce a dimensionful lattice spacing $d$ -- up to now this
has been equal to one -- and then let this spacing tend to zero while
keeping the `physical' width of the lattice, $N d$, finite.
This process -- which may have to be accompanied by a suitable tuning,
or {\em renormalisation}, of parameters to ensure that the objects of
interest retain finite values -- is known as taking the continuum limit.
It has the additional feature that the limiting theory can
often be studied using techniques from quantum field theory.

The six-vertex models lie at a phase transition of the more general
eight-vertex model for all $0<\eta<\pi/2$, and so are well-suited to
the taking of this limit. One place where universal behaviour can
then be detected is in the behaviour of the logarithm of the
dominant eigenvalue of the transfer matrix, $t_0$. As
$N\to\infty$\,,
\eq
\ln t_0(N)=-f\,N+\frac{\pi c_{\rm eff}}{6 N} +\dots\,.
\label{tf}
\en
The constant $f$ is simply the large-$N$ limit of the free energy
per site (\ref{freeen}), and such a term is expected on general
grounds. The next term is the signal of a phase transition: it
depends only algebraically on the system size, and its general form
is a consequence of the scaling symmetry characteristic of
(second-order) phase transitions. If we now introduce the lattice
spacing $d$, replace $N$ by $L:= Nd$  and define the
subtracted/rescaled free energy to be
\eq
F :=-\ln t_0(L) - fL
\en
then the `$\dots$' terms in (\ref{tf}) give vanishing  contributions
as $d \rightarrow 0$ with $L$ held fixed and in this limit
\eq
F(L)=-\frac{\pi c_{\rm eff}}{6 L}\,,
\label{tfc}
\en
where $L$ is now a continuous (positive) number.  This is the
expected  behaviour  of the free energy for a conformal field theory
(CFT) on an infinite cylinder with circumference $L$.
In unitary theories with periodic boundary conditions, the
proportionality constant $c_{\rm eff}$ coincides with the standard
Virasoro conformal central charge $c$.
The continuum limit of the six-vertex
model and the $XXZ$ spin chain is described by a unitary CFT with central
charge  $c^{\rm
  6V}=1$. The effective central charge  is
$c_{\rm eff}^{\rm 6V}=c^{\rm 6V}=1$ in the periodic  case, or
\eq
c_{\rm eff}^{\rm T6V} =1-{6 \phi^2 \over \pi(\pi - 2 \eta)}  <1
\en
in the twisted cases~\cite{Hamer:1987ei,Destri:1989dh}.

The eigenvector $\Psi^{(0)}$ corresponding to the dominant
eigenvalue $t_0$ just discussed is called the ground state of the
model, and most of the results to be described later concern the
Bethe roots for this state. However it is worth noting that the
conformal field theory dictates the behaviour of {\em all} states in
the continuum limit of the model \cite{Cardy:1984rp,Cardy:1986ie}. The
remaining states are sometimes called `excited states'; they can be
assigned an `energy' as minus the rescaled logarithm of the
corresponding eigenvalue, just as was done for the ground state  to
obtain the free energy $F$. The spectrum of the six-vertex model or
$XXZ$ model with periodic boundary conditions then consists of
states with energies that behave as
\cite{Cardy:1986ie,Alcaraz:1987ix,Woynarovich:1987ge,Karowski:1988kz,
Alcaraz:1988zr}
\eq
F_{|\{m_i \},\{m'_i\}, k, k' \rangle}(L)=\xi \lf ( -\frac{\pi c_{\rm
eff}}{6 L} +\frac{ 2\pi}{L}(x_{k,k'} +m+m') \ri)~,
\label{tfc2}
\en
where $k$, $k' \in \ZZ$,  $m=\sum_i m_i$,  $m'=\sum_i m'_i$  with
$m_i$ and $m'_i$  non-negative integers, and
$x_{k,k'}=k^2x+k'^2/(4x)$ with $x=(\pi-2\eta)/2\pi$. The quantity
$\xi$ is a model-dependent parameter (the `velocity of light'). It
is $1$ for the six-vertex model and, in the notation used here,
${\pi \sin 2\eta} /2\eta$ for the $XXZ$ model. (Alternatively, it
could be made equal to $1$ simply by multiplying the $XXZ$
Hamiltonian by an overall factor~\cite{vg}.) For each pair $k,k'$,
the state with $m=m'=0$ is associated in the CFT with what is called
a `primary field' with scaling dimension $x$. The states with other
values of $m$ and $m'$ are called `descendants' of this field, and
the even spacing of the energies of these descendants is a
characteristic feature of the spectrum of a CFT. A similar
tower-like structure also arises when twisted boundary conditions
are imposed \cite{Alcaraz:1987ix}. The terminology of primary fields
and descendants relates to an underlying symmetry of the conformal
field theories, the infinite-dimensional Virasoro algebra. For more
on these topics, see, for example, \cite{FMS}.

It turns out that the universal behaviour described above
corresponds to a special limit of the TQ and Bethe ansatz equations,
in which their forms simplify. For  this, it will be more convenient
to re-express the results obtained in sections \ref{aba} and
\ref{baxter} using an alternative set of variables.  Setting
\eq
E'_i=e^{2\nu_i}~~,~~~~ \omega=-e^{-2i\eta}= e^{2i\teta}\,,
\label{enu}
\en
the BAE become
\eq \fl \qquad \qquad
\prod^n_{l=1}\left(\frac{E'_l-\omega^{2}E'_i}{E'_l-\omega^{-2}E'_i}\right)
= -\omega^{2n-N} e^{-i 2 \phi}
\left(\frac{1+\omega\,E'_i}{1+\omega^{-1}E'_i}\right)^N \,,\qquad
i=1\dots n~.
\en
{}If we continue to concentrate on the ground state, then, as
already mentioned, $n=N/2$, and all of the $\nu_i$ lie on the real
axis. This translates into all of the $E'_i$ being real and
positive, and eliminates the factor $\omega^{2n-N}$ from the RHS of
the BAE. In the limit $N\to\infty$, the number of roots needed to
describe the ground state diverges. This complication is to some
extent compensated by the fact that the Bethe ansatz  equations for
the `extremal' roots, those $\nu_i$ lying to the furthest left or
right along the real axis, simplify in this limit, at least for
$\eta>\pi/4$. Since the  left and right sets  of extremal roots are
constrained  by  the   symmetry
\eq
q_0(-\nu,\phi) =q_0(\nu,-\phi) \leftrightarrow \nu_i(\phi) =
-\nu_{N/2+1-i}(-\phi)~,
\en
without loss of generality we shall concentrate on the  left
edge only.

The left edge of the root distribution tends to $-\infty$, as
$-\frac{2\eta}{\pi}\log N$. (This behaviour can be extracted, for
example, from the results of \cite{KBP}.) Hence the lowest-lying
$E'_i$ scale to zero as
\eq
E'_i \sim E_i N^{-4\eta/\pi}~.
\en
To capture their behaviour, replace each $E'_i$
with $N^{-4\eta/\pi}E_i$ in the BAE, and then hold the
$E_i$ finite as $N\to\infty$. The BAE simplify to the following:
\eq
\prod^{\infty}_{l=1}
\left(\frac{E_l-\omega^{2}E_i}{E_l-\omega^{-2}E_i}\right)
= - e^{- 2i\phi} ~,\qquad i=1\dots \infty~.
\label{simplebae}
\en
(For $\eta\le\pi/4$, the product must be regulated to ensure
convergence in the $N\to\infty$ limit, which complicates the story.
We won't discuss this any further here, except to remark that it
corresponds to the leaving of the semiclassical domain mentioned
below.)

A similar limit can be performed on $q_0(\nu)$, taking care to
adjust its normalisation to ensure a finite and non-zero result as
$N\to\infty$\,:
\eq \fl \quad
q_0(\nu) \rightarrow q_0(E) :=
\lim_{N \rightarrow \infty} \Big{[} e^{N\nu/2} q_0(\nu)\Big{]}_{\nu=
\fract{1}{2} \ln(E N^{-4\eta/\pi})}
=
\prod^{\infty}_{l=1} \left( 1 - {E \over E_l} \right)~.
\en
Taking the same limit with $t_0(\nu)$ and in
general on the functions $t^{(n)}(\nu)$ results in a
simplification of the TQ relation to
\eq
t_0(E)q_0(E)=e^{i \phi} q_0(\omega^2 E)+ e^{-i \phi} q_0(\omega^{-2}
E)~,
\label{ctqlat}
\en
 and the fusion relations (\ref{fus1}) and (\ref{fus2}) become
\eq
t^{(m)} (\omega^{-1} E ) t^{(m)} (\omega E )= 1+ t^{(m-1/2)} (E)
t^{(m+1/2)} (E)
\label{cfus1}
\en
and
\eq
t^{(1/2)}(E) t^{(m)}(\omega^{2m+1} E) = t^{(m+1/2)} (\omega^{2m}E)
+  t^{(m-1/2)} (\omega^{2m+2}E)~.
\label{cfus2}
\en

These equations control the distribution of the extremal roots in
the large-$N$ limit. Physically, they are important because they
turn out to determine the constant $c_{\rm eff}$ which controls the
leading finite-$N$ corrections to the ground state energy -- see
appendix~\ref{num} and  for example, \cite{KBP}.

As already mentioned, for $\omega$ a root of unity the fusion
relations truncate. For $\eta=\pi M/(2M{+}2)$ with
$2M \in \ZZ^+$, and $\phi=\pi/(2M+2)$,  the
truncated set of $t$-equations can  elegantly be written as:
\eq
\fl \quad t^{(m)}( \omega^{-1}E)t^{(m)}( \omega E)= 1+
\prod_{j=1/2}^{(h-1)/2}
\lf ( t^{(j)}(E) \ri)^{G_{2j,2m}}\quad,\quad m=1/2,1,\dots,(h-1)/2
\label{twl}
\en
where $h=2M$, $\omega=e^{\pi i/(M{+}1)}$
 and  $G_{ab}$ is the incidence matrix of the $A_{h-1}$
Dynkin diagram:
\vspace{0.5cm}
\[
\begin{array}{c}
\epsfxsize=.50\linewidth\ \epsfbox{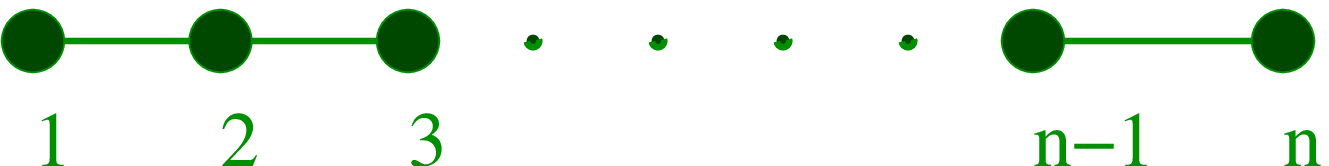}
\end{array}
\]
In the case of  $M=3/2$ and  $M=2$ the equations are respectively
\eq
t^{(1/2)}(\omega^{-1} E)  t^{(1/2)}(\omega E)= 1+ t^{(1/2)}(E)~;
\label{lyt}
\en
and
\bea
t^{(1/2)}(\omega^{-1} E)  t^{(1/2)}(\omega E) &=& 1+ t^{(1)}(E) \nn \\
t^{(1)}(\omega^{-1} E)  t^{(1)}(\omega E) &=& 1+ (t^{(1/2)}(E))^2 ~.
\label{qart}
\eea
A first hint of a link with the theory of ordinary differential equations
comes on comparing
equation~(\ref{lyt}) with the title and content of Sibuya's
paper~\cite{Si0}:
 `On the functional equation $f(\lambda)+
f(\omega\lambda)f(\omega^{-1}\lambda)=1$, ($\omega^5=1$)'. A precise
ODE/IM equivalence was  first established in~\cite{DTa} by mapping
(\ref{qart}) into a  functional relation which had previously been
associated with the quartic anharmonic oscillator by
Voros~\cite{Voros}, and more importantly by showing an exact
equivalence between the functions -- and not just the functional
relations -- involved in the two setups. As explained in
appendix~\ref{num}, techniques developed in the study of integrable
models allow functional equations of the type (\ref{twl}) to be
transformed into sets of nonlinear integral equations known as
thermodynamic Bethe ansatz (TBA) equations~\cite{ZamTBA}.

In conclusion: starting from the six-vertex model with twisted
boundary conditions and using the algebraic Bethe ansatz approach we
have derived  sets of functional relations: the  Baxter's TQ
relation, the quantum Wronskian and the fusion hierarchy.
Anticipating the correspondence with the theory of ordinary
differential equations the continuum limit was  taken, with the
final set of equations  encoding information about the $c=1$
conformal field theory defined  on a cylinder with twisted boundary
conditions, with the value of the twist depending on the twist in
the original six-vertex model.

The precise way to extract information such as the effective central
charge $c_{\rm eff}$ and the scaling dimensions goes through the
transformation of functional relations into nonlinear integral
equations (see  appendix~\ref{TBAsection}). However, it turns out
that there are other   ways to derive the same sets of functional
and integral equations rather than starting from the six-vertex
model. One  possibility is to work directly in field theory and
exploit  the fact that $c=1$ CFT also  corresponds to  the
ultraviolet limit of  the sine-Gordon model. The derivation of the
integral equations makes use of the sine-Gordon scattering matrix
description and as mentioned before goes under the name of the
thermodynamic Bethe ansatz~\cite{ZamTBA}.

Another approach also directly based on a CFT was proposed by
Bazhanov,  Lukyanov and Zamolodchikov in~\cite{BLZ1} and
further developed in \cite{BLZ2,BLZ3}.
The starting point of \cite{BLZ1,BLZ2,BLZ3} is not the unitary
$c=1$ conformal field theory defined  on a strip geometry with different
boundary
conditions as above, but a  CFT with central charge
\eq
c=1-(\beta-\beta^{-1})^2<1 \quad , \quad 0<\beta<1~,
\label{cblz}
\en
 with periodic boundary conditions.
This   theory is neither unitary nor minimal and  at
fixed values of
$\beta$ the Hilbert space still  depends on a  free parameter $p$
A brief summary
of  results relevant for the
ODE/IM correspondence is reported  in the next section.

\subsection{TQ equations in continuum CFT: the BLZ approach}
\label{blztq}

In \cite{BLZ1,BLZ2,BLZ3}, Bazhanov, Lukyanov and
Zamolodchikov showed how for integrable models
the  structures such as
Baxter's $\TT$ and $\QQ$ matrices may also be studied directly using
field-theoretic methods.   BLZ considered a CFT
with central charge $c$ parameterised in terms of $\beta$ according
to (\ref{cblz}).
The description of the conformal spectrum of the $XXZ$ model
 in terms of towers of states,
each tower consisting of a highest-weight state $|p\rangle$  and its
descendents, 
applies to  all CFTs. In this  case, the highest-weight states
have conformal dimension  $\Delta_p =(p/\beta)^2
+(c-1)/24$, where $p$ is a continuous parameter.

For each tower of states, BLZ define  a continuum analogue of the
lattice transfer matrix $\TT$, an
operator-valued entire function $\T(s,p)$.
 In analogy with (\ref{qtdef}), they also define
 a pair of operator-valued  functions
 $\Q_{\pm}(s,p)$. Together, these operators
mutually commute and
satisfy a TQ relation
\eq
\T(s,p) \Q_{\pm}(s,p)= \Q_{\pm}(q^2s,p)+\Q_{\pm}(q^{-2}s,p)
\label{conttq}
\en
with $q =\exp(  i \pi \beta^2)$.
Within each tower the
  highest-weight eigenvalues
\eq
T(s,p) = \langle p|  \T(s,p) | p \rangle
\en
and
\eq
Q_{\pm}(s,p)= \langle p | s^{\mp \Pop/\beta^2 } \Q_{\pm}(s,p) | p \rangle~,
\en
  satisfy the  TQ relation
\eq
T(s)Q_{\pm}(s)=
 e^{\mp 2\pi ip}Q_{\pm}(q^{-2}s)+
e^{\pm 2\pi ip}Q_{\pm}(q^2s)~,
\label{ctq}
\en
where  $\Pop$  is an operator such that
$\Pop |p \rangle = p | p\rangle $.
If we set  $\eta=\frac{\pi}{2}(1-\beta^2)$,
 this relation between the six-vertex anistropy $\eta$ and the coupling
 $\beta$ ensures that  $q^2=\exp(2i\pi\beta^2)$ is equal to
$\omega^2=\exp(-4i\eta)$ and the BLZ and continuum six-vertex
TQ relations match perfectly.
Moreover,
for  $\beta^2$ in the so-called semiclassical domain $0<
  \beta^2 <1/2 $, the eigenvalue $Q_{+}(s)$ can be written as a
convergent product   over its zeros $\{ s_k\}$:
\eq
Q_{+}(s)= \prod_{k=1}^\infty
\lf(1-\frac{s}{s_k}\ri)  \quad , \quad (Q_{+}(0)=1)~.
\label{aexp}
\en
Thus a set of  Bethe ansatz equations follows from the TQ relation
and the entirety in $s$ of the
 eigenvalues:
\eq
 \prod_{l=1}^\infty \left( \frac{s_l -q^2 s_i}{
s_l -q^{-2}s_i} \right) =-e^{4\pi i p} \quad,\quad i=1\dots\infty~.
\en

The other elements of the lattice picture also appear
 directly in the continuum context.
{}From the identity operator  $\T_0(s)$ and
$\T_{\hf}(s)\equiv \T(s)$, an infinite set of mutually
commuting operators are built using the  fusion relations:
\eq \fl \quad \quad
\T_j(q s)\T_j(q^{-1} s)= 1+ \T_{j-\hf}( s)
 \T_{j+\hf}( s) \quad , \quad j=\fract{1}{2},1,\fract{3}{2}\dots ~.
\label{tsys}
\en
At rational values of the parameter $\beta^2$, the hierarchy
 truncates to a finite set of operators,
 just as in the lattice case.
Alternatively,
the operators $\T_j(s)$ are given directly in terms of the $\Q$'s:
\eq
\fl \quad \quad 2 i \sin (2 \pi \Pop) \T_j(s) =  \Q_{+}(q^{2j+1}
s)   \Q_{-}(q^{-2j-1}
s)  -  \Q_{+}(q^{-2j-1} s)
\Q_{-}(q^{2j+1} s)~.
\label{taa}
\en
Evaluating on the  state $|p\rangle$
with $j=0$, we find  the continuum  version of the quantum
Wronskian
\eq
q^\frac{2p}{\beta^2} \,Q_{+}(q s ) Q_{-}(q^{-1} s )
-q^\frac{-2p}{\beta^2} \,Q_{+}(q^{-1} s ) Q_{-}(q  s )  = 2 i
\sin(2 \pi p)~,
\label{cqw}
\en
from which we  deduce $Q_{-} (s,p) = Q_{+} (s ,-p)$.

Since much of the current interest in the functional-relations
approach to integrable models is focused on the continuum field
theory applications, we concentrate on the above version of the
TQ relation  in the remainder of these notes. However it should be
remembered that the link with the theory of ordinary differential
equations applies equally to lattice models, so long as a large-$N$
limit  is taken in a suitable way. To be more precise, by setting
\eq
\beta^2=1-{ 2 \eta \over \pi}~,~~ p={\phi \over 2 \pi}
\en
one gets a full match between the  six-vertex and the BLZ
functional relations.
Furthermore, for a non-unitary CFT on a cylinder with
periodic boundary conditions the effective central charge is given by
\eq
c_{\rm eff}= c-24 \Delta_{min}~.
\en
We therefore see that $c_{\rm eff}= c-24 \Delta_{p=0}=1$ corresponds to
the effective central charge for the  untwisted six-vertex model.
And, within a single $p$ sector, the effective
central charge associated to the highest-weight  state
$\ket{p}$  is
\eq
c_{\rm eff}^{(p)}= c-24 \Delta_{p}=1-24 \left(\frac{p}{\beta}
\right)^2~,
\en
which matches the effective central charge for the six-vertex model
with twisted boundary conditions $\phi$.

Bazhanov, Lukyanov and Zamolodchikov also
discovered a relationship between the $T$ and $Q$ operators and
(perturbed) boundary conformal field theory~\cite{BLZ1},
which was followed up in later
work~\cite{Bazhanov:1998za,Dorey:1999cj,Lukyanov:2003nj,Lukyanov:2003rt}.
In the long term this probably corresponds to the most 
fertile ground for applications
of the ODE/IM correspondence. But as a   description of this
would take us too far from the themes of this review,
we
address the interested readers to the
papers~\cite{BLZ1,Dorey:1999cj,Lukyanov:2003nj,Lukyanov:2003rt}.


\subsection{Summary}
\label{summary}
We conclude our survey of integrable lattice models with a
summary of the vocabulary
introduced thus far, specialising to the simplified case of the
continuum limit.
In order to solve the six-vertex model, it suffices to diagonalise the

{\bf transfer matrix} $\TT$\\
which depends on the

{\bf spectral parameter} $\nu$\\
and has

{\bf eigenvalues} $t(\nu)$\,.\\
 These can be given in terms
of the

{\bf Bethe roots} $\{E_i\}$\\
 which solve

{\bf Bethe ansatz equations}
$\prod\left(\frac{E_l-\omega^{2}E_i}{E_l-\omega^{-2}E_i}\right)
= -e^{-i2 \phi}$\\[2pt]
 and these can be neatly encapsulated in Baxter's

{\bf TQ relation} $t(\nu)q(\nu)=e^{-i \phi} q(\nu+2i\eta)+ e^{i
\phi} q(\nu-2i\eta)$\,.



%


\resection{Ordinary differential equations and functional relations}
\label{funode}

Surprisingly, the functional equations found in the last section
also govern the problems in
$\PT$-symmetric quantum mechanics discussed in section~\ref{prelude}.
To understand how this comes about, we must first return to the subject
of $\PT$-symmetric eigenvalue problems and their generalisations
in a little more depth.

\subsection{General eigenvalue problems in the complex plane}
\label{compeig}
We begin with one piece of unfinished business
from section~\ref{prelude}:
what goes wrong with the Bender-Boettcher problem at
$M{=}2$, and what can be done to resolve it?
In figures \ref{fig1} and \ref{fig2}, the energy levels
continued smoothly past $M{=}2$, but in fact this can only be
achieved by implementing a suitable distortion of the problem as
originally posed. Consider the situation precisely at $M{=}2$\,:
the Hamiltonian is $p^2-x^4$, an `upside-down' quartic oscillator, and a
simple WKB analysis (about which more shortly) shows, instead of
the exponential growth or decay more generally found, wavefunctions
behaving as $x^{-1}\exp(\pm ix^3/3)$ as $x$ tends to plus or minus
infinity along the real axis.
{\em All}\/ solutions thus decay, albeit algebraically, and this
complicates matters significantly. The problem moves from what is called
the limit-point to the limit-circle case (see \cite{REL,rich}),
and additional boundary conditions should be imposed at infinity if
the spectrum is to be discrete.

While interesting in its own right, this is clearly
not the right eigenproblem
if we wish to find a smooth continuation from the region $M<2$.
Instead, it is necessary to enlarge the perspective and
treat $x$ as a genuinely complex variable. This
has been discussed by many authors, and is particularly emphasised in
the book by Sibuya \cite{Sha}\,, though
the treatment which follows is perhaps closer to that of
\cite{BT,BB}.

\begin{figure}[ht]
\begin{center}
\includegraphics[width=0.55\linewidth]{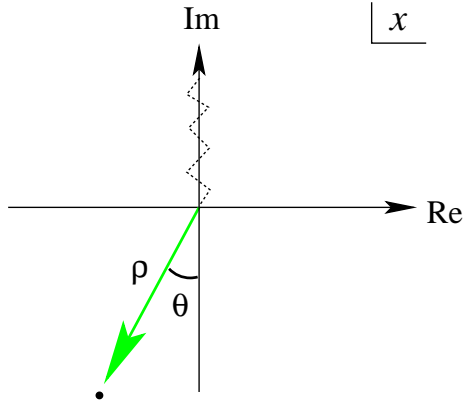}
\end{center}
\caption{A ray in the complex $x$-plane.\label{ray}}
\end{figure}

The key is to examine the behaviour of solutions as $|x|\to\infty$
along a general ray in the complex plane, in spite of the fact that
the only rays involved in the problem as initially posed were
the positive and negative real axes.
The WKB approximation tells us that
\eq
\psi(x)\sim P(x)^{-1/4}\,e^{\pm\int^x\!\sqrt{P(t)}dt}
\label{wkba}
\en
as $|x|\to\infty$, with $P(x)=-(ix)^{2M}+l(l{+}1)x^{-2}-E$. (This is
easily derived by
substituting $\psi(x)=f(x)e^{g(x)}$ into the ODE.)
Since the problem was set up with a branch cut running up the
positive-imaginary axis, it is natural to define general rays
in the complex plane by setting $x=\rho e^{i\theta}\!/i$ with $\rho$
real, as illustrated in figure~\ref{ray}.
\smallskip

\noindent
For $M>1$, the leading asymptotic predicted by (\ref{wkba})
is not changed if $P(x)$ is replaced by $-(ix)^{2M}$, and
substituting into the WKB formula we see two possible
behaviours, as expected of a second-order ODE:
\eq
\psi_{\pm}\sim P^{-1/4}\exp\left[\pm\fract{1}{M{+}1}e^{i\theta(1{+}M)}
\rho^{1{+}M}\right]\,.
\en
For most values of $\theta$, one of these solutions
will
be exponentially growing, the other exponentially decaying. But whenever
$\Re e[e^{i\theta(1{+}M)}]=0$, the two solutions swap r\^oles and there
is a moment when both oscillate, and neither dominates the other.
The relevant values of $\theta$ are
\eq
\theta=
\pm\frac{\pi}{2M{+}2}~,~
\pm\frac{3\pi}{2M{+}2}~,~
\pm\frac{5\pi}{2M{+}2}~,~\dots~.
\en
(Confusingly, the rays that these values of $\theta$ define
are sometimes called `anti-Stokes lines', and sometimes `Stokes
lines'. See, for example, \cite{berry}.)

\begin{figure}[ht]
\begin{center}
\includegraphics[width=0.55\linewidth]{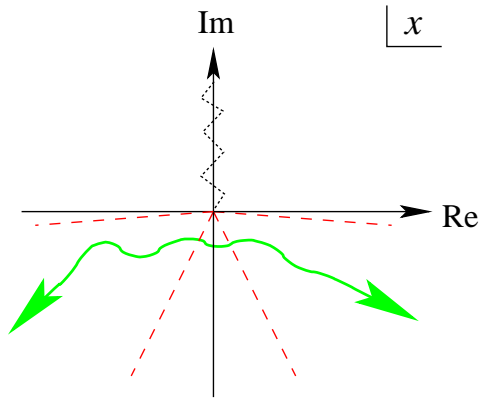}
\end{center}
\caption{A possible wavefunction contour for $M>2$.\label{sectors}}
\end{figure}

Whenever one of these lines lies along the positive
or negative real axis, the eigenvalue
problem as originally stated becomes much more delicate.
Increasing $M$ from $1$, the first time that this happens
is at $M=2$, the case of the
upside-down quartic potential discussed above. But now we see that the
problem arose because the line along which the
wavefunction was being considered, namely the real axis, happened to
coincide with an anti-Stokes line\footnote{as just mentioned, some
would have called this a
Stokes line.}.
We also see how the problem can be averted. Since all functions
involved are analytic, there is nothing to stop us from examining
the wavefunction along some other contour in the complex plane.  In
particular, before $M$ reaches $2$, the two ends of the contour can be bent
downwards from the real axis without changing the spectrum, so long as
their asymptotic directions do not cross any anti-Stokes lines in the
process.
Having thus distorted the original problem, $M$ can be increased
through $2$ without any difficulties.
The situation for $M$ just bigger
than $2$ is illustrated in figure \ref{sectors},
with the anti-Stokes lines shown dashed and the wiggly line a
curve along which the wavefunction $\psi(x)$ can be
defined.

The wedges between the dashed lines
are called {\em Stokes
sectors}, and in directions out to infinity which lie inside these
sectors, wavefunctions either grow or decay exponentially, leading to
eigenvalue problems with straightforward, and discrete, spectra.
Note that once $M$ has passed through $2$, as in figure~\ref{sectors},
the real axis is once again a `good' quantisation contour --
but for a {\em different}\/
eigenvalue problem, which is {\em not}\/ the analytic continuation of the
original $M<2$ problem to that value of $M$. (For the analogue of
figure~\ref{fig1} for this new problem, see figure~20 of \cite{BBN}.)
Going further, we could choose {\em any}\/ pair of Stokes sectors for the
start and finish of our contour. A priori, each pair of sectors defines a
different problem, though we shall see later that some of these problems
are related by simple variable changes.

\begin{figure}[ht]
\begin{center}
\includegraphics[width=0.55\linewidth]{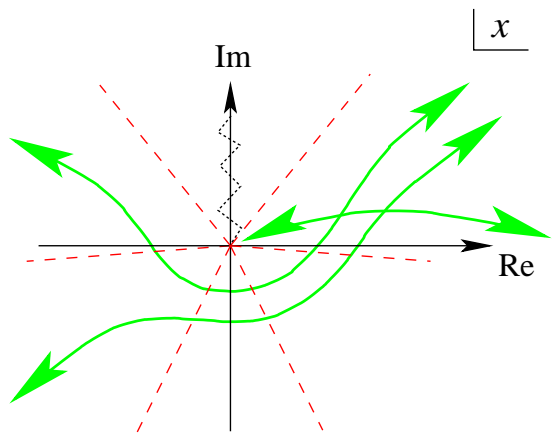}
\end{center}
\caption{Some further quantisation contours. \label{sectorsPT}}
\end{figure}

All of the problems do share one feature -- their quantisation contours
begin and end in the neighbourhood of the point $x=\infty$. In the
terminology of the WKB method,
they are related to `lateral' connection problems~\cite{OLV}.
There is one other special point for the ordinary differential
equation, namely the origin, and this provides another natural place where
quantisation contours can end.
Contours which join $x=0$ to $x=\infty$ lead to what are called `radial'
(or `central') connection problems, and with suitable boundary conditions
they can also have interesting, discrete,
spectra. However, if
both ends of
the contour are placed at $x=0$, the resulting eigenvalue problem
is always trivial. Some sample quantisation contours are shown in
figure~\ref{sectorsPT}.

We should pause for a moment to consider which boundary conditions
can be imposed at the origin, in order to understand why $x=0$ and
$|x|=\infty$ behave differently as end points of quantisation
contours. Even with the angular-momentum-like term $l(l{+}1)x^{-2}$
included, the singularity at the origin is much milder than that at
$|x|=\infty$, and irrespective of the direction in which it is
approached, solutions there behave algebraically, as $x^{l+1}$ or
$x^{-l}$. For this reason the complications associated with Stokes
sectors do not arise in the neighbourhood of the origin, and there
are just two natural boundary conditions to impose there -- we can
either demand that the solution behaves as $x^{l+1}$, or as $x^{-l}$
(the more singular of these two boundary conditions being defined by
analytic continuation). This contrasts with the situation near
$|x|=\infty$, where we can ask that a solution be subdominant in any
one of the potentially infinitely-many different Stokes sectors.
Expressed more technically, the ordinary differential equation has
two singular points, one at the origin and one at infinity. The
singular point at the origin is {\em regular}, and solutions have
straightforward series expansions in its vicinity. These converge in
the full neighbourhood of the origin, and can be analytically
continued in a simple  way\footnote{Cases with $2M$ not an integer
fall just outside the treatments in the standard texts, but they
behave in essentially the same way -- see, for example,
\cite{Cheng}. We have also glossed over some details, such as the
logarithms which can arise in certain situations. More background on
these issues can be found in \cite{ince}.}. Infinity, on the
other hand, is an {\em irregular} singular point, in the
neighbourhood of which solutions have asymptotic expansions which
only hold in selected Stokes sectors. This makes analytic
continuation around the point at infinity much more subtle, and
indeed this will be a major theme in the subsequent development.

To summarize:
associated with an ODE
of the type under consideration there
are many natural
eigenvalue problems, which fall into two classes.
Problems in the first, lateral, class are defined
by specifying a {\em pair} of Stokes sectors at infinity,
and then asking for the values of $E$ at which there exist solutions
to the equation which decay exponentially
in both sectors simultaneously. Problems in the second, radial, class
are defined by demanding decay in
a single Stokes sector at infinity, and imposing one of the
two simple boundary conditions at the origin.
The questions in $\PT$-symmetric quantum mechanics discussed in
section~\ref{prelude}
are all related to lateral
problems, with one particular pair of Stokes sectors selected.
Considerations of analytic
continuation have led us to put all pairs of sectors
on an equal footing, and we completed
the story by bringing in the radial problems as well.
But at this stage each eigenvalue problem sits on an isolated island,
each with its own private spectrum.

In the next subsection, we shall start to construct some bridges
between the
islands, using methods inspired by earlier work of Sibuya and of
Voros.
Remarkably, these
bridges turn out to be precisely the functional equations which had
previously arisen in the
context of integrable quantum field theory.

\subsection{A simple example}
\label{simp}
To illustrate the basic ideas in the simplest possible way,
for the time being we set $l(l{+}1)=0$, so we are dealing with the
original Bender-Boettcher family of eigenproblems:
\eq
-\frac{d^2}{dx^2}\psi(x)-(ix)^{2M}\psi(x)=E\psi(x)~~,\qquad\psi\in
L^2({\cal C})~.
\en

Now that the perspective has been widened to encompass
eigenvalue problems on general
contours, it is convenient to eliminate the factors of $i$ appearing
everywhere by making the variable changes
\eq
x\to x/i~,\quad E\to -E~
\label{varchange}
\en
so that the differential equation becomes
\eq
\left[
-\frac{d^2}{dx^2}+x^{2M}-E\right]\,\psi(x)=0\,.
\label{su2}
\en
This also
moves the branch cut onto the negative real axis, and the initial
quantisation contour onto the imaginary axis. (This final point
explains why the reality of
the spectrum remains a non-trivial question, despite the fact that the
coefficients in (\ref{su2}) are all real.)

Next, we
need to develop our treatment of ordinary differential equations in
the complex domain, relying largely on the
work of Sibuya and co-workers~\cite{HseihSha,Sha}.
The key result is the following:\\
$\bullet$~The ODE
(\ref{su2}) has a `basic' solution $y(x,E)$ such that\\
\noindent
{\bf (i)} $y$ is an entire function of $x$ and $E$;

\noindent
{}~~~~{\small [Though, because of the multivalued potential,
$x$ lives on a cover of $\CC \backslash \{0\}$
if $2M{\notin} \ZZ$
\footnote{The original work of Hseih and
  Sibuya~\cite{HseihSha} concerned only the case
$2M\in\mathbb{N}$, but the result also holds for the more general
situation  $2M\in\RR^+$  of eq.~(\ref{su2}), so long as the
branching at the origin is taken into account.  This  generalisation
was  explicitly discussed by Tabara in \cite{Tabara}.}.\,]}

\noindent
{\bf (ii)} as $|x|\to\infty$ with $|\arg\,x\,|<3\pi/(2M{+}2)$,
\bea
y\,&\sim&
{}~\frac{1}{\sqrt{2i}}\,
x^{-M/2}\exp\left[-\fract{1}{M{+}1}\,x^{M{+}1}\right]\,;
\label{yas}\\
y'&\sim& {-\frac{1}{\sqrt{2i}}}\,
x^{M/2}\exp\left[-\fract{1}{M{+}1}\,x^{M{+}1}\right]\,.
\eea
{}~~~~{\small [Though there are small modifications for $M\leq
1$ -- see, for example, \cite{DMST}\,.\,]}\\
Furthermore, properties {\bf (i)} and {\bf (ii)} fix $y$ uniquely.

\begin{figure}[ht]
\begin{center}
\includegraphics[width=0.55\linewidth]{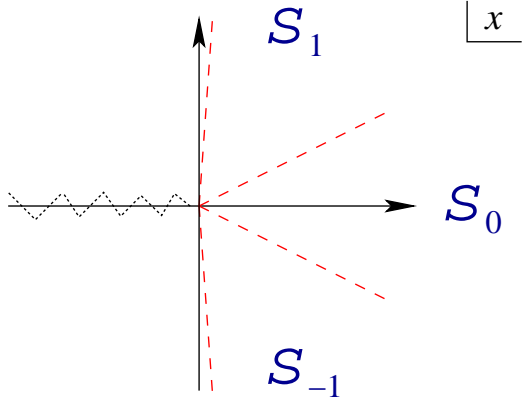}
\end{center}
\caption{Three Stokes sectors for the ODE (\ref{su2}), with $M=2.1$.
\label{sibsectors}}
\end{figure}

The second property can be understood via
the WKB discussion of section~\ref{compeig}.
With the shift from $x$ to $x/i$,
the anti-Stokes lines for (\ref{su2}) are
\eq
\arg(x)=
\pm\frac{\pi}{2M{+}2}~,~
\pm\frac{3\pi}{2M{+}2}~,~
\dots
\en
and in between them lie the Stokes sectors, which we label by defining
\eq
\CS_k:=\left|\arg(x)-
\frac{2\pi k}{2M{+}2}\right|<\frac{\pi}{2M{+}2}\,.
\en
Three of these sectors
are shown in the figure \ref{sibsectors}, a $90^{\rm o}$ rotation of
figure~\ref{sectors}.

The asymptotic given as property {\bf (ii)} then matches the result of
a WKB calculation in $\CS_{-1}\cup\CS_0\cup\CS_1$\,. The determination
of the large $|x|$ behaviour of the particular solution $y(x,E)$ beyond
these three sectors is a non-trivial matter, since the continuation
of a limit is not necessarily the same as the limit of a
continuation. This subtlety is related to the so-called
Stokes phenomenon, and it can be
handled using objects known as {\em Stokes multipliers}, to be introduced
shortly.

One more piece of terminology: an
exponentially-growing solution in a given sector is called {\em dominant}
(in that sector); one which decays there
is called {\em subdominant}. It is
easy to check that $y$ as
defined above is subdominant in $\CS_0$, and dominant in
$\CS_{-1}$ and $\CS_1$. A subdominant solution
to a second-order ODE in a sector is
unique up to a constant multiplier; this is why the quoted
asymptotics are enough to pin down $y$ uniquely.

Having identified one solution to the ODE, we can now generate
a whole family using a trick due to Sibuya. Consider the function
$
\hat y(x,E):=y(ax,E)
$
for some (fixed) $a\in\CC$. From (\ref{su2}), $\hat y(x,E)$ satisfies
\eq
\left[
-\frac{d^2}{dx^2}+a^{2M+2}x^{2M}-a^2E\right]\,\hat y(x,E)=0\,.
\en
(This is sometimes given the rather-grand name of
 `Symanzik rescaling'.)
If $a^{2M+2}{=}1$, shifting $E$ to $a^{-2}E$ shows that
$\hat y(x,a^{-2}E)$ again solves (\ref{su2}).
Defining
\eq
\omega:=e^{2\pi i/(2M{+}2)}
\en
and
\eq
y_k(x,E):=\omega^{k/2}y(\omega^{-k}x,\omega^{2k}E)
\label{ykdef}
\en
we then have the statements

\noindent
$\bullet$ $y_k$ solves (\ref{su2}) for all $k\in\ZZ$\,;
{}~~{\small [\,This follows since $(\omega^{-k})^{2M{+}2}=1$.\,]}

\noindent
$\bullet$ up to a constant, $y_k$ is the unique solution to (\ref{su2})
subdominant in $\CS_k$.
{}~~{\small [\,This follows easily from the asymptotic of $y$.\,]}

\noindent
$\bullet$ the functions $y_k$, $y_{k+1}$ are linearly independent for
all $k$, so
each pair $\{y_k,y_{k+1}\}$ forms a {\em basis} of solutions for
(\ref{su2}).
{}~~{\small [\,This follows on
comparing the asymptotics of $y_k$ and $y_{k+1}$
in either $\CS_k$ or $\CS_{k+1}$.\,]}

We have almost arrived at the TQ relation. Next, the fact that
$y_{-1}$ can be expanded in the $\{y_0,y_1\}$ basis shows that a
relation of the following form must hold:
\eq
y_{-1}(x,E)=C(E)y_0(x,E)+\widetilde C(E)y_1(x,E)~.
\label{stokesrel}
\en
This is an example of a {\em Stokes relation}, with
the coefficients $C(E)$ and $\widetilde C(E)$ {\em Stokes
multipliers}. They can be expressed in terms of Wronskians,
where~\cite{CL} the {\em Wronskian} of two functions $f$ and $g$ is
\eq
W[f,g]:=fg'-f'g\,.
\en
Given two solutions $f$ and $g$
of a second-order ODE with vanishing first-derivative
term, their Wronskian $W[f,g]$ is independent of $x$, and vanishes if
and only if
 $f$ and $g$
are proportional. As a convenient notation we set
\eq
W_{k_1,k_2}:=W[y_{k_1},y_{k_2}]=y_{k_1} y_{k_2}' -y_{k_1}' y_{k_2}
\label{wprops}
\en
and record the following two useful
properties\footnote{The normalisations in (\ref{yas}) and
(\ref{ykdef}), which differ from those adopted by Sibuya,
were expressly chosen so as to simplify these two formulae.}:
\eq
W_{k_1+1,k_2+1}(E)=W_{k_1,k_2}(\omega^2E)~,\quad W_{0,1}(E)=1\,.
\en
Now `taking Wronskians' of the Stokes relation (\ref{stokesrel})
first with $y_1$ and then with $y_0$
shows that
\eq
C=\frac{W_{-1,1}}{W_{0,1}}~~,\quad\tilde C=-\frac{W_{-1,0}}{W_{0,1}}=-1
\en
and so the relation itself can be rewritten as
\eq
C(E)y_0(x,E)=y_{-1}(x,E)+y_1(x,E)~,
\label{CYa}
\en
or, in terms of the original function $y$, as
\eq
C(E)y(x,E)
=\omega^{-1/2}y(\omega x,\omega^{-2}E)+
\omega^{1/2}y(\omega^{-1} x,\omega^2E)\,.
\label{CY}
\en
This looks very like a TQ relation. The only fly in the ointment is
the
$x$-dependence of the function $y$. But this is easily fixed: we just
set $x$ to zero. We can also take a derivative with respect to
$x$ before setting $x$ to zero, which swaps
the phase factors $\omega^{\pm1/2}$. So we define
\eq
D_-(E):=y(0,E)~~,\quad D_+(E):=y'(0,E)\,.
\en
Then the Stokes relation (\ref{CY}) implies
\eq
\fl
{}~~~~~~C(E)D_{\mp}(E)= \omega^{\mp 1/2}D_{\mp}(\omega^{-2}\!E)+
\omega^{\pm 1/2}D_{\mp}(\omega^2\!E)~,
\label{CD}
\en
and precisely matches the forms of the TQ equations (\ref{ctqlat})
and (\ref{ctq}) if the twist parameter is set to $\phi=2 \pi
p=\pi/(2M{+}2)$. Although the details are at this stage sketchy --
more will be provided in later sections -- we can already see
how some concepts in the two worlds of integrable models and
ordinary differential equations must be related:
$$
\begin{array}{|lcl|}
\hline
& & \\[-9pt]
{}~\parbox{5.1cm}{Six-vertex model with twist\\[-1pt]
$\phi=2 \pi p=\pi/(2M{+}2)$}&
&{}\parbox{4.9cm}{Schr\"odinger equation with\\[-1pt] homogeneous
potential $x^{2M}$}\\[12pt]
\hline
& & \\[-10pt]
\pppbox{Spectral parameter}&\lra&\mbox{Energy}\\[5pt]
\pppbox{Anisotropy}&\lra&\mbox{Degree of potential}\\[5pt]
\pppbox{Transfer matrix}&\lra&\mbox{The Stokes multiplier
$C$}\\[6pt]
\pppbox{Q operator}&\lra&\mbox{$D_-(E)$\,: the value of
$y(x,E)$ at $x=0$~~}\\[10pt]
\hline
\end{array}
$$
If $y$ on the last line is replaced by $y'$, then the twist changes
to $\phi=-\pi/(2M{+}2)$.

A small puzzle remains at this stage: why should one particular value
of the twist in the integrable model be singled out for a link with an
ordinary differential equation? This was resolved
shortly after the original observation in \cite{DTa}, when
Bazhanov, Lukyanov and Zamolodchikov
\cite{BLZa} pointed out that including an angular-momentum-like
term $l(l{+}1)/x^2$ allowed Q operators at other values of the
twist
to be matched. The details, and a further small generalisation of
the basic ODE (\ref{su2}), will be covered  in section~\ref{dict}.

\subsection{The spectral interpretation}
\begin{figure}[ht]
\begin{center}
\includegraphics[width=0.55\linewidth]{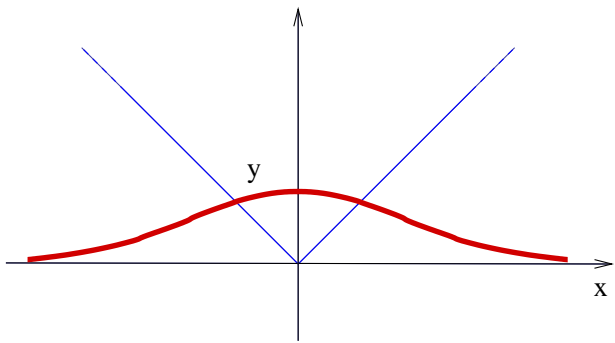}
\end{center}
\caption{$|x|$-potential, even sector: the ground state wavefunction.\label{even}}
\end{figure}

\begin{figure}[ht]
\begin{center}
\includegraphics[width=0.55\linewidth]{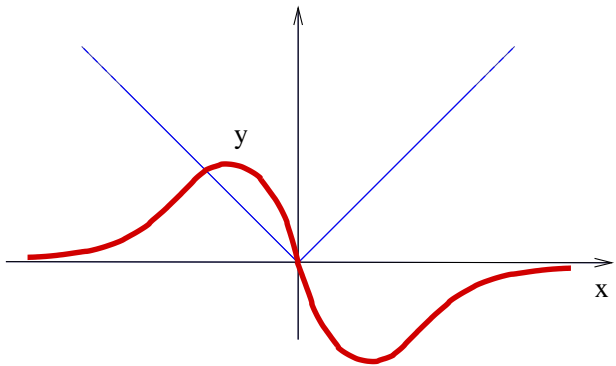}
\end{center}
\caption{$|x|$-potential, odd sector: the first excited state  wavefunction.\label{odd}}
\end{figure}

How should we think about $C$ and $D$? In fact they are spectral
determinants.
The {\em spectral determinant} of an eigenvalue problem is
a function which vanishes exactly at
the eigenvalues of that problem: it generalises
to infinite dimensions the characteristic polynomial
$\det(M-\lambda {\rm I})$
of a finite-dimensional matrix.
Recall that $C(E)$ is equal to the Wronskian $W_{-1,1}(E)$.
Thus $C(E)$ vanishes if and only if
$W[y_{-1},y_1]=0$, in other words if and only if
$E$ is such that $y_{-1}$ and $y_1$ are
linearly dependent. In turn, this is true
if and only if the ODE~(\ref{sh})
has a solution decaying in the two sectors $\CS_{-1}$ and $\CS_1$
simultaneously, which is exactly the lateral eigenvalue problem
discussed in section~\ref{prelude},
modulo the trivial redefinitions of $x$ and $E$.
This is enough to deduce that, up to a factor of an entire function
with
no zeros, $C(E)$ is the spectral determinant for the
Bender-Boettcher
problem\footnote{Since we performed
a variable change in this section compared with the
discussion in section~\ref{prelude},
it is in fact $C(-E)$ which provides the spectral determinant
for the Bender-Boettcher problem as originally formulated.}.
Even this ambiguity can be eliminated,
via Hadamard's factorisation theorem,
once the growth properties of the functions involved have been
checked;
see \cite{DTb} for details.
By its definition,
the zeros of $D_{-}(E)$ are the values of $E$ at
which the function $y(x)$, vanishing at
$x=\infty$, also vanishes at
$x=0$. Likewise,
the zeros of $D_{+}(E)$ are the values of $E$ at
which $y(x)$ has zero first derivative at $x=0$.
Thus $D_{\mp}(E)$ are also
spectral determinants.
Note that the vanishing of $D_{-}$ or $D_{+}$ corresponds
to there existing normalisable
wave functions for the equation on the full real axis,
with potential $|x|^{2M}$, which are
odd or even, respectively, as illustrated in figures \ref{even} and
\ref{odd}; this explains the labelling convention
adopted earlier.


This insight allows a gap in the correspondence to be filled.
We mentioned in section~\ref{aba} that
while the TQ relation is very restrictive, it does not have
a unique solution. So to claim that $D_-(E)$ is `equal' to
$A_+(\lambda,p)$
begs the question: which $A_+(\lambda,p)$?
To answer, we first note that the radial
problem for which $D_-(E)$ is the spectral determinant,
in contrast to the lateral
Bender-Boettcher problem,   {\em is} self-adjoint, and so all of its
eigenvalues are guaranteed to be real.
Back in the integrable model, we mentioned previously that
there is  only one solution to the BAE
with all roots real, so the question
is answered: the relevant $A_+(\lambda)$ is that corresponding to the
ground state in the spin-zero sector of the model.


%

\resection{Completing the dictionary}
\label{dict}

\subsection{Adding angular momentum}
\label{inhom}
We now restore the angular momentum term, and consider the
differential equation
\eq
\Bigl[-\frac{d^2}{dx^2}+x^{2M}+
\frac{l(l+1)}{x^2} \Bigr]\Phi(x)=E\,\Phi(x)\,.
\label{sh}
\en
This is the `$\pi/2$-rotated' version of the generalised
Bender-Boettcher problem (\ref{q3}).
In the process of discussing this ODE we shall fill in some
of the technical details skipped previously.
The $l=0$ case was the subject of the first ODE/IM
correspondence in \cite{DTa}, while the generalisation to
$l\neq 0$ was introduced in \cite{BLZa}. The initial treatment below
follows \cite{DTb}.

At the origin, solutions to (\ref{sh}) behave as a linear
combination of $x^{l+1}$ and $x^{-l}$. As mentioned in
section \ref{compeig}, a natural eigenproblem for this equation asks
for values of $E$ for which there is
a solution that vanishes as $x\to +\infty$,
and behaves as $x^{l+1}$ as $x\to 0$.
In the Stokes/WKB language this is a radial  problem.
For
$\Re e\,l>-1/2$, the condition at the origin
is equivalent to the demand that
the usually-dominant $x^{-l}$ behaviour there should be
absent; outside this region, more care is needed, but the problem
can be defined by analytic continuation. This issue will be discussed
in more detail in section \ref{qwronksec} below.

As for the simple example above we first employ Sibuya's trick.
{}From the uniquely-determined solution
$y(x,E,l)$, with   large,
positive  $x$   asymptotic \cite{Sha}
\footnote{
The result concerning the entirety  of  $y$ proved by Sibuya~(see
Ref.~\cite{Sha}) at $l=0$, $2M\in\mathbb{N}$  also holds for the
more general situation of eq.~(\ref{sh}) and eq.~(\ref{shg}) below
(so long as the branching at the origin is again taken into
account). The $l=0$,
$2M\in\RR^+$ case was discussed in \cite{Tabara}, while
the generalisation to a potential $P(x)/x^2$ with $P(x)$ a
polynomial in $x$ was studied by Mullin~\cite{Mullin}, and more
recently in~\cite{DeRa}. It is also worth noting that with a change
of variable it is possible to map (\ref{sh}) and (\ref{shg}) with
$l \in \RR$, $2M \in \mathbb{Q}^+$ onto particular cases of
those treated in~\cite{Mullin}.}
\eq
y(x,E,l) \sim
\frac{x^{-M/2}}{\sqrt{2 i}}
\exp \lf( - \fract{x^{M+1}}{M+1} \ri),
\label{yas2}
\en
we generate a set of functions
\eq
y_k(x,E,l)= \omega^{k/2}
y(\omega^{-k}x,\;\omega^{2k}E,\;l)
{}~~,\quad \omega = e^{\frac{2\pi i}{2M+2}}
{}~~,\quad k\in\ZZ
\label{yks}
\en
all of which solve~(\ref{sh}).
As before, any pair  $\{y_k,y_{k+1} \}$ forms a basis of the
two-dimensional space of solutions, and
so $y_{-1}$ can be written as a  linear combination of
$y_0$ and $y_{1}$. Rearranging the expansion and using the properties
(\ref{wprops}), which continue to hold with the addition of the
angular-momentum term,
\eq
C(E,l)\,y_0(x,E,l)=y_{-1}(x,E,l)+
y_{\;1}(x,E,l)\,,
\label{Ty}
\en
where Stokes multiplier $C(E,l)$ again takes a simple form in the
normalisations we have chosen:
\eq
C(E,l)=W[\,y_{-1}\,, y_1]/W[\,y_0,y_1]=
W[\,y_{-1}\,, y_1]\,.
\label{tform}
\en
There is one last complication: the angular-momentum term in
(\ref{sh}) means $x$ cannot simply be set to zero in (\ref{Ty})
to find a TQ relation.
Instead $y$ should be projected onto another solution,
defined via its asymptotics as $x \to 0$. Given that solutions
near $x=0$
behave as linear combinations of $x^{l{+}1}$ and $x^{-l}$,
a solution $\psi(x,E,l)$ can be defined by the requirement
\eq
\psi(x,E,l) \sim x^{l+1} + O(x^{l+3})\,.
\label{psidefn}
\en
This defines $\psi$
uniquely provided $\Re e\,l > -3/2$. A second solution
can be obtained from the first
by noting that the differential equation -- though not the boundary
condition -- is invariant under the analytic continuation
$l\to -1{-}l$. As a result, $\psi(x,E,-1{-}l)$
also solves (\ref{sh}). Near the origin, it behaves as
$x^{-l}$\,, and so at generic values of $l$ the pair of
 solutions
\bea
\psi_+(x,E)&:=&\psi(x,E,l)\,,\nn\\[2pt]
\psi_-(x,E)&:=&\psi(x,E,-1{-}l)
\label{psipmdef}
\eea
are linearly independent.
There are some subtleties to this procedure
at isolated values of $l$, to which
we shall return in section \ref{qwronksec}
below. However, they do not affect the initial
argument.
In discussions of the radial Schr\"odinger
equation (see, for example,  Ch.~4 of \cite{NEWT}),
$\psi_+$
is sometimes called the regular solution, if
$\Re e\,l>-1/2$.

We now take the Wronskian of both sides of (\ref{Ty})
with $\psi(x,E,l)$ to find an $x$-independent equation:
\eq
C(E,l) W[y_0,\psi](E,l)=W[y_{-1},\psi](E,l)+W[y_{1},\psi](E,l)\,.
\label{tqi}
\en
To relate the objects on the right-hand side of this equation back to
$W[y_0,\psi]$, we first define a set of `rotated' solutions by analogy
with (\ref{yks}):
\eq
\psi_k(x,E,l)= \omega^{k/2}
\psi(\omega^{-k}x,\;\omega^{2k}E,\;l)\,,\quad k\in\ZZ\,.
\label{psiks}
\en
These also solve (\ref{sh}), and a consideration of their behaviour as
$x\to 0$ shows that
\eq
\psi_k(x,E,l)= \omega^{-(l+1/2)k}\psi(x,E,l)\,.
\en
In addition,
\bea
W[y_k,\psi_k](E,l)
&=&
\omega^k\,W[y(\omega^{-k}x,\omega^{2k}E,l),
\psi(\omega^{-k}x,\omega^{2k}E,l)]\nn\\
&=&
W[y,\psi](\omega^{2k}E,l)\,.
\eea
Combining these results,
\eq
W[y_k,\psi](E,l)=\omega^{(l{+}1/2)k}W[y,\psi](\omega^{2k}E,l)
\en
and so, setting
\eq
D(E,l):=W[y,\psi](E,l)\,,
\label{dy}
\en
the projected Stokes relation (\ref{tqi}) is
\eq
C(E,l) D(E,l)=
\omega^{-(l+1/2)}D(\omega^{-2}E,l)+
\omega^{(l+1/2)}D(\omega^{2}E,l)\,.
\label{tq}
\en

\subsection{Matching TQ and CD relations}
\label{tqcd}
Finally we are ready to make the precise connection with the TQ relation
(\ref{ctq}). As a shorthand, define
\eq
D_{\mp}(E):=W[y,\psi_{\pm}](E,l)
\en
(so $D_-(E)\equiv D(E,l)$
and $D_+(E)\equiv D(E,-1{-}l)$). Then
(\ref{tq}) taken at $l$ and $-1{-}l$ becomes
\eq
C(E,l) D_{\mp}(E) =\omega^{\mp(2l+1)/2}
D_{\mp}(\omega^{-2}E)+
\omega^{\pm(2l+1)/2} D_\mp(\omega^{2}E)\,.
\label{CD2}
\en
If we set
\eq
\beta^2=\frac{1}{M{+}1}~~,\qquad
p=\frac{2l+1}{4M{+}4}
\label{params}
\en
then the match between the general TQ relation (\ref{ctq}) and  the Stokes
relation  (\ref{CD2}) is perfect, with the following
correspondences between objects
from the IM and ODE worlds:
\bea
T~&\leftrightarrow& ~C\nn\\
Q_{\pm}~&\leftrightarrow &~D_{\mp}~.\nn
\eea
The mapping could also have been made onto the limiting form of the
Bethe ansatz equations for the six-vertex model with twisted
boundary conditions that was obtained in section
\ref{baxter}.
In this case we have
\bea
t~&\leftrightarrow& ~C \\
q~&\leftrightarrow& ~D~, \nn
\eea  and, as mentioned before, the relationships
between the anisotropy parameter $\eta$ and the twist $\phi$
of the lattice model, and the parameters $M$ and $l$ appearing in
the potential of the Schr\"odinger equation are
\eq
\eta=\frac{\pi}{2}\frac{M}{M{+}1}\quad , \quad
\phi= \frac{(2l+1)\pi}{2M+2}~.
\en

The exact mapping between the functions appearing in these equations
can be found by examining the functions at $E=0$ and
their asymptotic behaviour at $E=\infty$.
Since the behaviour of
$D_{+}(E)$ can be deduced from that of $D_{-}(E)\equiv D(E,l)$,
we only need the
following results \cite{DTb}, which hold for $M>1$:
\begin{enumerate}
\item $C$ and $D$ are entire functions of $E$;
\item The zeros of  $D$ are all
real, and if $l>-1/2$ then they are all positive;
\item The zeros of $C$ are all real, and if
 $-1-M/2 < l < M/2$ then they are all  negative;
\item If
$M>1$ then the large-$E$ asymptotic of $D$ is
\eq
\ln D(E,l) \sim \frac{a_0}{2} (-E)^{\, \mu} \quad,\quad |E| \to
\infty \quad, \quad |\arg(-E) | <\pi~,
\label{das}
\en
where $\mu=(M{+}1)/2M $, and $ a_0 =  - \,\Gamma {(-\mu)}
\Gamma ({\mu +\hf)} /\sqrt \pi ~;$
\item At $E=0$
 \eq
\ D(E,l)|\phup_{E=0}=
\frac{1}{\sqrt{2i\pi}}
\Gamma{(1+\fract{2l+1}{2M+2})}
(2M{+}2)^{\frac{2l+1}{2M+2}
        +\frac{1}{2}}~.
\label{d0}
\en
\item The large-$E$ asymptotic implies that $D(E,l)$  has {\it
  order\footnote{Technically, the order of
an entire function $f(z)$ is defined to be equal to the lower bound
of all positive numbers $B$ such that $|f(z)|={\cal O}(e^{|z|^B}) $
as $|z|
\to \infty$.}}
equal to $\mu$, which is  strictly less than one for
$M>1$. Thus, Hadamard's
  factorisation applies in its simplest form and  $D(E,l)$ can  be represented
as:
\eq
D(E,l)=D(0,l) \prod_{k=1}^{\infty} \left(1 -{E \over E_k} \right)
\label{had}
\en
\end{enumerate}
Property (i)     follows from
the definition of $D$ as a Wronskian, since all functions
involved are entire functions of $E$. Properties (ii) and (iii) will
be proven in section~$\ref{realproof}$, while the proof of
properties (iv) and (v) can be found in appendix~\ref{asy_app}.

The relevant analytical properties of  $T(s)$ and $Q_{+}(s)$ are
given in \cite{BLZ1,BLZ2}. For $\beta^2$ in the semiclassical domain:
\begin{enumerate}
\item $T(s,p)$ and $Q_{\pm}(s,p)$ are entire functions of $s$ with an essential
  singularity  at infinity on the  real axis;
\item  All zeros of $Q_{+}(s,p)$ are real, and if  $2p>-\beta^2$ they
  are all strictly positive;
\item  All zeros of   $T(s,p)$ are real, and if
  $|p|<1/4$ they are all negative;
\item The large-$s$ asymptotics are
\eq \fl
\  \begin{array}{lcl}
\renewcommand{\arraystretch}{2.5}
\ln T(s,p)\ &\!\!\!\! \sim& 2 \sqrt { \pi} \, \frac{   \Gamma
  {(1-\mu)}}{\Gamma{( \frac{3}{2}- \mu)}} \, \Gamma{\lf (\frac{1}{2\mu}\ri
)}^{2\mu} \, (s)^{\, \mu} \\[10pt]
\ln Q_\pm (s,p)\ & \!\!\!\! \sim&   a_0 (M{+}1) \,
\Gamma {\lf (\frac{1}{2\mu} \ri)}  ^{2\mu}   \,(-s)^{\,\mu}
\end{array}
\quad,\quad
|s| \to \infty\quad,\quad \arg(-s)< \pi~,
\en
where $\mu=1/(2{-}2\beta^2)$ and $ a_0 $ is as defined above;

\item  If $s=0$
 \eq
Q_{+}(0)=1~.
\en
\item
The large-$s$ asymptotic implies that $Q_{\pm}(s)$  has
  order equal to
$1/(2-2\beta^2)$, which is again  strictly less than one for
$\beta^2$ in the semiclassical domain and we can write
\eq
Q_{\pm}(s)=\prod_{k=1}^{\infty} \left(1 -{s \over s_k} \right)
\en
The restriction of $\beta^2$ to the domain  $0<\beta^2<1/2$
 translates to the constraints $\pi/4<\eta<\pi/2$
  on the lattice model parameter $\eta$, and we see that the  point
at which the factorised products have to be regularised  coincide in
the two cases.  For convenience we have only considered
$\beta^2$
inside the semiclassical domain;
see \cite{BLZ2,BLZ3,DTb} for discussions on the
interesting case of $\beta^2 \ge 1/2 $.
\end{enumerate}

Given (\ref{params}), properties (i-iii) and (vi) match the
equivalent statements for $T(s,p)$ and $Q_{+}(s,p)$ with $E$
replaced by $s$. Noting that $\mu = (M{+}1)/2M = 1/(2{-}2\beta^2)$,
the asymptotic (iv) and the normalisations of $D_{-}$ and
$Q_{+}$ can be made to agree by setting
\eq \fl
\quad \quad \quad
s = v E
\quad, \quad v=(2M{+}2)^{-1/\mu} \,
\Gamma(\fract{1}{2\mu})^{-2} \quad , \quad \gamma_{\mp} =
D_\mp(0,2p/\beta^2{-}\fract{1}{2})^{-1}~.
\label{nudefn}
\en
The precise result is \cite{DTb}
\bea
Q_{\pm}(s,p)\Bigl |_{\beta^2}&=&\gamma_{\mp}\, D_{\mp}(\fract{s}{v}
,
\fract{2p}{\beta^2}{-}\fract{1}{2})\Bigl|_{M=\beta^{-2}{-}1}~;
\label{ad}
\\ [2pt]
T(s,p)\Bigl |_{\beta^2}&=&  C(
\fract{s}{v} ,
\fract{2p}{\beta^2}{-}\fract{1}{2})\Bigl|_{M=\beta^{-2}{-}1}  ~.
\eea

\subsection{The r\^ole of the fusion hierarchy}
\label{fusionrole}
In section \ref{fusion}, another set of functional relations found in
integrable models was described: the fusion hierarchy. Now that the TQ
relation has been mapped onto a Stokes relation, it is natural to ask
whether an analogue of the fusion hierarchy can also be found
in the differential equation world, and it turns out that this
is indeed possible \cite{DTb}.

Previously we
examined the expansion of $y_{-1}$ in the basis $\{y_{0},y_{1}\}$, but
one can equally ask about the expansion of  $y_k$ in any
other basis, such as $\{y_{k+r-1},y_{k+r}\}$:
\eq
y_{k-1} = C_k^{(r)} y_{k+r-1} + \tilde C_k^{(r)} y_{k+r}
\label{yk1}
\en

A change of   basis
from $\{y_{k+r-1},y_{k+r}\}$ to $\{y_{k-1},y_k\}$ can then
be encoded in a
$2 \times 2$ matrix ${\bf C}_k^{(r)}$ as
\eq
\lf(\matrix{y_{k-1}\cr y_{k}}\ri)=
{\bf C}^{(r)}_k\lf(\matrix{y_{k+r-1}\cr y_{k+r}}\ri)~~,
\quad
{\bf C}^{(r)}_k=
\lf(\matrix{C^{(r)}_k&\tilde C^{(r)}_k\cr
            C^{(r-1)}_{k+1}&\tilde C^{(r-1)}_{k+1}}\ri)\,.
\en
This matrix depends on $E$ and $l$, but not $x$.
The following properties are immediate:
\eq
{\bf C}^{(r)}_k(E,l)=
{\bf C}^{(r)}_{k-1}(\omega^2E,l)~,
\en

\eq
{\bf C}^{(0)}_k=
\lf(\matrix{1&0\cr
            0&1}\ri)~~,\quad
{\bf C}^{(1)}_k=
\lf(\matrix{C_k^{(1)} &-1\cr
            1&0}\ri)~.
\label{init}
\en
Further relations
reflect the fact that the change from the basis
$\{y_{k+r+n-1},y_{k+r+n}\}$ to
$\{y_{k+r-1},y_{k+r}\}$,
followed by the change from
$\{y_{k+r-1},y_{k+r}\}$ to
$\{y_{k-1},y_{k}\}$, has the same effect as
accomplishing the overall change in one go:
\eq
{\bf C}^{(r)}_k
{\bf C}^{(n)}_{k+r}
={\bf C}^{(r+n)}_{k}\,.
\label{monrel}
\en
(These express the consistency of the analytic continuations, and can
be thought of as monodromy relations.)
The case $r=1$  gives two non-trivial relations:
\eq
C^{(1)}_k C^{(n)}_{k+1}-C^{(n-1)}_{k+2}=C^{(n+1)}_k
\label{Crel}
\en
and
\eq
C^{(1)}_k\tilde C^{(n)}_{k+1}-\tilde C^{(n-1)}_{k+2}=\tilde C^{(n+1)}_k,
\label{Ctrel}
\en
which can be combined with the `initial conditions' (\ref{init}):
to deduce $\tilde C^{(2)}_k=-C^{(1)}_k\,$; and then the more general
equality
\eq
\tilde C^{(n)}_k=-C^{(n-1)}_k
\label{tceq}
\en
follows on comparing (\ref{Ctrel}) with (\ref{Crel}). If we now
set
\eq
C^{(n)}(E)=C_0^{(n)}(\omega^{-n+1}E)\,,
\en
then (\ref{Crel}) is equivalent to
\eq
C(E)C^{(n)}(\omega^{n+1\!}E)=C^{(n-1)}(\omega^{n+2\!}E)+
C^{(n+1)}(\omega^{n\!}E)\,,
\label{c0sys}
\en
and this matches the fusion relation (\ref{cfus2}). Since
$C^{(0)}(E)=1=T_0(E)$ and, from the last section,
$C^{(1)}(E)=C(E)=T_{1/2}(v  E)$, this establishes the basic equality
\eq
C^{(n)}(E)
=T_{n/2}(v E )~.
\label{beqq}
\en

To find the fusion relation (\ref{cfus1}), we start by
taking Wronskians of (\ref{yk1}) as before to obtain
\eq
C_k^{(r)} =  W_{k-1,k+r} \quad , \quad \tilde C_k^{(r)} =-
  W_{k-1,k+r-1}~,
\label{Cwron}
\en
from which we immediately recover (\ref{tceq}), and can also deduce
\eq
C_{k}^{(r)} = - C_{k{+}r{+}1}^{(-r-2)}~.
\en
This relation combined with the $n=-r$ case of (\ref{monrel}), namely
${\bf
  C}_k^{(r)} {\bf C}_{k+r}^{(-r)} =
1$, implies that
\eq
C^{(r-1)}(\omega^{-1\!}E)C^{(r-1)}(\omega E) -C^{(r)}(E)C^{(r-2)}(E)
=1~,
\label{csys}
\en
which reproduces the fusion relations (\ref{cfus1}), given the
identification (\ref{beqq}).

\begin{figure}[ht]
\begin{center}
\includegraphics[width=0.55\linewidth]{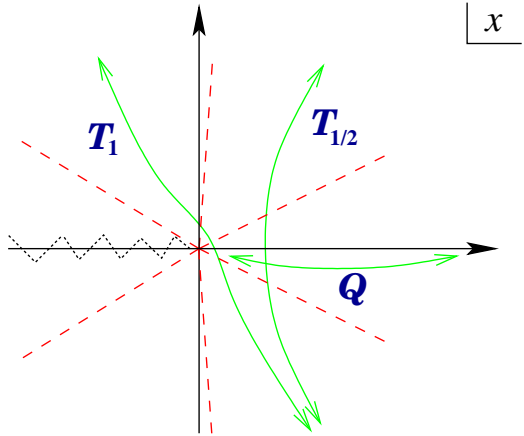}
\end{center}
\caption{General quantisation contours involved in the ODE/IM
correspondence. \label{sectorsQ}}
\end{figure}

Finally, combining (\ref{beqq}) and  (\ref{Cwron}) we have an
expression for $T_{n/2}(v E)$ in terms of a Wronskian:
\eq
T_{n/2}(v E) = C^{(n)}(E)= W_{-1,n}(\omega^{-n+1}E)~.
\label{cw}
\en
This result shows that the fused transfer matrices can also
be interpreted as spectral
determinants. The right hand side of (\ref{cw}) vanishes if and
only if $E$ is such that $y_{-1}$
and $y_n$ are linearly dependent, which in turn is true
if and only if the ODE has a
nontrivial
solution which simultaneously decays to zero as $|x|\to\infty$
in the sectors $\CS_{-1}$ and $\CS_{n}$. This is one
of  the lateral eigenvalue problems discussed in
section~\ref{compeig}, with the eigenvalues encoded in the zeros of
$C^{(n)}$.
The full story is illustrated in figure~\ref{sectorsQ}.

Once the fused transfer matrices have been understood in this way,
truncation of the fusion hierarchy can be
reinterpreted in terms of the (quasi-) periodicity
(in $k$) that the functions $y_k$
exhibit whenever $M$ is rational \cite{DTb}. In the simplest cases
(with $M$ rational and
$l(l{+}1){=}0$) this periodicity arises because the solutions to the
ODE live on a finite cover of $\CC\backslash\{0\}$; for other cases,
the monodromy around $x{=}0$ needs a little more care, but the story remains
essentially the same.

As a simple example, consider the Schr\"odinger problems
with $2M$ integer and $l(l+1)=0$.
In these cases all solutions of the ODE are single-valued
functions of $x$, and the sectors $\CS_{n+2M+2}$ and
$\CS_{n}$ coincide.  Thus both $y_n$ and $y_{n+2M+2}$ are
subdominant in $\CS_n$ and must be proportional. In fact, one can easily
show from the asymptotics that $y_{n+2M+2}(x,E,l)=-y_n(x,E,l)$,
from which we conclude from (\ref{cw}) that
\eq
C^{(2M)}(E) = 1  \quad , \quad  C^{(2M+1)}(E) = 0~.
\en
Thus the set (\ref{csys}) of functional relations truncates to
\eq
C^{(r)}(\omega^{-1}E) C^{(r)}(\omega E) =
1+ \prod_{n=1}^{2M-1} (C^{(n)}(E) ) ^{G_{nr}}~,
\label{ctba}
\en
perfectly matching the T-system (\ref{twl}).

\subsection{The quantum Wronskians}
\label{qwronksec}
There is one final set of functional relations to discuss: the quantum
Wronskian (\ref{cqw}) and its partner relations
(\ref{taa}) which express the $T$'s in terms
of the $Q$'s.  Recall
the solutions $\psi_{\pm}$\,, defined in equation (\ref{psipmdef})
by their behaviour at $x=0$. At generic values of $l$, these
give an alternative basis in which to expand $y(x,E,l)$.
Using the Wronskian
\eq
W[\psi_+,\psi_-]=2l+1
\label{psiwronk}
\en
(evaluated in the limit $x\to 0$), and the relations
$D_{\mp}=W[y,\psi_{\pm}]$\,,
\eq
(2l{+}1)\,y(x,E,l)=D_{-}(E)\,\psi_{-}(x,E)- D_{+}(E)\,\psi_{+}(x,E)\,.
\label{yexp}
\en
Notice that there is a problem with this expansion at $l=-1/2$\,,
which is easily understood: at this point the two solutions $\psi_+$
and $\psi_-$ coincide and no longer provide a basis,
a fact which is
reflected in the vanishing of their Wronskian (\ref{psiwronk}).
In fact this is not the only place where difficulties arise.
So long as $l$ lies in the right half-plane $\Re e\,l>-1/2$, the
solution $\psi_+(x,E)=\psi(x,E,l)$ can be proved to exist as the
limit of a convergent sequence of approximate solutions -- see
\cite{NEWT,EUAN,Cheng}. However, this does not work in the left
half-plane, which is another way to see why
the second solution $\psi_-(x,E)=\psi(x,E,-1{-}l)$
must initially be defined by analytic continuation. At isolated values
of $l$ in the left half-plane, poles may arise,
causing $\psi_-$ to be ill-defined. As discussed in Chapter 4 of
\cite{NEWT}, these poles can be removed
by multiplying $\psi_-$ by a regularising factor. However
this inevitably inserts extra zeros into the Wronskian
(\ref{psiwronk}), and the regularised $\tilde\psi_-$ fails to be
independent of $\psi_+$ at exactly the points where the previous
$\psi_-$ had failed to exist. For the simple power-law potential
$x^{2M}$ that
we are dealing with, the problem values of $l$ are best identified
via the iterative construction of \cite{Cheng}, and lead to the
conclusion -- see for example \cite{DTb} -- that $\{\psi_+,\psi_-\}$
fails to be a basis at the points
\eq
l+\fract{1}{2}=\pm(m_1+(M{+}1)m_2)\,,
\label{sing}
\en
where $m_1$ and $m_2$ are two non-negative integers.
Using (\ref{params}), for the integrable model this corresponds to the
twist values
\eq
2p=
\pm
(m_1\beta^2+m_2)~.
\label{probps}
\en
The set
\eq
2p=
\pm m_2~,~~(m_2=0,1,\dots)
\en
corresponds to the vanishing-points for the quantum Wronskian (see
eq. (\ref{cqw})) and  at
\eq
2p=- \beta^2 -m_2~,~~(m_2=0,1,\dots)
\en
 there is a normalisation problem for
$Q_+(s,p)$  as a consequence of the appearance of a
zero level ($s_k=0$ for some $k$)~\cite{BLZ2}.

The problem points (\ref{sing})
can be dealt with by a limiting procedure --
\cite{DTb} discusses the first case, $l=-1/2$. For now,
though, we will assume that $l$ has been picked so that
these subtleties do not arise. Then the pair
$\{\psi_+,\psi_-\}$ {\em does}\/ provide a basis for the ODE, and,
using (\ref{ykdef}), we define pairs of
solutions $\psi_k^\pm(x,E)$ as
\eq
\psi_{k}^{\pm}(x,E) = \omega^{k/2}
\psi_\pm(\omega^{-k}x,\omega^{2k}E,l)\,,
\quad k\in \ZZ\,.
\label{psipmkdef}
\en
These allow the rotated functions $y_k$ to be expanded as in
(\ref{yexp}):
\eq
(2l{+}1)\,y_k(x,E,l)= D_{-}(\omega^{2k}E)\,\psi_{k}^{-}(x,E) +
 D_{+}(\omega^{2k}E)\,\psi_{k}^{+}(x,E)\,.
\label{ypsi}
\en
The Wronskians of these solutions
are very simple:
\bea
W[\psi_{k}^{+} , \psi_{p}^{+}] &=& W[\psi_{k}^{-} , \psi_{p}^{-}]
=0\,, \\[3pt]
W[\psi_k^{-} , \psi_p^{+}] &=& (2l{+}1)\,\omega^{(k-p)(l+1/2)}\,.
\eea
The fundamental quantum Wronskian is now almost immediate: in our
normalisations,
$W[y_{-1},y_0]=1$\,, and substituting the expansion (\ref{ypsi})
into this formula and using bilinearity of the Wronskian
yields
\eq
 (2l+1) = \omega^{-(l+1/2)} D_{-}(\omega^{-1} E) D_{+}(\omega E)
-  \omega^{l+1/2}  D_{-}(\omega E)D_{+}(\omega^{-1} E)~.
\label{odeqwronk}
\en
The fused Wronskians are equally straightforward.
Taking the Wronskian of $y_{-1}$ and $y_{n}$, again using the expansion
(\ref{ypsi}), shifting $E$
to $\omega^{-n+1 }E$ and using the relation (\ref{cw}) for $C^{(n)}$
shows that
\bea
 (2l+1) C^{(n)}(E) &=&
\omega^{-(n+1)(l+1/2)} D_{-}(\omega^{-n-1}E)
 D_{+}(\omega^{n+1}E) \nn \\ && \qquad -
  \omega^{(n+1)(l+1/2)} D_{-}(\omega^{n+1}E)
 D_{+}(\omega^{-n-1}E)~,
\label{cdd}
\eea
which is precisely (\ref{taa}).

Next we evaluate (\ref{cdd}) at $l=0$ and $n=M$ for integer
$M$, and replace
$C^{(M)}$ with
$T_{M/2}$ to  find
\eq
\frac{i}{2} \, T_{M/2}(E) = D_{-}(-E,0) D_{+}(-E,0) \equiv D(-E)~.
\label{tmd}
\en
The final equality follows from recalling that $D_{-}(E,0)$ and
$D_{+}(E,0)$  are the even and odd subdeterminants for the spectral
problem defined on the full real axis. This, up to a factor of $i/2$
arising from our normalisation of $D(0)$, is precisely the original result
conjectured in \cite{DTa}.

In fact, the quantum Wronksian relation (\ref{odeqwronk}) was the
initial source of the ODE/IM correspondence \cite{DTa}. As a bridge
between the ODE and IM worlds, it has some advantages over the `TQ'
approach we have stressed in the foregoing. In particular, it allows
for a complete proof of the correspondence \cite{BLZa}.

\subsection{Numerical techniques}
One of the
 byproducts of the ODE/IM
correspondence
was the realisation  that energy levels for Schr\"odinger
problems  can be calculated using nonlinear integral equations
\cite{DTa} of either TBA type or an alternative type known as the Destri
 de Vega or Kl\"umper-Batchelor-Pearce  equations 
 (or simply NLIEs).
In  appendix~\ref{deriv} we derive the TBA equations and
NLIEs associated with the spectral problems above, and
explain how to obtain the spectrum  by solving these
equations iteratively.

These methods of solving
such eigenvalue problems
appear to be new, though iterative solution methods based
on functional relations for spectral determinants had previously been
employed by Voros~\cite{Voros}, in work which was an important input to the
initial observation of the ODE/IM correspondence in \cite{DTa}.
Numerically, the method is rather efficient --
integral equations tend to be easier to solve than differential ones,
and spectral determinants encode {\em all} the eigenvalues at once.

\subsection{The full dictionary}
\label{full}
All good correspondences need a dictionary, and the table below
summarizes the mapping between objects seen by the integrable model
and the differential equation.
\smallskip
$$
\begin{array}{|lcl|}
\hline
& & \\[-9pt]
{}\parbox{3cm}{Integrable\\[-1pt] Model}&
&{}\parbox{3.3cm}{Schr\"odinger\\[-1pt] equation}\\[10pt]
\hline
& & \\[-10pt]
\pppbox{ Spectral parameter}&\lra&\ppbox{Energy}\\[5pt]
\pppbox{ Anisotropy}&\lra&\ppbox{Degree of potential}\\[5pt]
\pppbox{ Twist parameter}&\lra&\ppbox{Angular momentum}\\[7pt]
\pppbox{ (Fused) transfer\\   matrices}&\lra&\ppbox{Lateral spectral
problems defined at $|x|{=}\infty$}\\[14pt]
\pppbox{ $Q$ operators}&\lra&\ppbox{
Radial spectral problems linking $|x|{=}\infty$ and $|x|{=}0$}\\[14pt]
\pppbox{ Truncation of the\\  fusion hierarchy}&\lra
&\ppbox{ Solutions on finite covers of $\CC \backslash
\{0\}$}\\[10pt]
\hline
\end{array}
$$

\vspace{0.2cm}

\noindent
Armed with the dictionary,  the horizontal axis of figure \ref{fig1}
can be annotated
to indicate which integrable models correspond to the various values
of $M$ in the Bender-Boettcher problem. Thus for $M=\hf
,1,\frac{3}{2},2$ and $3$,
the relevant
integrable models are the $N{=}2$ SUSY point of the sine-Gordon
model
the free-fermion
point, the Yang-Lee model, $\ZZ_4$ parafermions and the $4$-state Potts
model respectively.
It is amusing that the $x^3$ potential is related by the
correspondence to the Yang-Lee model
(or, strictly speaking, to the sine-Gordon
model at the value of the coupling which allows for a reduction to Yang-Lee),
thus returning by a very indirect route to
the original thought of Bessis and Zinn-Justin.  Note that while the
various results discussed here are derived for $M>1$,
the  `Airy case' ($M=1/2 ,l=0$) has already been studied in \cite{vorair},
and has been shown to be consistent with integrable field theory
predictions \cite{DTa,fend}.

We finish this section with the table below, which records the notation
used for the related objects appearing in the six vertex model, the
continuum integrable model and the differential equation picture.
\smallskip

$$
\begin{array}{|lll|}
\hline
& & \\[-9pt]
{}\parbox{3cm}{Six vertex\\[-1pt] model}
&{}\parbox{3.3cm}{Integrable\\[-1pt] model}
&{}\parbox{3.3cm}{Schr\"odinger\\[-1pt] equation}\\[10pt]
\hline
& & \\[-10pt]
{}~\nu & s & E\\[5pt]
{}~\eta & \beta & M\\[5pt]
{}~\phi& \phi & l \\[5pt]
{}~t^{(m)}& T_{m} & C^{(2m)} \\[5pt]
{}~q_0(\nu,\pm \phi) & Q_{\pm} & D_{\mp}\\[5pt]
\hline
\end{array}
$$


%
\resection{Applications and generalisations}
\label{spepro}
The correspondence between ordinary differential equations and
integrable models is intriguing, but is it good for anything? One
answer comes from the fact that the properties of ordinary
differential equations and their solutions are rather better
understood than those of the $T$ and $Q$ functions from integrable
quantum field theory. The correspondence therefore allows a number of
conjectured properties of these functions to be proven for the first
time -- analyticity, duality\dots
~\cite{Bazhanov:1998za,BLZa,BHK,Bazhanov:2003ua,Lukyanov:2003rt,
Lukyanov:2003nj,
  Lukyanov:2005nr}.

On the other hand, ideas from the theory of integrable models
have led to new insights into the spectral properties of a number of
quantum-mechanical problems. This section begins with some
examples of this aspect, starting with the reality properties in
$\PT$-symmetric quantum mechanics that were described back in
section~\ref{prelude}. First we sketch how the most
general class of problems discussed there can be linked with
integrable models.

\subsection{Inhomogeneous potentials}
\label{sps}
A key feature of the proof of the TQ relation in section
\ref{funode} was
the fact that the differential equation mapped to itself under the
`Sibuya' variable change $x\mapsto\omega^{-1}x$, $E\mapsto\omega^2
E$, where $\omega{=}\exp(\pi i/(M{+}1))$. In turn, this relied on
the `potential' $V(x)=x^{2M}$ being invariant under $V(x)\mapsto
\omega^{-2}\,V(\omega^{-1}x)$. A natural generalisation, first
observed in \cite{Sc} and further explored in \cite{DDTb}, is to add
a term which is exactly anti-invariant under the same
transformation. This leads to the consideration of the inhomogeneous
potential $V(x)=x^{2M}+\alpha\,x^{M-1}$ and the ordinary
differential equation
\eq
\Bigl[-\frac{d^2}{dx^2}+x^{2M}+\alpha x^{M-1}+
\frac{l(l+1)}{x^{2}} \Bigr]\Phi(x)=E\,\Phi(x)\,,
\label{shg}
\en
where $\alpha$ is a free parameter.
Again the radial spectral problem asks for those values of $E$
admitting solutions which vanish
 as $x\to +\infty$,
and behave as $x^{l+1}$ as $x\to 0$.  With the inclusion
of the additional term the subdominant large-$x$ asymptotic, defining
the fundamental Sibuya solution, becomes
\eq
y(x,E,\alpha,l) \sim
\frac{x^{-M/2 -\alpha/2}}{ \sqrt{2 i}}
\exp \lf( - \fract{x^{M+1}}{M+1} \ri).
\label{yas3}
\en
For integer $k$ we then generate a set of solutions to (\ref{shg})
\eq \fl
{}~~~~
y_k(x,E,\alpha,l)= \omega^{k/2 +  k\alpha/2}
y(\omega^{-k}  x,\;\omega^{2k} E,\; \omega^{(M+1)k}\alpha,\;l)
\quad,\quad \omega = e^{\frac{\pi i}{M+1}}~,
\en
with each pair $\{y_k,y_{k+1} \}$ forming a basis of solutions.
Therefore, just as before, there must be a Stokes relation of the form
\eq
C(E,\alpha,l) y_0(x,E,\alpha,l)=y_{-1}(x,E,\alpha,l)+
y_{1}(x,E,\alpha,l)\,,
\label{Tya}
\en
with Stokes multiplier $C(E,\alpha,l)$
\eq
C(E,\alpha,l)=W[\,y_{-1}\,, y_1]/W[\,y_0,y_1]=
W[\,y_{-1}\,, y_1]~.
\label{tform2}
\en
Since $\omega^{M+1}=-1$, the only values of $\alpha$ to arise in the
set of functions $y_k(x,E,\alpha,l)$, related by (\ref{Tya}),
are $\alpha$ itself, and $-\alpha$. This
makes it convenient to define
\eq
C^{(\pm)}(E) := C(E, \pm \alpha,l)~,\quad y^{(\pm)}(x,E):=
y(x,E,\pm \alpha,l)
\en
after which taking (\ref{Tya}) at  positive and
negative $\alpha$ implies the
pair of coupled equations
\eq \fl
{}~~C^{(+)}(E) y^{(+)}(x,E) =\omega^{-(1+\alpha)/2}
y^{(-)}(\omega  x, \omega^{2 M}  E)+
 \omega^{(1+\alpha)/2} y^{(-)}(\omega^{-1}  x,\omega^{-2 M}  E)\,,
\label{peqpa}
\en
\eq \fl
{}~~C^{(-)}(E) y^{(-)}(x,E) =\omega^{-(1-\alpha)/2}
y^{(+)}(\omega  x, \omega^{2 M}  E)+
 \omega^{(1-\alpha)/2} y^{(+)}(\omega^{-1}  x,\omega^{-2 M}  E) \,.
\label{peqp}
\en
To remove the $x$-dependence we project onto solutions
$\psi_\pm(x,E,\alpha,l)$ which are defined at the
origin by
\eq
\psi(x,\alpha,l)=\psi_{+}(x,\alpha,l)  \sim x^{l+1} +\dots \quad ,
\quad x\to 0\,
\en
and $\psi_{-}(x,\alpha,l) = \psi(x,\alpha,-1{-}l)$.
The subscripts $\pm$ should not be
  confused with the bracketed superscripts $(\pm)$, which are used to
  differentiate  positive and negative values of $\alpha$.
We set
\eq
D(E, \alpha,l) = W[y,\psi_+](E,\alpha,l)~.
\en
Then, reasoning as before, the  Stokes relations
(\ref{peqpa}) and (\ref{peqp}) imply
\begin{eqnarray}
\fl
{}~~ C^{(+)}(E) D^{(+)}(E) =\omega^{-(2l+1+\alpha)/2}
D^{(-)}(\omega^{2M} E)+
\omega^{(2l+1+\alpha)/2} D^{(-)}(\omega^{-2M}  E)\, ,
\label{tqa}    \\[4pt]
\fl
{}~~  C^{(-)}(E) D^{(-)}(E) =\omega^{-(2l+1-\alpha)/2}
D^{(+)}(\omega^{2M} E)+ \omega^{(2l+1-\alpha)/2}
D^{(+)}(\omega^{-2M}  E)\, ,
\label{tqb}
\end{eqnarray}
where  $D^{(\pm)}(E)=D(E,\pm \alpha,l)$ are easily identified with the
spectral determinants for the radial problems (\ref{shg}) taken at
$\pm \alpha$.  These equations entwine the spectral problem at
$+\alpha$ with that at $-\alpha$, and reduce to two copies of the
six-vertex TQ relation for $\alpha=0$.
At general $\alpha$ \cite{Sc}
they match equations satisfied by the $\TT$- and
$\QQ$-operators for the 3-state
Perk-Schultz lattice model
\cite{PS,Schula,Schulb}, a model with
$U_q(\widehat {gl}(2 |1))$ symmetry.

\subsection{$\PT$-symmetry and reality proofs}
\label{realproof}
For maximal generality, we
consider  the three-parameter family of Hamiltonians
$ \CH_{M,\alpha,l}$
and look for solutions of the corresponding Schr\"odinger
equation
\eq
\lf [ -\frac{d^{2}}{dx^{2}} -(ix)^{2M} - \alpha
(ix)^{M-1}+\frac{l(l+1)}{x^{2}} \ri] \psi_k(x) = \lambda_k
\psi_k(x)
\label{PTg}
\en
that decay simultaneously at both ends of a contour $\CaC$ which can
be taken to be the real axis if $M{<}2$, suitably distorted to pass
below the origin if $l(l+1) \neq 0$. For larger values of $M$ the
contour should be further distorted so as to remain in the pair of
Stokes sectors which includes the real axis at $M{=}1$, as
illustrated in figure \ref{sectors} of  section \ref{funode}.
We shall prove \\
\noindent {\bf Theorem:} The spectrum of $\CH_{M,\alpha,l}$ is
\begin{itemize}
\item {\it real} for $M>1$ and $\alpha < M+1+|2l+1| $~;
\item {\it positive} for $M>1$ and $\alpha < M+1-|2l+1|$~.
\end{itemize}
This result
includes the cases considered in section~\ref{prelude}.

First we remove the factors of $i$ by setting $\Phi(x)=
\psi(x/i)$, so that the Schr\"odinger problem becomes
\eq \fl \ \
\Bigl[-\frac{d^{2}}{dx^{2}}+x^{2M}+\alpha x^{M-1}+
\frac{l(l+1)}{x^{2}} \Bigr]\Phi_k(x)=-\lambda_k\,\Phi_k(x)\,,
{}~~~\Phi_k(x) \in L^2(i\CaC)\, ,
\label{shl}
\en
and has the same form as (\ref{shg}) with $E=-\lambda_k$, though
with different boundary conditions: to qualify as an eigenfunction,
$\Phi$ must decay as $|x|\to\infty$ along the contour $i\CaC$.
However, it is an easy generalisation of the discussion in section
\ref{dict} that  the  function $C(-\lambda,\alpha,l)$ defined in
(\ref{tform2}) is the spectral determinant associated to the
spectral problem (\ref{shl}). This identification will allow us to
prove the theorem using techniques inspired by the Bethe ansatz.

We start from   equation (\ref{tqa}):
\eq
\fl \qquad
C^{(+)}(E) D^{(+)}(E) =\omega^{-(2l+1+\alpha)/2} D^{(-)}(\omega^{-2}
E)+ \omega^{(2l+1+\alpha)/2} D^{(-)}(\omega^{2} E)\, ,
\label{tq11}
\en
and  define the zeros of $C^{(+)}(E)=C(E,\alpha,l)$ to be the set
$\{ -\lambda_k\}$, and the zeros of $D^{(\pm)}(E)$ to be the set
$\{E^{(\pm)}_k\}$. Setting $E=E^{(\pm)}_k$ in (\ref{tq11}) leads
back to a (coupled) set of Bethe ansatz equations for the zeros of
$D^{(\pm)}$, much as before. However we can also investigate the
effect of setting $E=-\lambda_k$. The left-hand side of (\ref{tq11})
is again zero, so the equation can be rearranged to read
\eq
D^{(-)}(\omega^{2} \lambda_k)/
D^{(-)}(\omega^{-2} \lambda_k)
= -\omega^{-2l-1-\alpha}\,.
\en
For $M>1$, WKB estimates show that the function $D^{(-)}(E)$ has
order less than one, so that it can be written as a simple product
over its zeros, as in (\ref{had}) for the
homogeneous potentials.  Using this factorised
form gives the following set of constraints on the  $\lambda_k$'s:
\eq
\prod_{n=1}^{\infty} \lf( { E^{(-)}_n + \omega^{2} \, \lambda_k
\over E^{(-)}_n +  \omega^{-2} \, \lambda_k} \ri) = - \omega^{-2l-1
- \alpha }\, ,\qquad k=1,2,\dots~.
\label{bb}
\en
Notice what has been achieved here -- the little-understood set of
numbers $\{\lambda_k\}$, eigenvalues of a non-Hermitian eigenvalue
problem, are being constrained by the much better-understood set
$\{E_n^{(-)}\}$\,.
Since the original eigenproblem (\ref{PTg}) is invariant under $l\to
-1{-}l$, we can assume $l\ge -1/2$ without any loss of generality.
Then each $E^{(-)}_n$ is an eigenvalue of an Hermitian operator
$\CH_{M,-\alpha,l}$, and hence is real. Furthermore a Langer
transformation~\cite{La} (see also~\cite{BLZa,DTb}) shows that the
$E_n^{(-)}$ solve a generalised eigenproblem with an
everywhere-positive `potential', and so are all positive, for
$\alpha<1{+}2l$. This can be sharpened by considering the value of
$D^{(-)}(E)|\phup_{E=0}$\,, found by an easy generalisation
of the result obtained in
appendix~\ref{asy_app} for the cases with $\alpha=0$. From
\eq
D^{(\pm)}(E)|\phup_{E=0}=
\sqrt {\frac{2}{i}}\Bigl(\fract{M{+}1}{2}\Bigr)^{\frac{2l+1\mp\alpha}{2M+2}
         + \frac{1}{2}}
\frac{\Gamma\left(\frac{2l+1}{M+1}+1\right)}
{\Gamma\left(\frac{2l+1\pm\alpha}{2M+2}+\frac{1}{2}\right)}~.
\label{dalp0}
\en
we see that  $D^{(-)}(E)|\phup_{E=0}$ first vanishes when
$\alpha=M{+}2l{+}2$. Until this
point is reached, no  eigenvalue $E^{(-)}_n$ can have passed the
origin, and all must be positive.
(It might be worried that negative eigenvalues could appear from
$E=-\infty$\,, but this possibility can be ruled out by a consideration
of the Langer-transformed version of the equation.)

Taking the modulus${}^2$ of (\ref{bb}), using
the reality of the $E^{(-)}_k$, and writing the eigenvalues of
(\ref{PTg}) as $\lambda_k=|\lambda_k|\exp(i\,\delta_k)$, we have
\eq
\prod_{n=1}^{\infty} \lf ( { (E^{(-)}_n)^{2} + |\lambda_k|^2  + 2  E^{(-)}_n
 |\lambda_k| \cos(\fract{2 \pi}{M+1} + \delta_k)
 \over
(E^{(-)}_n)^{2} + |\lambda_k|^2  + 2  E^{(-)}_n
 |\lambda_k| \cos(\fract{2\pi}{M+1} - \delta_k)}
\ri)
= 1\, .
\label{abs}
\en
For $\alpha<M+2l+2$\,, all the $E^{(-)}_n$ are positive,
and each single term in the product on the LHS of (\ref{abs}) is either
greater than, smaller than, or equal to one
depending  only on the relative values of the cosine terms in the numerator
and denominator. These are
independent of the index $n$.
Therefore the only possibility to match the RHS is for each term in the
product to be individually equal to one, which for $\lambda\phup_k\neq 0$
requires
\eq
\cos(\fract{2\pi}{M+1}+\delta_k)=\cos(\fract{2\pi}{M+1}-\delta_k)
\quad
\mbox{or}
\quad
\sin(\fract{2\pi}{M{+}1})\sin(\delta_k)=0\,.
\en
Since  $M > 1$, this latter condition implies
\eq
\delta_k= m \pi\, ,~~~~m \in \ZZ
\en
and this establishes the reality of the
eigenvalues of (\ref{PTg})
 for $M{>}1$ and $\alpha{<}M+2l+2$ or, relaxing the condition on
$l$, $\alpha{<}M+1+|2l{+}1|$.

The plots and discussion in section \ref{prelude} indicate that
most of the $\lambda_k$ become complex as $M$ falls below $1$,
at least for $\alpha=0$.  We now see that this coincides with the
point at which the
order of
$D^{(-)}(E)$ is greater than $1$,
the factorised form of $D^{(-)}(E)$ provided by Hadamard's theorem
no longer has such a simple form, and the proof just given breaks down.

The borderline case $M=1$ is the simple harmonic oscillator,
exactly solvable for all
$l$ and $\alpha$\ in terms of confluent hypergeometric functions.
The case $\alpha=0$ is discussed in detail in appendix \ref{app:SHO};
since a nonzero value of $\alpha$ can be absorbed in a simple shift of
$E$, it is easily seen that the correctly-normalised solution
for the general case is
\eq
y(x,E,\alpha,l)= \frac{1}{\sqrt{2i}}x^{l+1} e^{-x^{2}/2}
U(\half(l+\fract{3}{2}) - \fract{
E-\alpha}{4},l+\fract{3}{2} , x^{2})\,,
\en
and that
\eq
C(E,\alpha,l)|\phup_{M=1}=\frac{2\pi}
{\Gamma\Bigl(\frac{1}{2} +\frac{2l+1+E-\alpha}{4}\Bigr)
\Gamma\Bigl(\frac{1}{2} -\frac{2l+1-E+\alpha}{4}\Bigr)}\,.
\en
Thus the eigenvalues of (\ref{PTg}) for $M=1$ are
$\lambda_k=4k-2-\alpha\pm(2l{+}1)$, $k=1,2,\dots$~. They are all
real for {\em all}\/ real values of $\alpha$ and $l$, and are all
positive for $\alpha{<}2-|2l{+}1|$.

To discuss positivity at general values of $M>1$, we can continue in
$M$, $\alpha$ and $l$ away from a point in the region $\{M{=}1,\,
\alpha{<}2-|2l{+}1|\}$.
So long as $\alpha$ remains less than $M+1+|2l{+}1|$, all eigenvalues will
be confined to the real axis during this
process, and the first passage of an eigenvalue from positive
to negative values will be signalled by the presence of
a zero in $C(-\lambda,\alpha,l)$ at $\lambda=0$.
Using (\ref{dalp0}) we have
\eq
C(E,\alpha,l)|\phup_{E=0}=
\Bigl(\fract{M{+}1}{2}\Bigr)^{\frac{\alpha}{M+1}}
\frac{2\pi}
{\Gamma\left(\frac{1}{2} +\frac{2l+1-\alpha}{2M+2}\right)
\Gamma\left(\frac{1}{2} -\frac{2l+1+\alpha}{2M+2}
\right)}~,
\label{Calpha}
\en
and so the first zero arrives at $E=-\lambda=0$ when
$\alpha=M+1-|2l{+}1|$. Thus for all $\alpha{<}M+1-|2l{+}1|$\,, the
spectrum is entirely positive, as claimed.

\begin{figure}[t]
\begin{center}
\includegraphics[width=0.33\linewidth]{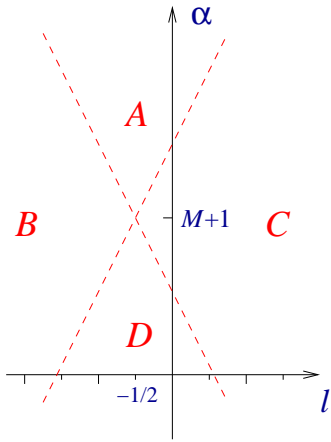}
\end{center}
\caption{The approximate `phase diagram' at fixed $M$. The proof in
the text shows that the spectrum is
entirely real in regions $B$, $C$ and $D$, and positive in
region $D$.\label{regions}
}
\end{figure}

Referring to the regions $A$, $B$, $C$ and $D$ shown on
figure~\ref{regions}, the proof establishes reality
for $(\alpha,l) \in  B \cup C\cup D$, and  positivity  for
$(\alpha,l) \in D$. This doesn't mean that the spectrum immediately
acquires complex elements when the region $A$ is entered, though.
In \cite{DDTc} the shape of the domain of unreality was
investigated in more detail, for the special case $M=3$\,. (This value
was chosen because the $\PT$-symmetric problem is there equivalent to
a radial problem, making the numerics a little easier to handle. This
equivalence will be explained in the next subsection.)

\begin{figure}[h]
\begin{center}
\includegraphics[width=0.7\linewidth]{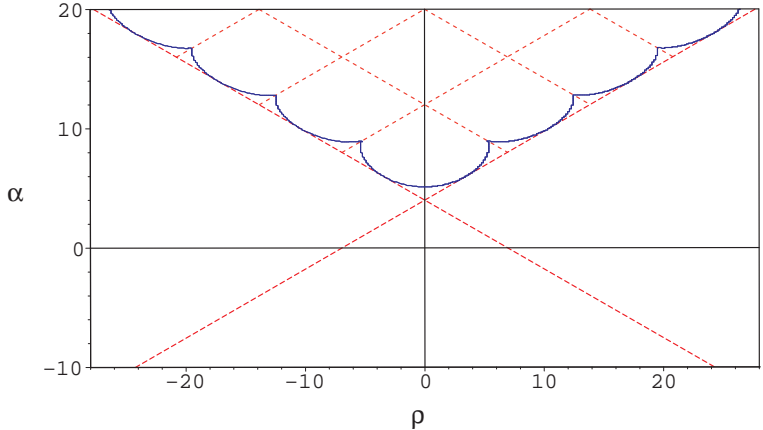}
\end{center}
\caption{The domain of unreality of $\CH_{3,\alpha,\rho}$, where
$\rho=\sqrt{3}(2l{+}1)$.  Also shown are segments
of the lines along which the problem has a protected zero-energy
level.\label{fign}}
\end{figure}

\noindent
{}Figure~\ref{fign} repeats the plot from \cite{DDTc} already shown in
the prelude, this time superimposing the boundaries of regions $A$,
$B$, $C$ and $D$.  The full domain of unreality (the
interior of the cusped line),  only touches the borders of region
$A$ at isolated points.  There is also a small,
approximately-triangular region inside $A$
near $\rho=0$ within which the spectrum is entirely real and
positive, despite being outside region $D$.

In fact, the condition $\alpha{<}M+1+|2l+1|$, which arose as a
technical point in the proof, has a rather more physical explanation
\cite{DDTc}.  Along the lines $\alpha=M+1\pm(2l+1)$ the problem can
be reformulated in terms of a $\PT$-symmetric version of
supersymmetric quantum mechanics. On the supersymmetric lines the
model has a protected zero-energy state which permits eigenvalues to
become degenerate and pair off into complex conjugate pairs. For
$M=3$ this occurs at the points where the cusped line touches the
borders of $A$ on figure \ref{fign}.

The idea of a protected zero-energy level
also leads to a partial explanation for the
pattern of cusps seen in figure~\ref{fign}. In addition to the
supersymmetric lines $\alpha=M{+}1\pm(2l{+}1)$,
there is a protected zero-energy state along {\it all}\/
line segments  shown in figure~\ref{fign}: this follows from
(\ref{Calpha}), which shows that $C(E,\alpha,l)$ vanishes whenever
$\alpha=(M{+}1)(2n+1)\pm (2l{+}1)$, $n=0,1,\dots$~. For $n\ge 1$
these are
precisely the lines on which cusps would be expected to be found,
as follows from a consideration of the mergings of the relevant
levels, and the fact -- special to $M=3$ --
that the merging levels enjoy an $E\to -E$
symmetry.

\begin{figure}[h]
\begin{center}
\includegraphics[width=0.72\linewidth]{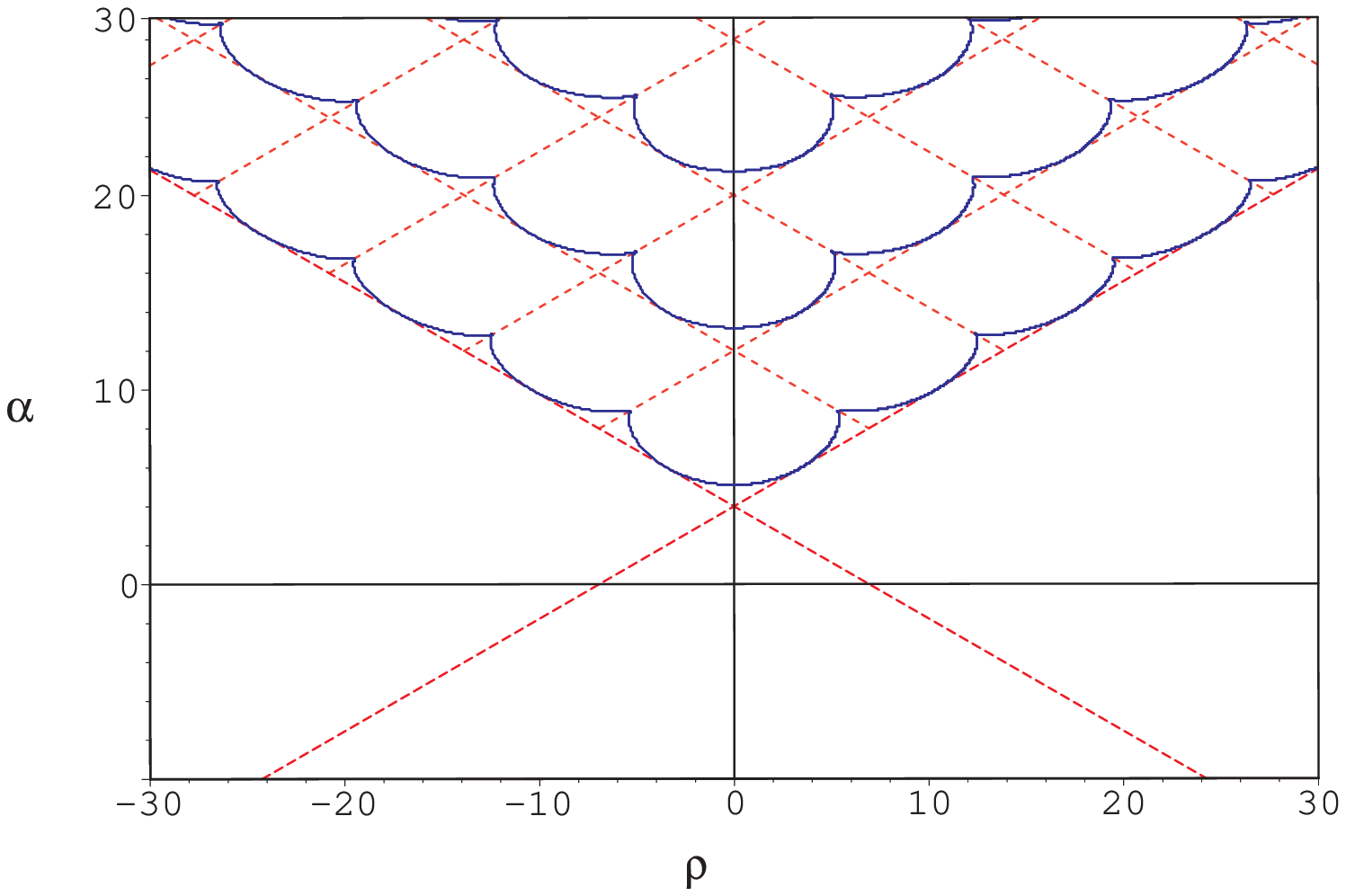}
\end{center}
\caption{An expanded view of figure \protect\ref{fign}, showing
the frontiers of the
regions with zero, two, four and six complex levels.
\label{fignl}}
\end{figure}

The cusped line marking the frontier of the zone of reality is in fact
only the first of infinitely-many such lines -- figure \ref{fignl}
shows some further lines across which pairs of energy levels become
complex.

Finally, we remark that Shin has extended the Bethe-ansatz-inspired
ideas of the main proof given above to establish spectral
reality for a somewhat wider class of non-Hermitian
potentials~\cite{Shin:2002vu,Shin:2002ay}.

\subsection{Curiosities at $M=3$}
\label{spec_cur}
The correspondence with Bethe ansatz systems can also reveal
unexpected relationships between the spectral properties of
apparently very different problems. One set of examples, obtained in
\cite{DDTb}, connects quantum-mechanical problems via a third-order
equation. Consider the radial problem studied in section~\ref{sps}
at $M=3$. For later convenience, we also swap $l$ for the variable
$\rho:=\sqrt{3}(2l{+}1)$ already used in figure \ref{fign}, so that
the differential equation is
\begin{equation}
\left[-\frac{d^{2}}{dx^{2}}+x^{6}+\alpha x^{2}+
\frac{\rho^2-3}{12\,x^{2}}\,\right]\Phi(x) = E\,\Phi(x)\,.
\label{eqe2}
\end{equation}
The spectrum is encoded in the spectral determinant
$D(E,\alpha,l)|_{M=3}$.

\begin{figure}[h]
\begin{center}
\includegraphics[width=0.28\linewidth]{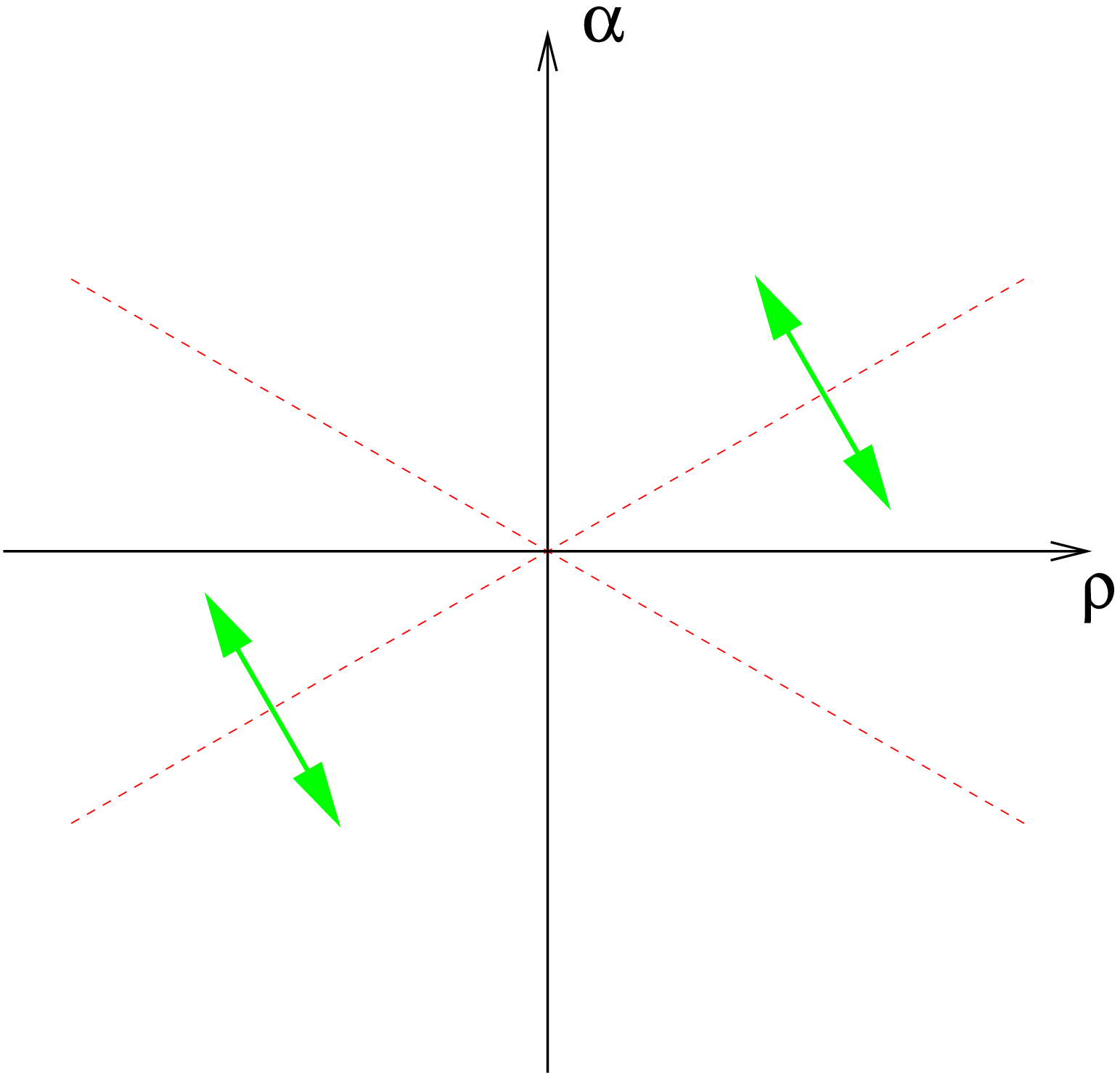}
{}~~~~~~~
\includegraphics[width=0.28\linewidth]{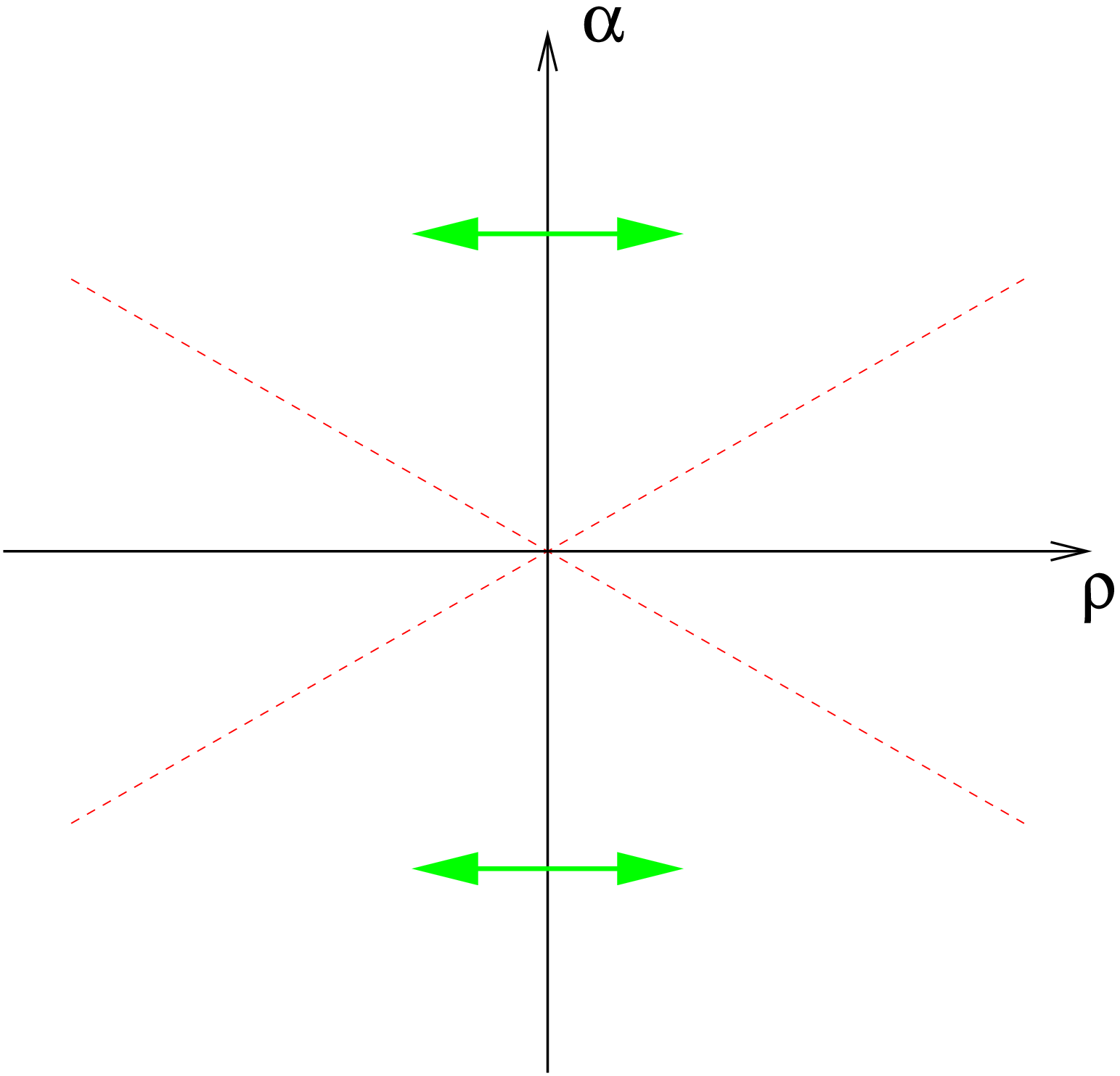}
{}~~~~~~~
\includegraphics[width=0.28\linewidth]{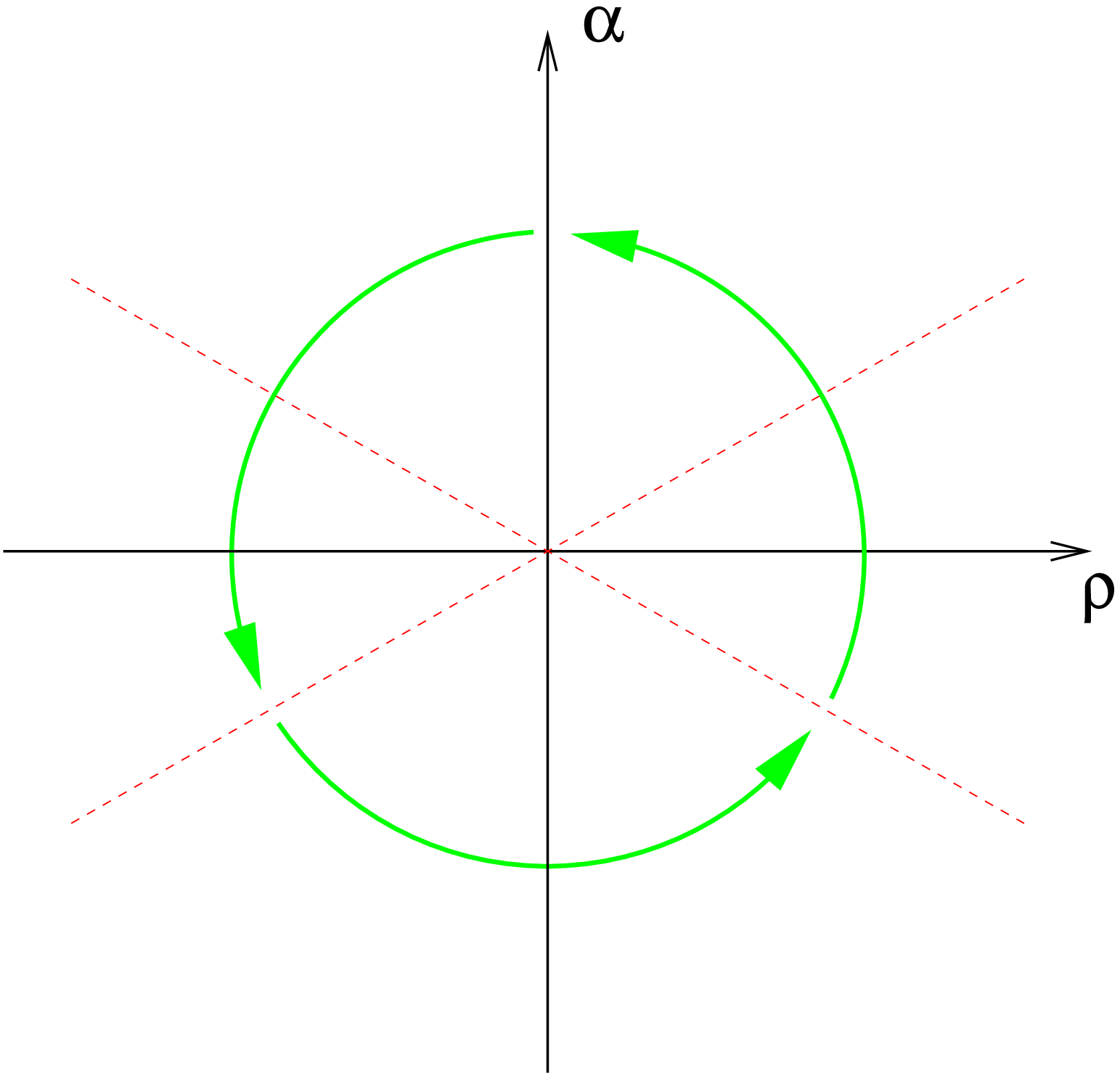}
\\[1pt]
\parbox{0.28\linewidth}{~~\small {\bf (a)} $\balpha\mapsto\TT\balpha$}
{}~~~~~~~
\parbox{0.28\linewidth}{~~\small {\bf (b)} $\balpha\mapsto\LL\balpha$}
{}~~~~~~~
\parbox{0.28\linewidth}{~~\small {\bf (c)} $\balpha\mapsto\LL\TT\balpha$}
\end{center}
\caption{The three mappings involved in the spectral
equivalences.\label{equivs}}
\end{figure}

On the other hand, we can consider a {\it third}\/-order ordinary
differential equation:
\begin{equation}
\left[ D(g_2{-}2)D(g_1{-}1)D(g_0) + x^{3}\,\right]\phi(x)=
\frac{3\sqrt{3}}{4}\,E\,\phi(x)
\label{thord}
\end{equation}
where $D(g)\equiv (\frac{d}{dx}-\frac{g}{x})$. Third-order equations
of this type are generally associated with $SU(3)$-related Bethe
ansatz equations, as shown in \cite{DTc,DDTa}, but for the
particular `potential' $x^{3}$ that appears in (\ref{thord}), they
collapse onto $SU(2)$-type equations
\cite{DDTb}. With the choice
\begin{equation}
g_0=1{+}(\alpha{+}\sqrt{3}\rho)/4~,~~
g_1=1{+}\alpha/2~,~~
g_2=1{+}(\alpha{-}\sqrt{3}\rho)/4~,
\end{equation}
the collapse is onto the same
set of equations as govern the problem (\ref{eqe2}).
This leads to a rather remarkable spectral equivalence between
pairs of second- and third- order ordinary differential equations.
Furthermore, the
third-order equation is symmetrical in $\{g_0,g_1,g_2\}$, a
feature which is completely hidden in the original second-order
equation. By playing with this symmetry, some
novel spectral equivalences can be established between different
second-order radial problems, and also between these and certain
lateral problems \cite{DDTb}.
The parametrisation in terms of $\alpha$ and $\rho$ allows these
relations to be expressed in a compact way
in terms of certain $2\times 2$ matrices in the Weyl
group of $SU(3)$ \cite{DDTcz}. With
$\balpha=
\Big(\!\begin{array}{c}\alpha\\[-4pt] \rho\end{array}\!\Big)$\,, let
\begin{eqnarray*}
\mbox{Radial}(\balpha)&=&{\rm Spect}\left(\,\mbox{(\ref{eqe2}) with radial
boundary conditions}\,\right) \\
\mbox{Lateral}(\balpha)&=&{\rm Spect}\left(\,\mbox{(\ref{eqe2}) with
lateral boundary conditions}\,\right)
\end{eqnarray*}
and define matrices $\LL$\,, $\TT$ and $\LL\TT$ by
\begin{equation}
\fl\qquad
{\mathbb L}=\left(\begin{array}{cc}
1&0\\
0&-1
\end{array}\right)~
 , \quad
\TT=\frac{1}{2}\left(\begin{array}{cc}
-1&\sqrt{3}\\
\sqrt{3}&\,1
\end{array}\right)~
 , \quad
\LL\TT=\frac{1}{2}\left(\!\begin{array}{cc}
-1&\sqrt{3}\\
-\sqrt{3}&\,-1
\end{array}\right).
\end{equation}
The reflections $\LL$ and $\TT$
together generate the Weyl group of $SU(3)$, with their product $\LL\TT$
being an anticlockwise rotation in the $(\rho,\alpha)$ plane by $2\pi/3$.
The three mappings are illustrated in figure \ref{equivs}.
With this notation in place, three simply-stated spectral
equivalences hold:
\begin{eqnarray*}
{\bf (a)}~~~\mbox{Radial}(\balpha)&=& \mbox{Radial}(\TT\balpha)\\[3pt]
{\bf (b)}~~~\mbox{Lateral}(\balpha)&=& \mbox{Lateral}(\LL\balpha)\\[3pt]
{\bf (c)}~~~\mbox{Lateral}(\balpha)&=& \mbox{Radial}(\LL\TT\balpha)~.
\end{eqnarray*}

Of these, the second is trivial -- it just encodes the $l\to -1-l$
symmetry of the lateral problem -- but the other two are not. The
third, {\bf (c)}, gives a direct insight into spectral reality for
$M=3$: it demonstrates that the $\PT$-symmetric problem for this
case is equivalent to an explicitly Hermitian problem. Notice that
the domain of unreality for the $\PT$-symmetric situation -- shown
in figure \ref{fign} above the wavy line -- is rotated under {\bf
(c)} into a region of negative $\rho$, where $l$ is sufficiently
negative that the radial problem $\mbox{Radial}(\LL\TT\balpha)$ is
also non-Hermitian and so no contradiction with unreality arises.
(Interestingly, there are similar equivalences
\cite{Andrianov:1982,bg,Jones:2006qs} for a subset of the
non-Hermitian $M=2$ problems, though the methods of proof in those
papers are very different.)

In the $(\alpha,\rho)$ coordinates adapted to $M=3$ the
lines with a protected zero energy level shown on figure \ref{fign}
are $\alpha = 4J\pm\rho/\sqrt 3$, $J=1,3,5,\dots$~, and they have
an extra significance \cite{DDTc}: they correspond to points at
which the model exhibits a $\PT$-symmetric version of the quasi-exact
solvability discussed in \cite{Tur,Ush}, with an {\em odd}\/ number
$J$ of `QES' eigenfunctions and eigenvalues which can be found
algebraically. In fact, a further spectral equivalence from
\cite{DDTb}, related to a higher-order supersymmetry
\cite{KP0,Ao0,DDTb},
can be used to show that if $\balpha$ lies on the line
$\alpha=4J+\rho/\sqrt{3}$, with $J$ an
integer, then the {\em non}-QES part of the
$\PT$-symmetric spectrum $\mbox{Lateral}(\balpha)$ is the same
as the {\em full}\/ $\PT$-symmetric
spectrum $\mbox{Lateral}(\TT\balpha)$.
(Likewise, if $\balpha$ lies on $\alpha=4J-\rho/\sqrt{3}$ then there
is a mapping of the non-QES spectrum onto
$\mbox{Lateral}(\LL\TT\LL\balpha)$\,.)
If $\balpha$ is inside the region $A$, so that some levels of
$\mbox{Lateral}(\balpha)$ might be
complex, then, looking at figure \ref{equivs}a,
$\TT\balpha$ must be outside this region. Thus the
non-QES part of $\mbox{Lateral}(\balpha)$ is real, and so the
levels which become complex
as $\balpha$ moves into region $A$ through QES values
must lie in the solvable part of the spectrum \cite{DDTc}.
This complements a similar, but as yet unproven, conjecture
concerning quartic QES potentials \cite{BB4}.

\subsection{Further generalisations}
\label{Gen}
The ODE/IM correspondence introduced here for $SU(2)$ Bethe-ansatz
systems has been extended to a number of other integrable models and
ordinary differential equations.
To finish, we present a very
schematic summary of the integrable models and differential
equations involved in correspondences to date. The cited references
should be consulted for more details of the models and
parameters involved in
each case.
\begin{itemize}
\item{
{\bf Six-vertex model, $\boldsymbol{su(2)}$
  spin-$\boldsymbol{\fract{1}{2}}$ XXZ quantum  chains and the
  perturbed 
boundary sine-Gordon model}~\cite{DTa,BLZa,DTb,Sa}:
\newline As already described,
the ground state of the six-vertex model is related to the spectral
properties of the ordinary differential equation
\eq
\Bigl[-\frac{d^2}{dx^2}+x^{2M}+ \frac{l(l+1)}{x^2} -E
\Bigr]\Psi(x)=0~.
\label{ge1}
\en
This correspondence has been generalised \cite{BLZhigh} to
map the excited
states of the integrable model to the following differential
equations:
\eq
\fl \
 \Bigl[-\frac{d^2}{dx^2}+x^{2M}+ \frac{l(l+1)}{x^2} -2 {d^2 \over
dx^2} \sum_{k=1}^L \ln(x^{2M+2}-z_k)-E \Bigr]\Psi(x)=0~,
\label{ge2}
\en
where the constants $\{z_k\}$ satisfy
\eq
\fl
\sum_{{j=1\atop j \ne k}}^{{\atop L}}
{z_k(z_k^2+(M{+}3)(2M{+}1) z_k z_j +M(2M{+}1)
z_j^2) \over (z_k -z_j)^3} - {M z_k \over 4(M{+}1)} + \Delta=0\,,~
\en
and
\eq
\Delta= {(2 l+1)^2 - 4 M^2 \over 16(M{+}1)}~.
\en
 }
These equations were also obtained in \cite{Fioravanti:2004cz}. Very
recently, they and the simpler equation for the ground state have 
been interpreted as special cases of Langlands duality
\cite{feiginfrenkel}.
\item{{\bf Perk-Schultz  models and the perturbed
hairpin model of boundary
interaction}\cite{Sc,DDTb,
Fateev:2005kx}:
\newline
Adding a inhomogeneous term to (\ref{ge1}) gives the ordinary
differential equation discussed in section \ref{sps}:
\eq
{\Big [}-\frac{d^2}{dx^2}+x^{2M}+\alpha x^{M-1}+ \frac{l(l+1)}{x^2}
-E {\Big]}\Psi(x)=0\,.
\label{sh2}
\en
Via a change of variables this is equivalently 
\eq
\Bigl[ - \frac{d^2}{dx^2} - p^2  +2q \kappa e^x  + \kappa^2
(e^{2x}+e^{-nx}) \Bigr]\Psi(x)=0\,,
\label{hairp}
\en
the form that was used in \cite{Fateev:2005kx}.
} 

\item{{\bf Spin-$\boldsymbol{j}$ $\boldsymbol{su(2)}$ quantum
      chains and the boundary parafermionic sinh-Gordon
      model}\cite{Luky,zampf,lukypf}:
\newline
The higher spin $su(2)$ chains are found by a process known as fusion,
and their ground state Bethe ansatz roots group into characteristic
`strings' in the imaginary direction of length $2j$. This behaviour is
captured by generalised eigenvalue problems of the form
\eq
\Bigl[-\frac{d^2}{dx^2}+(x^{2M}-E)^{2j}+ \frac{l(l+1)}{x^2}
\Bigr]\Psi(x)=0
\label{spinj}
\en
or equivalently 
\eq
\lf[-\frac{d^2}{dx^2} + \kappa^2(e^{-\frac{bx}{Q}
}+e^{\frac{x}{bQ}})^{2j}-\xi^2\ri]\Psi (x)=0
\label{bpf}
\en
where $Q=b+b^{-1}$. The formation of strings for this equation is
discussed in \cite{ddmst}.}

\item{ {\bf $\boldsymbol{su(n)}$ vertex
      models}~\cite{DTc,Sb,DDTa,BHK,Luky,ddmst}:
\newline
To encode Bethe ansatz systems related to $su(n)$ with $n>2$, it turns
out to be necessary to go to higher-order ordinary differential
equations:
\eq
\Bigl[(-1)^{n}D_n({\bf g})-P_K(x,E) \Bigr]\Psi(x)=0\,,
\label{sun}
\en
where ${\bf g}=\{g_0,g_1\,\dots\,g_{n-1}\}$ is a collection of
`twist' parameters, 
\eq
\fl \
D_n({\bf g})=D(g_{n-1}-(n{-}1))\,\dots\, D(g_1-1)\,D(g_0)~,
{}~~~D(g)=\left(\frac{d}{dx}-\frac{g}{x}\right)
\en
and 
\eq 
P_K(x,E)=(x^{h M/K} -E)^K
\label{PKdef}
\en
with $h=n+1$, the dual Coxeter number of $su(n)$, and the integer $K$
gives the degree of fusion of the vertex model.
}
\item{{\bf Finite spin-$\boldsymbol{j}$ XXZ quantum chains at
      $\boldsymbol{\Delta=\pm \fract{1}{2}}$}\cite{dst}:
\eq
\Bigl[
   -\frac{d^2}{dx^2} - \frac{N(N+1)}{ \cosh^2 x}
+  \frac{M(M+1)}{ \sinh^2 x} +\sigma^2  \Bigr] \Psi(x)
 =0\,,
\en
with $\sigma=(m+1)/(2j+2)$. }

\item{{\bf Coqblin-Schrieffer model}\cite{Bazhanov:2003ua}:
\eq
\Bigl[\left(-i \frac{d}{dx} +h_1 \right) \dots \left(-i
\frac{d}{dx}+h_n \right) -e^{n \theta} e^{x}x   \Bigr]\Psi(x)=0\,.
\label{cs}
\en
Notice that, after a variable change, this 
is a particular case of (\ref{sun}).
}
\item{{\bf Circular 
Brane}\cite{Lukyanov:2003rt}:
\eq
\Bigl[ - \frac{d^2}{dy^2} 
+ h^2 e^y 
+ \kappa^2 \exp(e^y)
\Bigr]\Psi(y)=0\,.~
\label{circb}
\en
}
\item{{\bf Paperclip
      models}\cite{Lukyanov:2003nj,Lukyanov:2005nr}:
\eq
\Bigl[ - \frac{d^2}{dx^2} - p^2 {e^x \over 1+e^x} -(q^2-\frac{1}{4})
{e^x \over (1+e^x)^2} + \kappa^2 (1+e^x)^n \Bigr]\Psi(x)=0\,.~
\en
}
Note, setting $x=y-\log n$ and taking the limit $n\to\infty$ reduces 
this to (\ref{circb}), the circular brane. The paperclip models and
the perturbed boundary hairpins of \cite{Fateev:2005kx} both exhibit 
the conformal hairpin boundary condition in their ultraviolet limits, 
but the perturbations are different.

\item{{\bf Finite spin-$\boldsymbol{\fract{1}{2}}$ XYZ quantum chain
      (off-critical deformation of the
$\boldsymbol{\Delta{=}-\fract{1}{2}}$ XXZ
model)}\cite{bm}:
\eq
6 q {\partial \over \partial q} \Psi(x,q) = \Bigl[ -{ \partial^2
\over
\partial x^2} + 9 n(n+1) \Wp(3 x|q^3)+c(q,n) \Bigr] \Psi(x,q)
\en
where $\Wp$ is  the Weierstrass elliptic function,  $q=e^{i \pi \tau}$,
$\tau$ is the modular parameter,
\eq
c(q,n)= -3 n(n+1) {\vt'''_1(0,q^3) \over \vt'_1(0,q^3)}\,,
\en
and  $\vt_1$ is the  elliptic theta function. }
\item{ {\bf $\boldsymbol{so(2n)}$ vertex models}\cite{ddmst}:
\eq
\fl \ \left[ D_{n}({\bf g^{\dagger}}) \left( \frac{d}{dx}
\right)^{-1}\!
D_{n}({\bf g}) -\sqrt{P_K (x,E)} \left(\frac{d}{dx} \right)
\sqrt{P_K (x,E)}\,\right]\Psi(x,E,{\bf g})=0\,,
\en
where ${\bf g}=\{g_0,g_1\,\dots\,g_{n-1}\}$ is a set of twist
parameters as for $su(n)$, \newline
${\bf g}^\dagger =  \{n{-}1{-}g_0 , n{-}1{-}g_1
,\dots,n{-}1{-}g_{n-1} \}$\,, and the generalised potential term
$P_K(x,E)$ is as in (\ref{PKdef}), but with $h=2n{-}2$.
}

\item{{\bf $\boldsymbol{so(2n+1)}$ vertex models}\cite{ddmst}:
\eq
\fl \ \left[ D_{n}({\bf g^{\dagger}}) D_{n}({\bf g}) + \sqrt{P_K (x,E)}
\left(\frac{d}{dx} \right) \sqrt{P_K (x,E)}\,\right]\Psi(x,E, {\bf
g})=0\,,
\en
where $P_K(x,E)$ is as in (\ref{PKdef}) with $h=2n{-}1$.
}

\item{{\bf $\boldsymbol{sp(2n)}$ vertex models}\cite{ddmst}:
\eq
\fl \  \left[ D_{n}({\bf g^{\dagger}}) \left(\frac{d}{dx}
\right)D_{n}({\bf g})   -P_K(x,E)  { \left(d
\over dx \right)^{-1}}P_{K} (x,E)\,\right]\psi(x,E, {\bf g})=0\,,
\en
where $P_K(x,E)$ is as in (\ref{PKdef}) with $h=n{+}1$.
}

\end{itemize}

\newpage

\resection{Conclusions}
\label{dev}

The main conclusion of this review can be stated very simply: it is
that the $\T$ and $\Q$ operators that arise in certain integrable
quantum field theories encode spectral data. This gives a novel
perspective on the Bethe ansatz, and also a new way to treat
spectral problems via the solution of nonlinear integral equations.
As an application, we have shown how Bethe ansatz ideas led to a
proof of a reality property in $\PT$-symmetric quantum mechanics.
Techniques from integrability also shed light on the way that
reality is lost
\cite{DMST}.
The correspondence has proved useful in the converse
direction as well \cite{Bazhanov:1998za,Bazhanov:2003ua,
Lukyanov:2003rt,Lukyanov:2003nj}.

There are many further problems to be explored,
and here we list just a few. First,
it will be interesting to find out how many other BA systems can be
brought into the correspondence, beyond those listed in section
\ref{Gen}.

Second, most of the correspondences established to date have 
concerned massless integrable lattice models, in a field theory
limit where the number of sites, and of Bethe ansatz roots, tends to
infinity.  The first match to a finite lattice system has been
reported in \cite{dst}, linking the P\"oschl-Teller (Heun) equation
in complex domain and spin-$\frac{L-2}{2}$ XXZ spin chains at
special choices of $\Delta$. It would be of interest to extend this
result to other models.

Third, while we have concentrated on describing the
correspondence for the ground state eigenvalues of the $\T$ and $\Q$
operators, a spectral interpretation for the excited state
eigenvalues $t_1(\nu)$, $t_2(\nu)$ and so on has now been found for
the massless $SU(2)$-related system in the continuum limit
\cite{BLZhigh}.  The
complicated distributions of Bethe roots is reflected in the
significantly more complicated nature of the  Schr\"odinger
problems.  In particular the potentials are no longer real, even for
the radial problems. Since all eigenvalues satisfy the {\em same}
functional relations as for the ground state, it is expected that a
correspondence for the excited state eigenvalues of all models
should exist.

Fourth, it would be of interest
to extend the correspondence to encompass  massive integrable
models, a first step towards this goal having  very recently been
made in \cite{bm}.

Finally, we should
admit that our observations remain at a rather formal
and mathematical level.
It is natural to ask whether there is a more physical explanation for the
correspondence, but perhaps this question will be easier to answer
once some of the other open problems  have been resolved.

\medskip
\noindent{\bf Acknowledgements --}
%

%
We would like to thank Carl Bender, Christian Korff, Sergei
Lukyanov, Davide Masoero, Rafael Nepomechie,  Adam Millican-Slater,
Junji Suzuki, Andr\'e Voros and Miloslav Znojil for useful
discussions. This project was  partially supported by the European
network EUCLID (HPRN-CT-2002-00325), INFN grant TO12, NATO grant
number PST.CLG.980424,  The Nuffield Foundation grant number
NAL/32601, and a grant from the Leverhulme Trust.


\appendix

%
\resection{The algebraic Bethe ansatz}
\label{aba_app}
In this appendix, we fill in some details of the Bethe ansatz for
the six-vertex model. The technique we describe is called the
algebraic Bethe ansatz, an elegant formulation developed by the
`Leningrad school', which reveals the r\^ole of the Yang-Baxter
equation in a particularly transparent way. Our presentation borrows
heavily from
\cite{abarefdv}, fixing one or two typos; other useful references
are
\cite{abareff} and \cite{abarefkbi}. To keep the discussion
self-contained, we begin by recalling some of the definitions from
section \ref{sect:fnrel}.

\begin{figure}[ht]
\begin{center}
\includegraphics[width=0.6\linewidth]{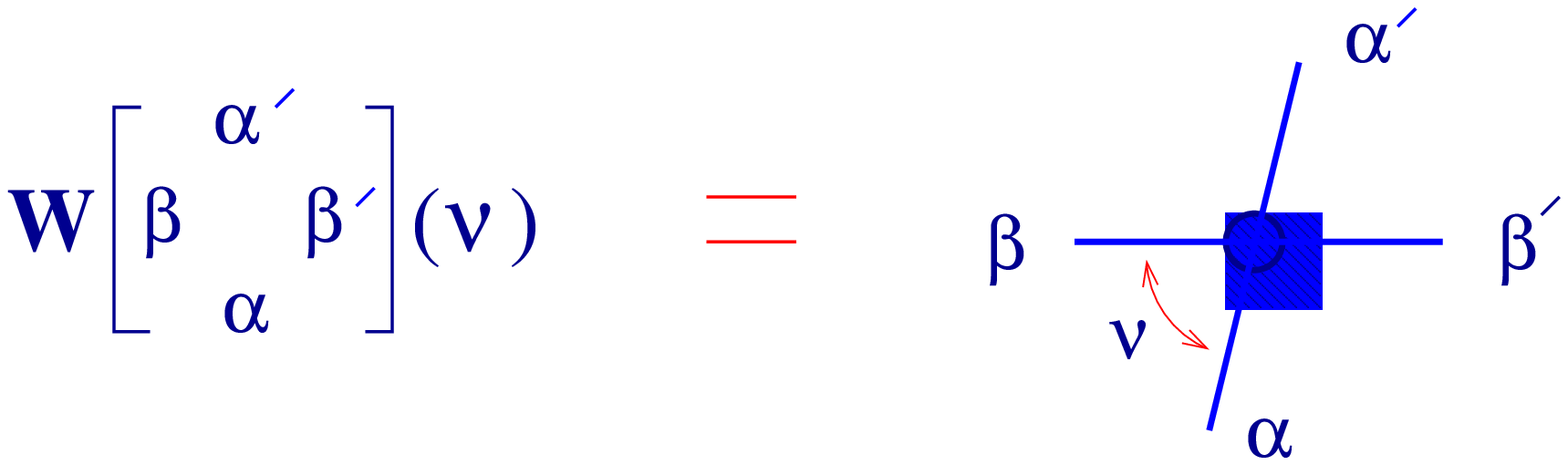}
\end{center}
\caption{The local Boltzmann weights.\label{boltz2}}
\end{figure}

The local Boltzmann weights for the model are parametrised in terms of
the spectral parameter $\nu$ and the anisotropy $\eta$ as
\eq
a(\nu,\eta)= \sin(\eta+i\nu)=
W\!\!\left[\mbox{$\rightarrow$\raisebox{1.5ex}{$\uparrow$}%
\raisebox{-1.5ex}{\hspace{-6pt}$\uparrow$}$\rightarrow$} \right]
=
W\!\!\left[\mbox{$\leftarrow$\raisebox{1.5ex}{$\downarrow$}%
\raisebox{-1.5ex}{\hspace{-6pt}$\downarrow$}$\leftarrow$} \right]~;
\label{b1}
\en
\eq
b(\nu,\eta)=  \sin(\eta-i\nu)=
W\!\!\left[\mbox{$\rightarrow$\raisebox{1.5ex}{$\downarrow$}%
\raisebox{-1.5ex}{\hspace{-6pt}$\downarrow$}$\rightarrow$} \right]
=
W\!\!\left[\mbox{$\leftarrow$\raisebox{1.5ex}{$\uparrow$}%
\raisebox{-1.5ex}{\hspace{-6pt}$\uparrow$}$\leftarrow$}\right]~;
\label{b2}
\en
\eq
c(\nu,\eta)=\sin(2 \eta)~~~~~=
W\!\!\left[\mbox{$\rightarrow$\raisebox{1.5ex}{$\uparrow$}%
\raisebox{-1.5ex}{\hspace{-6pt}$\downarrow$}$\leftarrow$} \right]
=
W\!\!\left[\mbox{$\leftarrow$\raisebox{1.5ex}{$\downarrow$}%
\raisebox{-1.5ex}{\hspace{-6pt}$\uparrow$}$\rightarrow$}\right]~,
\label{b3}
\en
representing the weights $W$  as in
figure~\ref{boltz2}.  

{}From these local weights, a $2^N\times 2^N$ transfer matrix
$\T^{\Bbalpha'}_{\Bbalpha}(\nu)$ is constructed:
\eq
\fl
{}~~\TT^{\Bbalpha'}_{\Bbalpha}(\nu)
=
\sum_{\{\beta_i\}}
W \!\!\left[\mbox{$\beta_1$\raisebox{1.5ex}{$\alpha'_1$}%
\raisebox{-1.5ex}{\hspace{-14pt}
$\alpha_1$}$\beta_2$} \right]\!(\nu)\,
W \!\!\left[\mbox{$\beta_2$\raisebox{1.5ex}{$\alpha'_2$}%
\raisebox{-1.5ex}{\hspace{-14pt}
$\alpha_2$}$\beta_3$} \right]\!(\nu)\,
W \!\!\left[\mbox{$\beta_3$\raisebox{1.5ex}{$\alpha'_3$}%
\raisebox{-1.5ex}{\hspace{-14pt}
$\alpha_3$}$\beta_4$} \right]\!(\nu)
\dots
W \!\!\left[\mbox{$\beta_N$\raisebox{1.5ex}{$\alpha'_N$}%
\raisebox{-1.5ex}{\hspace{-17pt}
$\alpha_N$}$\beta_1$} \right]\!(\nu)
\label{transfdef}
\en
where $\Balpha=(\alpha_1,\alpha_2\dots\alpha_N)$ and
$\Balpha'=(\alpha'_1,\alpha'_2\dots\alpha'_N)$ are multiindices.

The task of the Bethe ansatz is to find eigenvectors $\Psi$ and
eigenvalues $t$ of $\TT$\,, so that
\eq
\sum_{\Bbalpha'}
\TT^{\Bbalpha'}_{\Bbalpha}\Psi_{\Bbalpha'}=t\,\Psi_{\Bbalpha}~.
\en

\begin{figure}[ht]
\begin{center}
\includegraphics[width=0.9\linewidth]{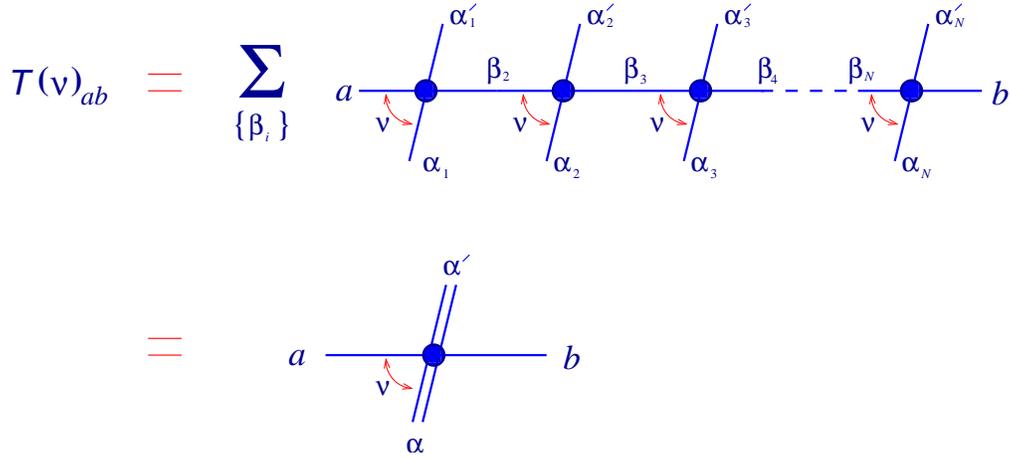}
\end{center}
\caption{The monodromy matrix.\label{monod}}
\end{figure}

As mentioned in section \ref{sect:fnrel}, the first step is to make
a well-informed guess as to the form of a putative eigenvector. The
algebraic Bethe ansatz makes this guess with the help of an
additional piece of machinery, the {\em monodromy matrix}~$\CT$:
\eq
\fl
{}~~
\CT^{\Bbalpha'}_{\Bbalpha}(\nu)_{ab}
=
\sum_{\{\beta_i\}}
W \!\!\left[\mbox{$a$\raisebox{1.5ex}{$\alpha'_1$}%
\raisebox{-1.5ex}{\hspace{-14pt}
$\alpha_1$}$\beta_2$} \right]\!(\nu)\,
W \!\!\left[\mbox{$\beta_2$\raisebox{1.5ex}{$\alpha'_2$}%
\raisebox{-1.5ex}{\hspace{-14pt}
$\alpha_2$}$\beta_3$} \right]\!(\nu)\,
W \!\!\left[\mbox{$\beta_3$\raisebox{1.5ex}{$\alpha'_3$}%
\raisebox{-1.5ex}{\hspace{-14pt}
$\alpha_3$}$\beta_4$} \right]\!(\nu)
\dots
W \!\!\left[\mbox{$\beta_N$\raisebox{1.5ex}{$\alpha'_N$}%
\raisebox{-1.5ex}{\hspace{-17pt}
$\alpha_N$}$b$} \right]\!(\nu)
\label{monoddef}
\en
The definition of $\CT_{ab}$ differs from that of
$\TT$ in the omission of the
sum over one of the horizontal
spins, $\beta_1$\,.
Of course,
$\TT$ can immediately be recovered from $\CT_{ab}$\,, simply by
performing this
sum:
\eq
\TT(\nu)= \sum_{a}\CT(\nu)_{aa}= \CT(\nu)_{\rightarrow
\,\rightarrow} + \CT(\nu)_{\leftarrow \,\leftarrow}~.
\en
However, it turns out that important data is also hidden in
the off-diagonal elements of $\CT$.
As for the transfer matrix, illustrated in figure \ref{transf}
of section \ref{sect:fnrel}, the
definition $\CT$
is most easily digested with the aid of a picture, figure
\ref{monod}, where in the second line
we used the double lines as a shorthand for the
entire collections of lines carrying the multiindices $\Bbalpha$ and
$\Bbalpha'$.

\noindent
For convenience, we write the components of $\CT$ as
\eq
\CT(\nu)_{ab}{}~{}={}~{} \left(
\matrix{\CT(\nu)_{\,\rightarrow\,\rightarrow}&\CT(\nu)_{\,\rightarrow
\,\leftarrow}
 \cr
\CT(\nu)_{\,\leftarrow\,\rightarrow}&\CT(\nu)_{\,\leftarrow\,\leftarrow}
} \right)
{}~{}={}~{}
\left(\matrix{A(\nu)&B(\nu) \cr
C(\nu)&D(\nu)}\right)
\en
so that
\eq
\TT(\nu)=A(\nu)+D(\nu)~,
\en
and, in pictures,
\bea
A(\nu)={}~\raisebox{-17pt}{\epsfxsize=46pt\epsfbox{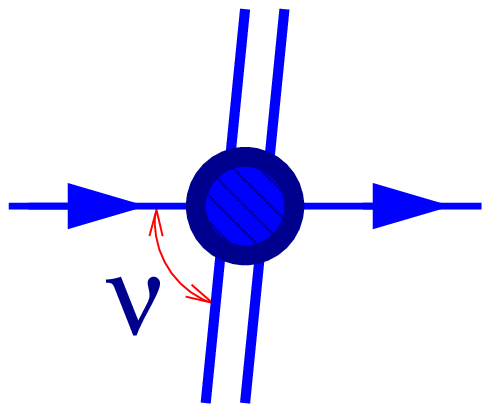}}
&\qquad&\qquad
B(\nu)={}~\raisebox{-17pt}{\epsfxsize=46pt\epsfbox{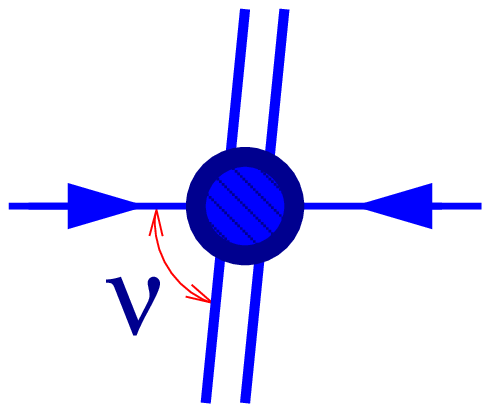}}
\nn\\[5pt]
C(\nu)={}~\raisebox{-17pt}{\epsfxsize=46pt\epsfbox{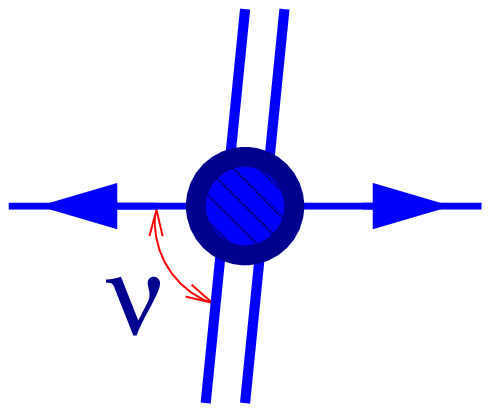}}
&\qquad&\qquad
D(\nu)={}~\raisebox{-17pt}{\epsfxsize=46pt\epsfbox{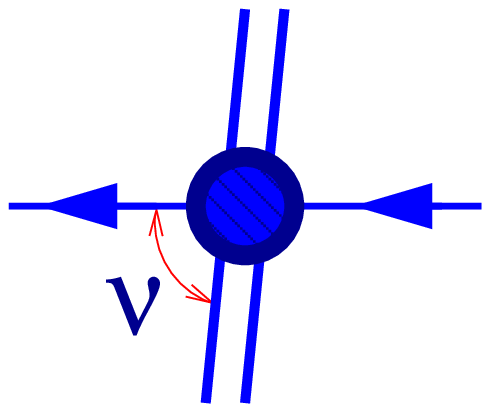}}
\eea
As matrices acting on $2^N$-dimensional vectors, $A$, $B$, $C$ and $D$
inherit an important property from the microscopic Boltzmann weights
by virtue of the ice rule, which preserves the total flux of arrows
through each vertex and hence through each collection of vertices.
If we define $S$ to be the total number of up arrows minus the total
number of down arrows (the `spin'),
then $A$ and $D$ act on vectors with a definite value of $S$ to give
vectors with the same value,
while $B$ decreases $S$ by two, and $C$
increases it by two.

\begin{figure}[ht]
\begin{center}
\includegraphics[width=0.47\linewidth]{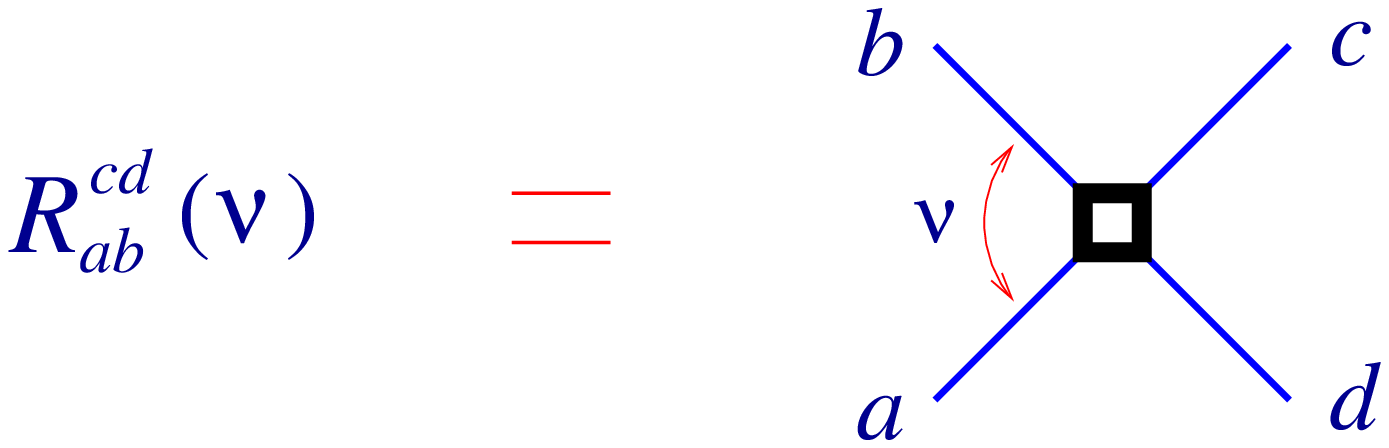}
\end{center}
\caption{The $R$ matrix.\label{Rmat}
}
\end{figure}

\begin{figure}[ht]
\begin{center}
\includegraphics[width=0.84\linewidth]{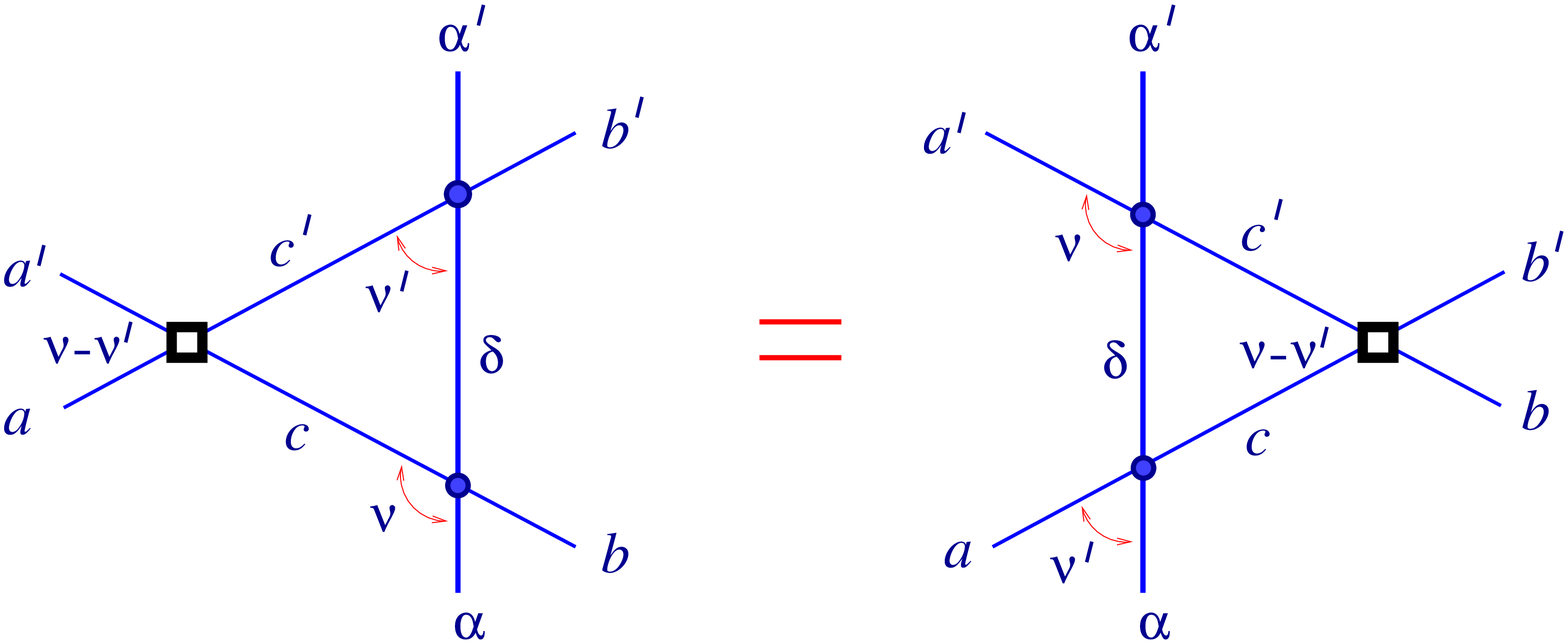}
\end{center}
\caption{Intertwining the local Boltzmann weights.\label{wcomm}}
\end{figure}

We now come to a key property of the Boltzmann weights of the six-vertex
model, a sufficient condition for integrability. There
exists a collection of numbers
$R^{ab}_{cd}(\nu)$, making up the so-called $R$ matrix and related to
the Boltzmann weights via $R(\nu)=W(\nu-i\eta)$, such that
\eq
\fl~ R^{c'\!c}_{aa'}(\nu{-}\nu')
W\!\!\left[\mbox{\small $c\,\,%
$\raisebox{1.5ex}{$\delta$}\raisebox{-1.5ex}{\hspace{-8.5pt}
$\alpha$}$\,\,b$} \right]\!(\nu)\,
W\!\!\left[\mbox{\small $c'\,\,%
$\raisebox{1.5ex}{$\alpha'$}\raisebox{-1.5ex}{\hspace{-11.5pt}
$\delta$}$\,\,\,b'$} \right]\!(\nu')
=
W\!\!\left[\mbox{\small $a\,\,%
$\raisebox{1.5ex}{$\delta$}\raisebox{-1.5ex}{\hspace{-8.5pt}
$\alpha$}$\,\,c$} \right]\!(\nu')\,
W\!\!\left[\mbox{\small
$a'\,\,$\raisebox{1.5ex}{$\alpha'$}\raisebox{-1.5ex}{\hspace{-11.5pt}
$\delta$}$\,\,\,c'$} \right]\!(\nu)
R^{b'\!b}_{cc'}(\nu{-}\nu')~.
\en
This, a version of the famous Yang-Baxter relation,
`intertwines' the local Boltzmann
weights at spectral parameters $\nu$ and
$\nu'$. It is best handled diagrammatically, representing
$R^{cd}_{ab}$ as in figure \ref{Rmat}.
The perhaps-awkward, but standard, placing of the indices reflects the
interpretation of $R$ as a matrix acting on the
tensor product $V_1\otimes V_2$, where $V_1$ and $V_2$ are
the two-dimensional vector
spaces, spanned by
$\rightarrow$ and $\leftarrow$,  that are
seen by the indices $c$ and $d$ respectively.
The resulting pictorial representation of the
intertwining relation is shown in figure \ref{wcomm}.
Apart from a shift in the spectral parameter, the entries of the $R$
matrix are just the original Boltzmann weights, so in particular they
respect the ice rule and the flux of arrows through the `$R$' vertex is
conserved. Explicitly, the
non-zero entries are
\bea
R^{\rightarrow\rightarrow}_{\rightarrow\rightarrow}(\nu)=
R^{\leftarrow\leftarrow}_{\leftarrow\leftarrow}(\nu) &=&
\raisebox{-17pt}{\epsfxsize=40pt\epsfbox{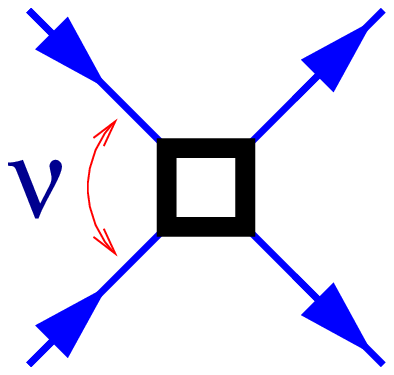}}=
\raisebox{-17pt}{\epsfxsize=40pt\epsfbox{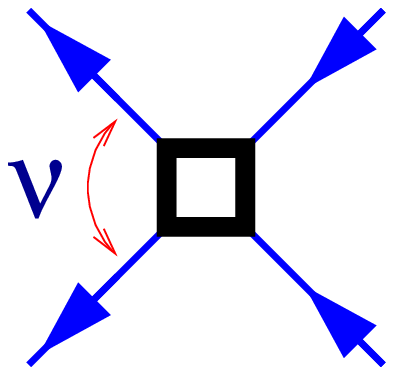}}=
a(\nu{-}i\eta,\eta) = \sin(2\eta{+}i\nu)
\nn\\
R^{\leftarrow\rightarrow}_{\leftarrow\rightarrow}(\nu)=
R^{\rightarrow\leftarrow}_{\rightarrow\leftarrow}(\nu)
 &=&
\raisebox{-17pt}{\epsfxsize=40pt\epsfbox{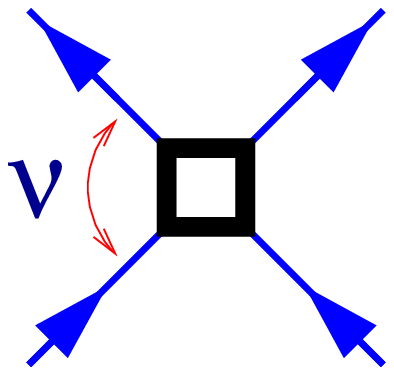}}=
\raisebox{-17pt}{\epsfxsize=40pt\epsfbox{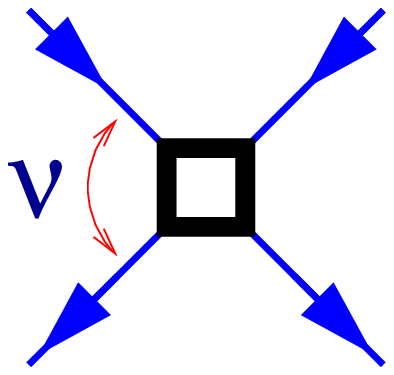}}=
b(\nu{-}i\eta,\eta)= -\sin(i\nu) \nn\\
R^{\rightarrow\leftarrow}_{\leftarrow\rightarrow}(\nu)=
R^{\leftarrow\rightarrow}_{\rightarrow\leftarrow}(\nu)
 &=&
\raisebox{-17pt}{\epsfxsize=40pt\epsfbox{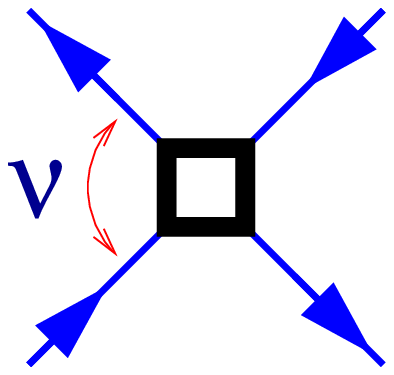}}=
\raisebox{-17pt}{\epsfxsize=40pt\epsfbox{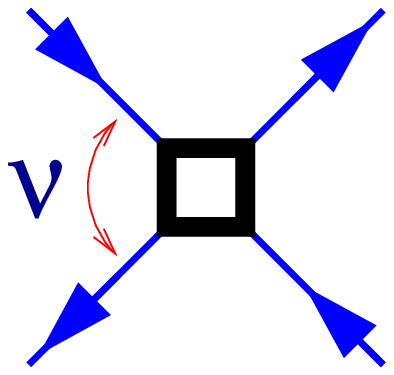}}=
c(\nu{-}i\eta,\eta)= \sin(2\eta)~.  \eea
It is also useful to write
$R^{cd}_{ab}(\nu)$ as a $4\times 4$ matrix, with the rows indexed by
$ab$ and the columns by $cd$\,, and both pairs of indices
running $\rightarrow\rightarrow$, $\rightarrow\leftarrow$,
$\leftarrow\rightarrow$, $\leftarrow\leftarrow$ : \eq
R^{cd}_{ab}(\nu)= \left(
\begin{array}{cccc}
\sin(2\eta{+}i\nu)&0&0&0\\
0&-\sin(i\nu)& \sin(2\eta)&0\\
0&\sin(2\eta)&-\sin(i\nu)&0\\
0&0&0& \sin(2\eta{+}i\nu)
\end{array}
\right)~.
\en
This notation makes it easy to check the identity
\eq
\sum_{e,f}R^{ef}_{ab}(\nu)
R^{cd}_{ef}(-\nu)= \sin(2\eta{+}i\nu)\sin(2\eta{-}i\nu)\,
\delta^c_a
\delta^d_b
\label{unit}
\en
which will be used shortly.
(In the parallel universe of integrable quantum field theory,
this is related to a property called `unitarity'.)

\begin{figure}[ht]
\begin{center}
\includegraphics[width=0.85\linewidth]{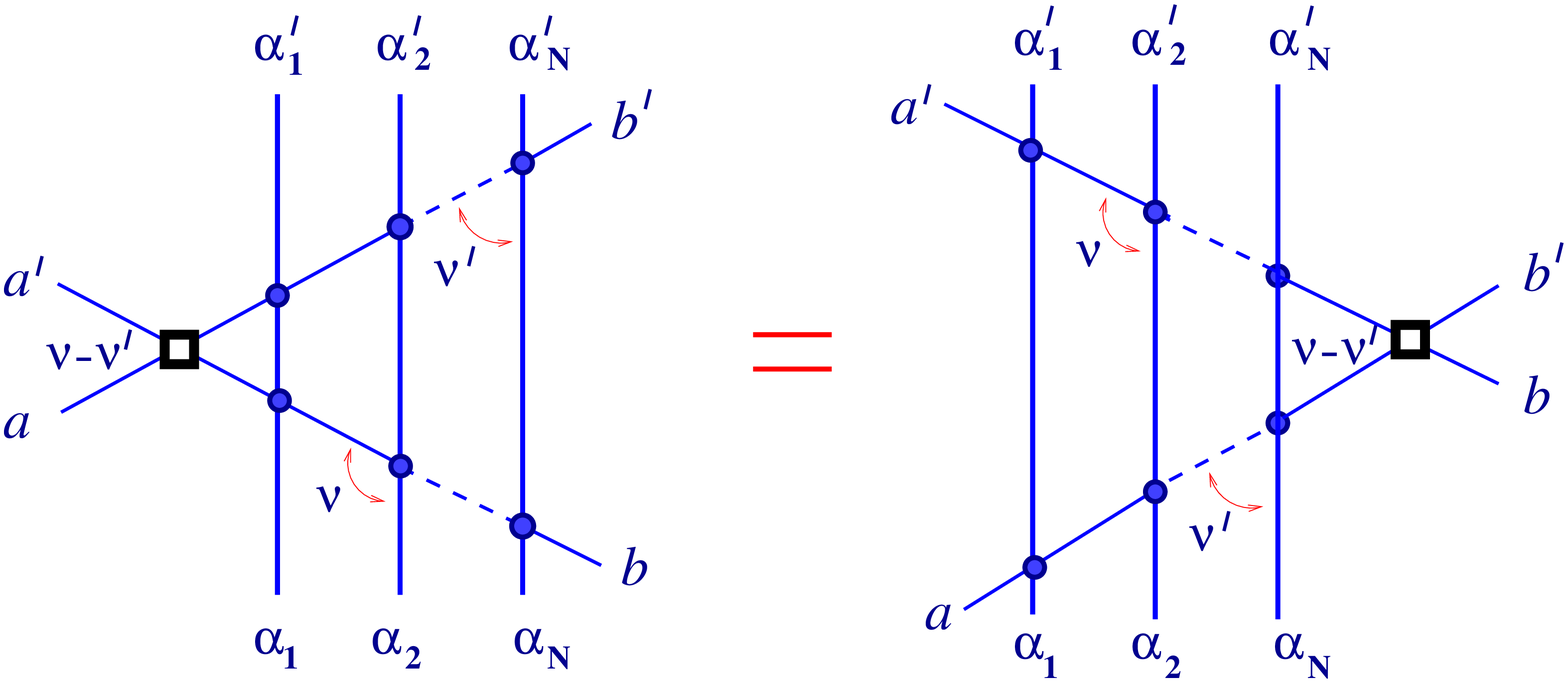}
\end{center}
\caption{Commuting the monodromy matrices.\label{tcomm}}
\end{figure}

Once the local intertwining relation is known, it is a simple matter to chain
together $N$ such equalities to find an analogous relation for the monodromy
matrix. This is illustrated in figure \ref{tcomm}.

In equations, with sums on repeated indices,
\eq
\fl\qquad~~
R^{c'\!c}_{aa'}(\nu-\nu') (\CT^{\Bbdelta}_{\Bbalpha})_{cb}(\nu)
(\CT^{\Bbalpha'}_{\Bbdelta})_{c'b'}(\nu') =
(\CT^{\Bbdelta}_{\Bbalpha})_{ac}(\nu')
(\CT^{\Bbalpha'}_{\Bbdelta})_{a'c'}(\nu) R^{b'\!b}_{cc'}(\nu-\nu')\,.
\en
It is usually convenient to leave the fact that each component of $\CT$
is itself a matrix implicit, so that this becomes
\eq
R^{c'\!c}_{aa'}(\nu-\nu') \CT_{cb}(\nu)
\CT_{c'b'}(\nu') =
\CT_{ac}(\nu')
\CT_{a'c'}(\nu) R^{b'\!b}_{cc'}(\nu-\nu')\,.
\en
Because of the way we have set things up, the ordering of matrices
right to left in this equation corresponds to the ordering top to
bottom of monodromy matrices in figure~\ref{tcomm}.

Two important consequences now follow. First,  multiplying both sides
by $R^{dd'}_{b'b}(\nu'{-}\nu)$, summing on $b$ and $b'$,
taking traces on $a,d$ and on
$a',d'$, and then finally using the cyclic property of the
trace and (\ref{unit})\,,
it can be seen that the transfer matrices at different values of the
spectral parameter {\em commute}:
\eq
[\TT(\nu),\TT(\nu')]=0\,.
\en
This establishes the claim made in section \ref{sect:fnrel}, and shows
that all of the
$\TT(\nu)$ can be diagonalised simultaneously, with
$\nu$-independent eigenvectors.

The second consequence is more technical, but crucial for the
algebraic Bethe ansatz approach to the construction of eigenvectors.
The key idea is to treat $B(\nu)$ as a creation operator, and for
this it is necessary to know how to pass the other components of the
monodromy matrix through it. Taking the relations implied by
figure~\ref{tcomm} over all possible values of the vector
$(a,a',b,b')$ of free indices gives a total of $16$ quadratic
relations. These can be rearranged to give, amongst others, the
following exchange relations: 
\bea
(\rightarrow,\rightarrow,\leftarrow,\leftarrow)&:&
[B(\nu),B(\nu')]=0~;\label{Aexch}\\[3pt]
(\rightarrow,\leftarrow,\leftarrow,\leftarrow)&:&
D(\nu)B(\nu')=g(\nu{-}\nu')B(\nu')D(\nu)-h(\nu{-}\nu')B(\nu)D(\nu')~;
\label{Cexch}
\eea
and (after swapping $\nu$ and $\nu'$)
\eq
\fl
(\rightarrow,\rightarrow,\leftarrow,\rightarrow)~~:~~
A(\nu)B(\nu')=g(\nu'{-}\nu)B(\nu')A(\nu)-h(\nu'{-}\nu)B(\nu)A(\nu')~.
\label{Bexch}
\en
The `structure constants' $g$ and $h$ are related to the
components of the $R$ matrix:
\bea
g(\nu)&=&
\frac{R^{\leftarrow\leftarrow}_{\leftarrow\leftarrow}(\nu)}%
{R^{\rightarrow\leftarrow}_{\rightarrow\leftarrow}(\nu)}
=\frac{a(\nu-i \eta,\eta)}{b(\nu-i \eta,\eta)}
=-\frac{\sin(2\eta+i\nu)}{\sin(i\nu)}~;\\[3pt]
h(\nu)&=&
\frac{R^{\leftarrow\rightarrow}_{\rightarrow\leftarrow}(\nu)}%
{R^{\rightarrow\leftarrow}_{\rightarrow\leftarrow}(\nu)}
=\frac{c(\eta)}{b(\nu-i\eta,\eta)}
=-\frac{\sin(2\eta)}{\sin(i\nu)}~.
\eea
One example should illustrate the derivation of these relations.
Setting $(a,a',b,b')=
(\rightarrow,\leftarrow,\leftarrow,\leftarrow)$ in figure
\ref{tcomm} and recording only the nonvanishing terms in the sums on
left and righthand sides,
\eq
\fl~
\raisebox{-49pt}{\epsfxsize=120pt\epsfbox{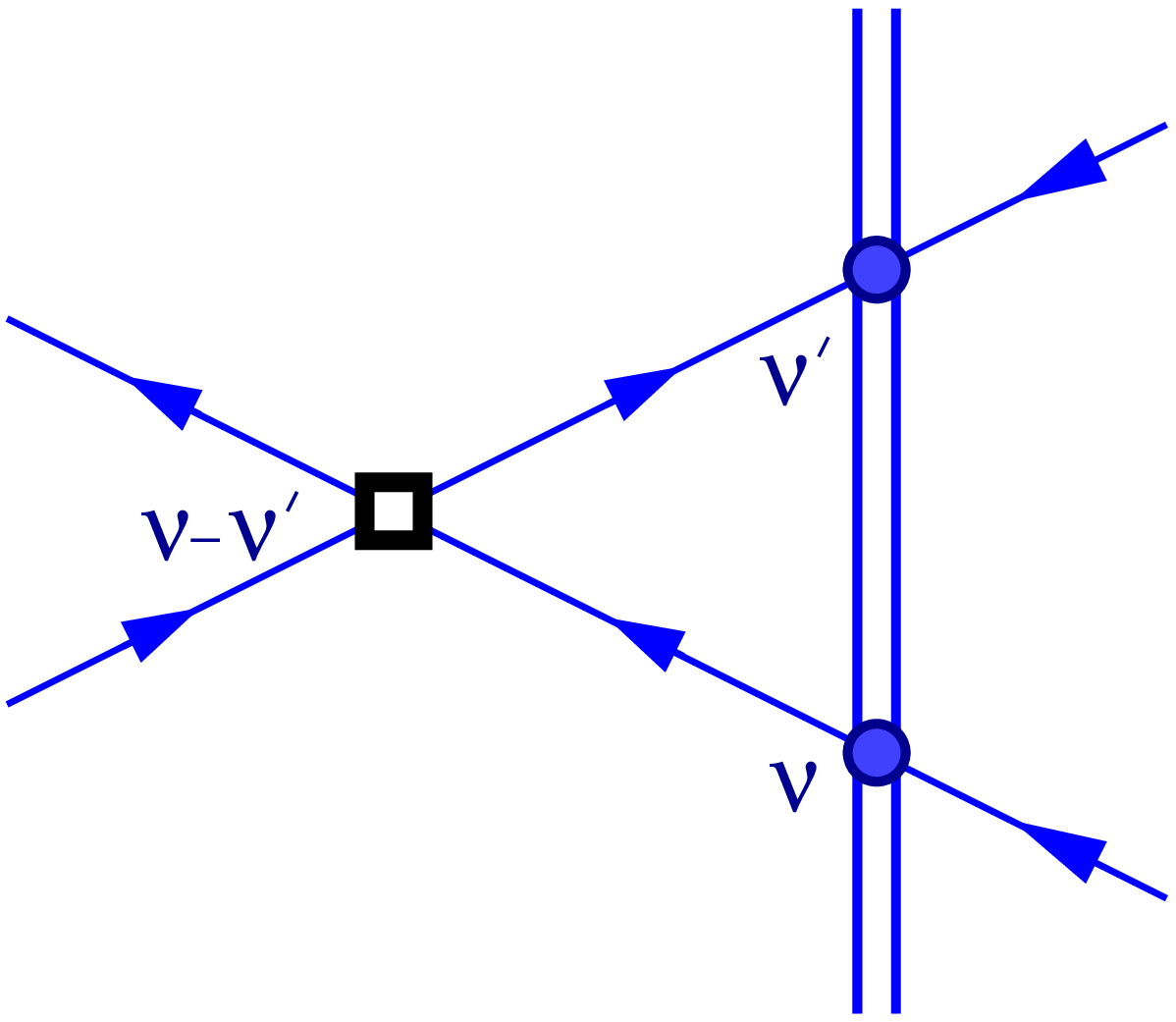}}~+~
\raisebox{-49pt}{\epsfxsize=120pt\epsfbox{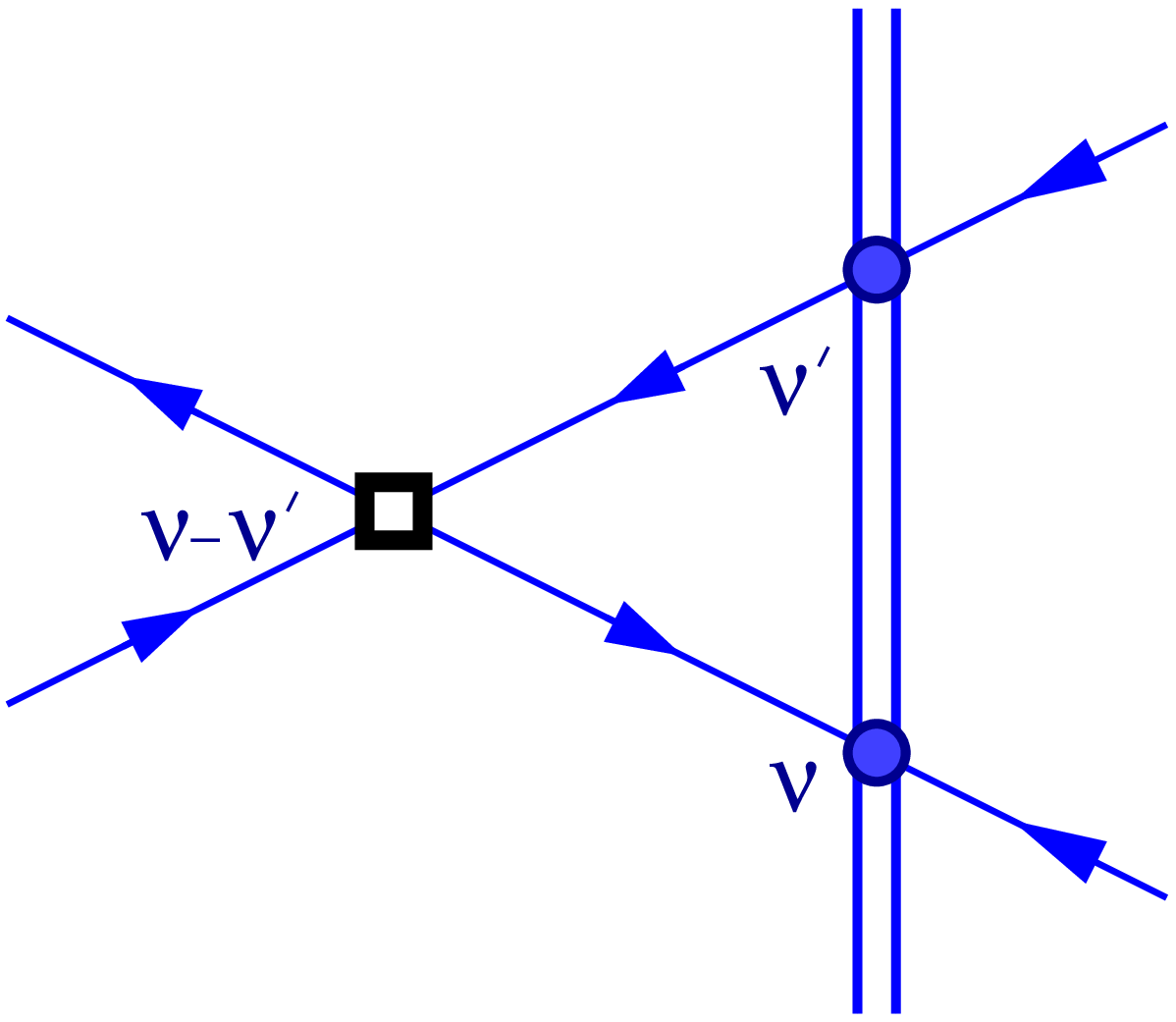}}~=~
\raisebox{-49pt}{\epsfxsize=120pt\epsfbox{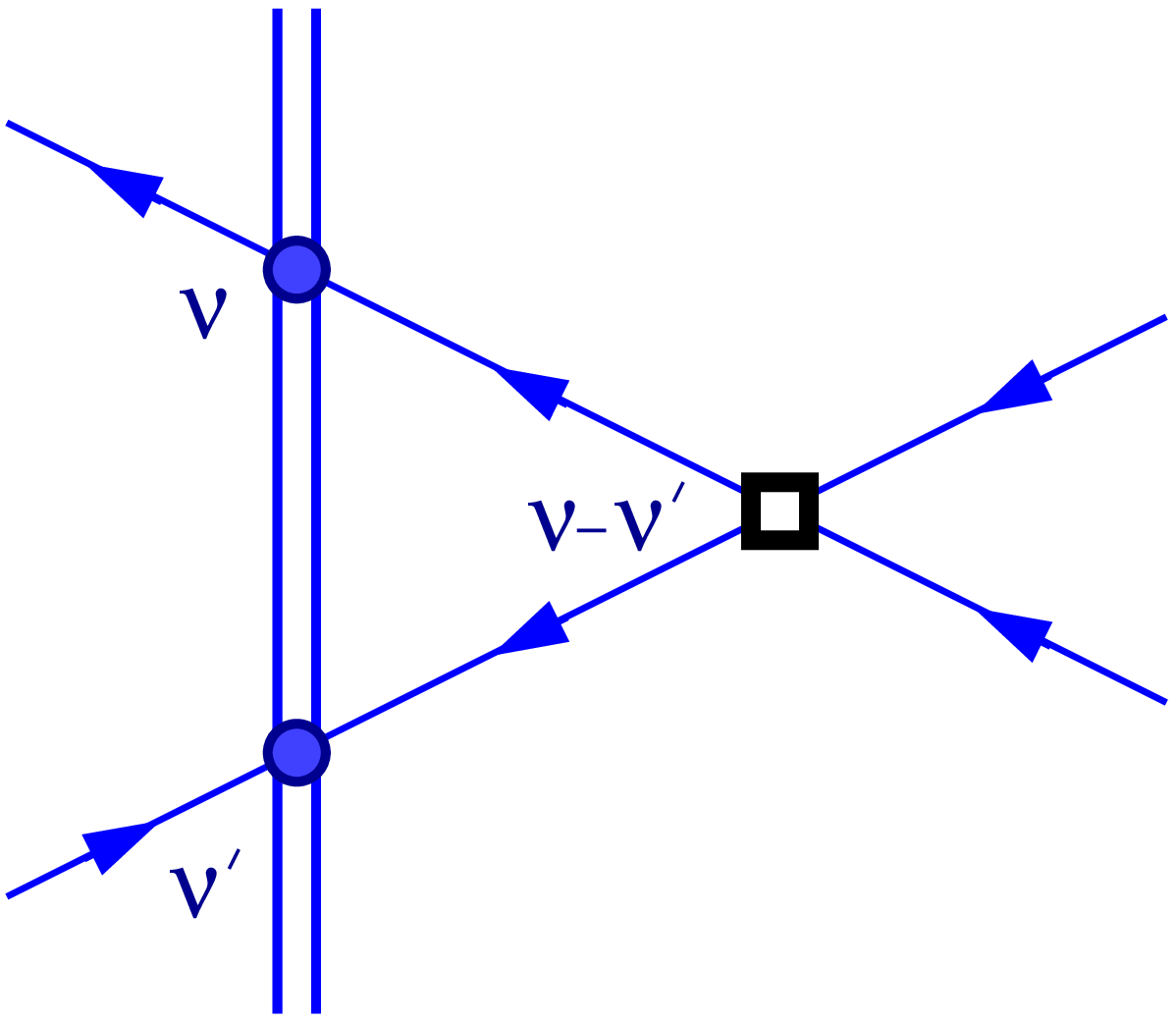}}
\en
Translated into symbols, this is
\eq
\fl~~
R^{\rightarrow\leftarrow}_{\rightarrow\leftarrow}(\nu-\nu')
D(\nu)B(\nu')+
R^{\leftarrow\rightarrow}_{\rightarrow\leftarrow}(\nu-\nu')
B(\nu)D(\nu')
=
B(\nu')D(\nu)
R^{\leftarrow\leftarrow}_{\leftarrow\leftarrow}(\nu-\nu')
\en
which can be quickly rearranged to give
(\ref{Cexch}).

At last we can write down some eigenvectors.
The first is straightforward: it is
the false ferromagnetic
ground state $\ket{\Omega}=\ket{\uparrow\,\uparrow\dots\uparrow\,}$.
In a vector notation,
\eq
\ket{\Omega}=
\left(\matrix{1 \cr 0}  \right)
\otimes
\left(\matrix{1 \cr 0} \right)
\otimes
\left(\matrix{1 \cr 0} \right)
\dots
\en
Since $A$ and $D$ cannot change the value of the spin $S$,
and $\ket{\Omega}$ is
the unique state with (maximal) spin $N$,
it is an eigenvector of both $A$ and $D$. In fact,
\eq
A(\nu)\ket{\Omega}=a^N\!(\nu,\eta)\ket{\Omega}\,,~~~
D(\nu)\ket{\Omega}=b^N\!(\nu,\eta)\ket{\Omega}\,.
\en
On the other hand $C(\nu)\ket{\Omega}=0$ (there is no state with a
larger value of $S$)\,, while
$B(\nu)\ket{\Omega}$ is, in general, a
new state.
For the simplest example, at $N=1$,
$\ket{\Omega}= \bigl({1\atop 0} \bigr)$ and
$A$, $B$, $C$ and $D$ are all $2{\times}2$ matrices:
\eq
A|\phup_{N=1}=
\left(\matrix{ a  & 0 \cr
0  & b } \right)
{}\,,~~~~
B|\phup_{N=1}=
\left(\matrix{ 0  & 0 \cr
c  & 0  } \right)\,,
\en
\eq
C|\phup_{N=1}=
\left(\matrix{ 0  & c \cr
0  & 0 } \right)
{}\,,~~~~
D|\phup_{N=1}=
\left(\matrix{ b  & 0 \cr
0  & a } \right)\,.
\en
Thus
$A(\nu)\ket{\Omega}=a(\nu)\ket{\Omega}$,
$D(\nu)\ket{\Omega}=b(\nu)\ket{\Omega}$,
$C(\nu)\ket{\Omega}=0$, and
$B(\nu)\ket{\Omega}=c(\nu)\bigl({0\atop 1} \bigr)$.
Notice also that
$B(\nu)\ket{\Omega}$ would serve as
a second
eigenvector of $\TT=A+D$.
In fact, in this rather-trivial case it could equally have been
$B(\nu')\ket{\Omega}$,
for (almost) {\em any} value of $\nu'$.
To generalise this observation, at larger values of $N$ one might search
for eigenvectors of $\TT$ of the form
\eq
\ket{\Psi}=B(\nu_1)B(\nu_2)\dots B(\nu_n)\ket{\Omega}\,.
\en
This is the guess mentioned in section \ref{sect:fnrel} as stage
 (i) of the Bethe ansatz. For $N>1$ this guess (or {\em
ansatz}) does not always work -- extra conditions need to be imposed
on the numbers $\nu_1\,\dots \nu_n$. To find these, we first compute
the separate actions of $A(\nu)$ and $D(\nu)$ on our would-be
eigenvector $\ket{\Psi}$. Using the $A{-}B$ exchange relation,
(\ref{Bexch}),
\bea
A(\nu)\ket{\Psi}
&=&A(\nu)B(\nu_1)B(\nu_2)\dots B(\nu_n)\ket{\Omega}\nn\\[3pt]
&=&g(\nu_1{-}\nu)B(\nu_1)A(\nu)B(\nu_2)\dots B(\nu_n)\ket{\Omega}\nn\\[3pt]
&&{~~}-h(\nu_1{-}\nu)B(\nu)A(\nu_1)B(\nu_2)\dots
B(\nu_n)\ket{\Omega}\,.
\eea
Continuing to pass $A$ through the $B$ operators, but only recording
explicitly those
terms resulting from the first term on the
right-hand side of (\ref{Bexch}),  yields
\bea
A(\nu)\ket{\Psi}
&=&\prod_{i=1}^{n}g(\nu_i-\nu)\,a^N\!(\nu,\eta)\ket{\Psi}\nn\\
&&{~~}-h(\nu_1{-}\nu)
\prod_{i=2}^{n}g(\nu_i-\nu_1)\,a^N\!(\nu,\eta)
B(\nu)B(\nu_2)\dots B(\nu_n)\ket{\Omega}\nn\\
&&{~~}+\mbox{further terms}~.
\eea
A nice way to
reconstruct the further terms uses the observation that $\ket{\Psi}$
depends on the parameters
$\{\nu_1\dots\nu_n\}$ in a symmetrical manner, since, from
(\ref{Aexch}), the $B(\nu_j)$ commute.
This enables us to deduce that
\eq
A(\nu)\ket{\Psi}=\Lambda^+\ket{\Psi}+\sum_{k=1}^n\Lambda^+_k\ket{\Psi_k}
\en
where
\eq
\fl\qquad
\Lambda^+=a^N\!(\nu,\eta)\prod_{j=1}^{n}g(\nu_j-\nu)~,~~~
\Lambda^+_k=-a^N\!(\nu_k,\eta)h(\nu_k-\nu)\prod_{j=1\atop j\neq k}^{n}g(\nu_j-\nu_k)
\en
and
\eq
\ket{\Psi_k}=B(\nu)\prod_{j=1\atop j\neq k}^{n}B(\nu_j) \ket{\Omega}~.
\en
(It is instructive to check this directly, at least for $n=2$.)

The same steps can be repeated to find the action of the matrix
$D(\nu)$ on $\ket{\Psi}$, using (\ref{Cexch}) and (\ref{Bexch}). The
result is:
\eq
D(\nu)\ket{\Psi}=\Lambda^-\ket{\Psi}+\sum_{k=1}^n\Lambda^-_k\ket{\Psi_k}
\en
with
\eq
\fl\qquad
\Lambda^-=b^N\!(\nu,\eta)\prod_{j=1}^{n}g(\nu-\nu_j)~,~~~
\Lambda^-_k=-b^N\!(\nu_k,\eta)h(\nu-\nu_k)\prod_{j=1\atop j\neq k}^{n}g(\nu_k-\nu_j)~.
\en
The ansatz $\ket{\Psi}$ will be an eigenvector of $\TT=A+D$ if
and only if all terms proportional to $\ket{\Psi_k}$ can be made to
cancel between $A(\nu)\ket{\Psi}$ and $D(\nu)\ket{\Psi}$\,. This requires
$\Lambda^+_k+\Lambda^-_k=0$ for $k=1\dots n$, or
\eq
(-1)^n\prod_{j=1}^n\frac{\sinh(2i\eta-\nu_k+\nu_j)}{\sinh(2i\eta-\nu_j+\nu_k)}=
-\frac{a^N\!(\nu_k,\eta)}{b^N\!(\nu_k,\eta)}~~,\qquad k=1\dots n\,.
\en
These are the {\em Bethe ansatz equations} (BAE) for the {\em roots}
$\{\nu_1\dots
\nu_n\}$.
It is important to reiterate
that these equations do not have a unique
solution, but rather a discrete set,
matching the fact that the matrix
$\TT$ has many eigenvalues.
For each self-consistent solution $\{\nu_i\}$ of the BAE,
the corresponding eigenvector $\ket{\Psi}$
of $\TT$ has eigenvalue
\bea
t(\nu)&=& \Lambda^++\Lambda^-\nn\\
&=&
a^N\!(\nu,\eta)\prod_{j=1}^{n}g(\nu_j-\nu)+
b^N\!(\nu,\eta)\prod_{j=1}^{n}g(\nu-\nu_j)~,
\eea
recovering the formula (\ref{tevalue}) quoted in the main text.

The method   can be  straightforwardly
extended  to the more general situation  of
twisted boundary conditions~\cite{KBP,Alcaraz:1988zr}.
The twist described earlier can be implemented
by  trading
\eq
W \!\left[\mbox{$\beta_N$\raisebox{1.5ex}{$\alpha'_N$}%
\raisebox{-1.5ex}{\hspace{-17pt}
$\alpha_N$}$\rightarrow$} \right]\!(\nu)
{}~\longrightarrow~
~e^{-i \phi}~W \!\left[\mbox{$\beta_N$\raisebox{1.5ex}{$\alpha'_N$}%
\raisebox{-1.5ex}{\hspace{-17pt}
$\alpha_N$}$\rightarrow$} \right]\!(\nu)
\en
and
\eq
W \!\left[\mbox{$\beta_N$\raisebox{1.5ex}{$\alpha'_N$}%
\raisebox{-1.5ex}{\hspace{-17pt}
$\alpha_N$}$\leftarrow$} \right]\!(\nu)
{}~\longrightarrow~
e^{i \phi}~W \!\left[\mbox{$\beta_N$\raisebox{1.5ex}{$\alpha'_N$}%
\raisebox{-1.5ex}{\hspace{-17pt}
$\alpha_N$}$\leftarrow$} \right]\!(\nu)
\en
in the definition (\ref{monoddef}) of  the monodromy matrix, so
that $\CT(\nu)_{ab}$ becomes
\eq
\CT_{ab}{}~{}={}~{} \left( \matrix{ e^{-i
\phi}~\CT_{\,\rightarrow\rightarrow}& e^{i \phi} \CT_{\,\rightarrow
\leftarrow}
 \cr
 e^{-i \phi}~\CT_{\,\leftarrow\rightarrow}&
e^{i \phi}\CT_{\,\leftarrow\leftarrow} } \right) {}~{}={}~{}
\left(\matrix{ e^{-i \phi}~ A(\nu)& e^{ i \phi } B(\nu)
\cr e^{-i \phi}~ C(\nu)& e^{i \phi} D(\nu)}\right)~.
\en
The more general transfer matrix
\eq
\TT(\nu)= e^{-i \phi}  A(\nu)+e^{i \phi} D(\nu)~,
\en
then has  eigenvalues
\bea
t(\nu)&=&  e^{-i \phi} \Lambda^++  e^{i \phi}\Lambda^-\nn\\
&=& e^{-i \phi} a^N\!(\nu,\eta)\prod_{j=1}^{n}g(\nu_j-\nu)+ e^{i \phi}
b^N\!(\nu,\eta)\prod_{j=1}^{n}g(\nu-\nu_j)~,
\label{Ntq}
\eea
provided that the  set of roots $\{\nu_1\dots \nu_n\}$  solves the
twisted Bethe ansatz equations
\eq \fl \qquad
(-1)^n\prod_{j=1}^n\frac{\sinh(2i\eta-\nu_k+\nu_j)}{\sinh(2i\eta-\nu_j+\nu_k)}=
-e^{-2i \phi} \frac{a^N\!(\nu_k,\eta)}{b^N\!(\nu_k,\eta)}~~,\qquad
k=1\dots n\,.
\en

Finally, note that the four-spin reversal invariance of the local
Boltzmann weights implies the
following property of the transfer matrix:
 \eq
\TT_{\alpha}^{\alpha'}(\nu,-\phi) =  \TT_{\bar \alpha}^{\bar
  \alpha'}(\nu,\phi)~,~~~
(\bar{\uparrow}=\downarrow,\bar{\downarrow}=\uparrow)~.
\label{inv0}
\en
Indeed, from the definitions (\ref{b1}), (\ref{b2}), (\ref{b3}) and
(\ref{monoddef}) we have
$\CT^{\Bbalpha'}_{\Bbalpha}(\nu)_{\,ab} =\CT^{\bar{\Bbalpha}'}_{\bar{\Bbalpha}}
(\nu)_{\bar{a}  \bar{b}}$  and therefore
\bea
\fl \qquad
\TT^{\Bbalpha'}_{\Bbalpha}(\nu,-\phi) &=&
 e^{i\phi} ~\CT^{\Bbalpha'}_{\Bbalpha}(\nu)_{\,\rightarrow\rightarrow}
+ e^{-i\phi}\CT^{\Bbalpha'}_{\Bbalpha}(\nu)_{\,\leftarrow\leftarrow}
= e^{i\phi}
~\CT^{\bar{\Bbalpha}'}_{\bar{\Bbalpha}}(\nu)_{\,\leftarrow\leftarrow}
+ e^{-i\phi}
\CT^{\bar{\Bbalpha}'}_{\bar{\Bbalpha}}(\nu)_{\,\rightarrow\rightarrow}
\nn \\
&=& \TT^{\bar{\Bbalpha}'}_{\bar{\Bbalpha}}(\nu,\phi)~~.
\eea
A  first  consequence of the equality (\ref{inv0})
is that  $\TT(\nu,\phi)$ and
$\TT(\nu,-\phi)$ have the same set of eigenvalues. Furthermore,
 (\ref{inv0})  also means that
\eq
\TT(\nu,-\phi)= U \, \TT(\nu,\phi) \, U^{-1}~~,~~U=
\overbrace{
\sigma_x \otimes \sigma_x \dots \otimes \sigma_x}^N
\en
where $\sigma_x$ is in the vector notation a Pauli matrix. Hence
$U^2=1$ and
the set of eigenvectors of
$\TT(\nu,\phi)$ splits into $U$-invariant singlets
and doublets. Invoking the Perron-Frobenius theorem in the spin $0$
sector at $\phi=0$ and continuity
we conclude that the ground state is a singlet
and its associated eigenvalue $t_0(\nu,\phi)$ depends on
$\phi$ only through its square:
\eq
t_0(\nu,\phi)=t_0(\nu,-\phi)~.
\label{inv}
\en



\resection{ODE results}
 \label{asy_app}
This appendix collects some properties of a particular
solution of the ordinary differential equation
\eq
\Bigl[-\frac{d^2}{dx^2}+x^{2M}+
\frac{l(l+1)}{x^2} \Bigr]y(x)=E\,y(x)\,,
\label{fsh}
\en
defined for $M>1$ by the asymptotic
\eq
y(x)\,\sim
{}~\frac{1}{\sqrt{2i}}\,
x^{-M/2}\exp\left[-\fract{1}{M{+}1}\,x^{M{+}1}\right]
\label{appas}
\en
as $x$ tends to infinity in any closed subsector of
$\CS_{-1}\cup\CS_0\cup\CS_1$\,, where
\eq
\CS_k:=\left|\arg(x)-
\frac{2\pi k}{2M{+}2}\right|<\frac{\pi}{2M{+}2}\,.
\en
This is the `Sibuya solution', subdominant in $\CS_0$\,,
relevant to the basic ODE/IM correspondence connecting the six-vertex model
to an ordinary differential equation.

\subsection{The special case of $M=1$}
\label{app:SHO}

Setting $M=1$ gives the simple harmonic
oscillator, exactly solvable for all
$l$ in terms of the confluent hypergeometric functions
$M(a,b,z)$ and $U(a,b,z)$.
The correctly-normalised solution,
subdominant in $S_0$, is
\eq
y(x,E,l)= \frac{1}{\sqrt{2i}}\,x^{l+1} e^{-x^2/2}
U(\half(l+\fract{3}{2}) - \fract{E}{4},l+\fract{3}{2},x^2)\,,
\en
and has the large $x$ asymptotic
\eq
y(x,E,l) \sim
\frac{1}{\sqrt{2 i}}\,
x^{-1/2+E/2}[1+{\cal O}(x^{-2})]\,e^{-\hf x^2}\,.
\en
Notice the appearance of an extra factor of $x^{E/2}$
compared with the formula (\ref{appas}). This has an interesting
effect on the TQ relation \cite{DTb}\footnote{The argument here is
a slight streamlining of that given in \cite{DTb}.}.
First, the modified asymptotic means that the most natural definition
of the shifted solutions $y_k(x,E,l)$ is now
\eq
y_k(x,E,l)= \omega^{k/2 -  kE/2}
y(\omega^{-k}  x,\;\omega^{2k} E,\;l)~,\qquad \omega = e^{i\pi/2}~,
\en
since this preserves the property $W_{0,1}(E,l)=1$.
The basic Stokes relation (\ref{CYa}) remains
\eq
C(E,l)y_0(x,E,l)=y_{-1}(x,E,l)+y_1(x,E,l)\,,
\label{Bstokes}
\en
where $C(E,l)=W_{-1,1}(E,l)$. This must be projected onto a solution
defined through its behaviour at the origin to eliminate $x$. Define
$\psi(x,E,l)$ as in (\ref{psidefn}) through the small-$x$ asymptotic
\eq
\psi(x,E,l)\sim x^{l+1}+O(x^{l+3})
\en
and shifted solutions
\eq
\psi_k(x,E,l)=\omega^{k/2+kE/2}\psi(\omega^{-k}x,\omega^{2k}E,l)\,.
\en
The prefactor ensures that
$W[y_k,\psi_k](E,l)=W[y,\psi](\omega^{2k}E,l)$\,, while a consideration
of the small-$x$ behaviour of $\psi_k$ shows that
$\psi_k(x,E,l)=\omega^{-k(l{+}1/2)+kE/2}\psi(x,E,l)$. Hence
$W[y_k,\psi](E,l)=\omega^{k(l{+}1/2)-kE/2}\,W[y,\psi](\omega^{2k}E,l)$.
Taking the Wronskian of (\ref{Bstokes}) with $\psi$ and setting
$D(E,l)=W[y,\psi](E,l)$,
\eq
\fl\qquad
C(E,l)D(E,l)=\omega^{-(l{+}1/2)+E/2}D(\omega^{-2}E,l)+
\omega^{(l{+}1/2)-E/2}D(\omega^{2}E,l)\,.
\en
The modified asymptotic has resulted in a couple of
extra $E$-dependent factors.  On the integrable models side of the
correspondence, $M=1$ is the free-fermion point and explicit
constructions of the $T$ and $Q$ functions had led
Bazhanov, Lukyanov and Zamolodchikov to
precisely the same `renormalisation' of the TQ relation \cite{BLZ2},
and so the correspondence survives.

The functions $C$ and $D$ can also be calculated directly from the
ordinary differential equation, of course.
Taking from \cite{AS} the analytic continuation formula
\eq \fl\qquad
U(a,b,ze^{2\pi i n}) = (1-e^{-2 \pi i b n }) \frac{\Gamma
(1-b)}{\Gamma (1+a-b)} M(a,b,z) + e^{-2\pi i b n} U(a,b,z)\,,
\en
the Wronskian
\eq
W[U(a,b,z) , M(a,b,z)]=\frac{\Gamma(b)}{\Gamma(a)} z^{-b} e^z\,,
\en
and, for $b>1$, the $|z|\to 0$ asymptotic
\eq
U(a,b,z)\sim\frac{\Gamma(b-1)}{\Gamma(a)} z^{1-b}+\dots
\en
one finds
\eq
C(E,l)|\phup_{M=1}=\frac{2\pi}
{\Gamma\Bigl(\frac{1}{2} +\frac{2l+1+E}{4}\Bigr)
\Gamma\Bigl(\frac{1}{2} -\frac{2l+1-E}{4}\Bigr)}
\label{csho}
\en
and
\eq
D(E,l)|\phup_{M=1}=\frac{2\,\Gamma\Bigl(l+\frac{3}{2}\Bigr)}%
{\sqrt{2i}\,\Gamma\Bigl(\frac{2l+3-E}{4}\Bigr)}~.
\en
The zeros of $D(E,l)$ are at $E=2l-1+4k$, $k=1,2,\dots$\,, and reproduce
the well-known eigenvalues of the radial simple harmonic oscillator.
Zeros of $C(E,l)$ correspond to there being normalisable
wavefunctions on a contour which runs along the imaginary axis, save
for a diversion to the right of the singularity at $x=0$. A
variable-change $x\to x/i$ maps this to the real axis and results in
the so-called $\PT$-symmetric simple harmonic oscillator,
discussed in \cite{Znojil:1999qt} and \cite{DTb}. Since this variable
change also negates $E$,
the eigenvalues of this problem, from (\ref{csho}), are
$4k-2\pm(2l{+}1)$, $k=1,2,\dots$~.

\subsection{The values of $C(0,l)$ and $D(0,l)$}
For general values of $M$, the ordinary differential equation
cannot be solved in closed form.
However it does simplify at $E=0$, and this allows the
value $D(0,l)$ to be found.
Observe first that the function
\eq
\varphi(x)=\left(\fract{M{+}1}{2}\right)^{\frac{M}{2M{+}2}}
x^{\frac{M{-}1}{2M{+}2}}\,y
\left(\left(\fract{M{+}1}{2}\right)^{\frac{1}{M{+}1}}
x^{\frac{2}{M{+}1}},E,l\right)
\en
solves the Schr\"odinger equation
\eq
\Bigl[\,-\frac{d^2}{dx^2}+x^{2}-\sigma x^{\frac{2-2M}{M+1}}
+ \frac{\gamma(\gamma+1)}{x^2}
\,\Bigr]
\varphi(x)=\Lambda \;\varphi(x) \, ,
\label{spectiii}
\en
where
\eq
\sigma=\Bigl(\fract{2}{M{+}1}\Bigr)^{\frac{2M}{M+1}}E\,,~~~
\gamma=\frac{2l{+}1}{M{+}1}-\frac{1}{2}\,,~~~
\Lambda=0~.
\en
At $E{=}0$, $\sigma$ is zero and
(\ref{spectiii}) is the simple harmonic oscillator, which as just
described is exactly
solvable in terms of the confluent
hypergeometric functions $U(a,b,z)$ and $M(a,b,z)$.  The
correctly-normalised  solution, subdominant in the sector ${\cal S}_0$, is
\eq
\varphi(x)|\phup_{E=0}=\frac{1}{\sqrt{2i}}\,x^{\gamma+1}\,e^{-x^2/2}
\,U\left(\,\fract{1}{2}(\gamma{+}\fract{3}{2}),
\gamma{+}\fract{3}{2},x^2\,\right).
\en
Reversing the variable changes, extracting the leading behaviour as
$x\to 0$ and comparing with
(\ref{dy}), we find
 \eq
\ D(E,l)|\phup_{E=0}=
\frac{1}{\sqrt{2i\pi}}
\Gamma{(1+\fract{2l+1}{2M+2})}
(2M{+}2)^{\frac{2l+1}{2M+2}
        +\frac{1}{2}}\,.
\label{Dzero}
\en
The value of $C(0,l)$ is easier to find: from
(\ref{tq}) at $E=0$,
\eq
C(E,l)|\phup_{E=0}=
2 \cos \left( \frac{2l{+}1}{2M{+}2} \pi \right).
\label{Tzero}
\en

\subsection{The large-$E$ behaviour of $D(E,l)$}
Given that $\psi(x,E,l)$ is defined to have small-$x$ behaviour
$\psi(x,E,l)\sim x^{l+1}$, the Wronskian
$D(E,l)\equiv W[y(x,E,l),\psi(x,E,l)]$ can be evaluated for
$\Re e\,l>-1/2$
as
\eq
D(E,l) = \lim_{x \rightarrow 0}\left[(2l{+}1)x^l\,y(x,E,l)\right],
\label{newD}
\en
where
\eq
\lf[-\frac{d^2}{dx^2} +x^{2M}
 +\frac{l(l+1)}{x^2}-E\ri] y
=0\,.
\en
For this subsection the aim is to find the leading behaviour of
$D(E,l)$ as $|E|\to\infty$.  As observed by Langer \cite{La},
the usual WKB approximation cannot be applied directly in this
situation because of the
singularity at $x=0$.
Instead, Langer suggested making the variable change
$x=e^z$\,,\ $y(x)= e^{z/2}\phi(z)$\,. Setting $\lambda = l+1/2$ the
equation becomes
\eq
\lf[-\frac{d^2}{dz^2} +R(z,E,\lambda)\ri] \phi =0
\label{sr1}
\en
where $R(z,E,\lambda) = e^{2z/g} 
 -Ee^{2z} +\lambda^2$ and we set $g=1/(M{+}1)$.
The general WKB approximation for $\phi$ is
\eq \fl
\ \frac{A}{R(z,E,\lambda)^{1/4}} \exp\!\lf[ \int_{z_0}^z
\!\!\sqrt{ R(u,E,\lambda)} \, du \ri ]  +
\frac{B}{R(z,E,\lambda)^{1/4}} \exp\!\lf[-\!\!\int_{z_0}^z \!\!
\sqrt{ R(u,E,\lambda)} \, du \ri ]~
   \en
where $z_0$ is an arbitrary constant. Provided $\arg(-E)<\pi$,
this WKB
expression is valid for all real $z$ (and thus for $x$ all the way
down to zero)
since
$R^{-3/4}(R^{-1/4})^{''}$ tends to zero uniformly in $z$ as $|E| \to
  +\infty$  (see \cite{OLV} for more
discussion of the validity of the WKB approximation).
The solution we aim to approximate is subdominant as $x \to
+\infty$, which requires $A=0$. The value of $B$ is fixed by
the large-$x$ asymptotic (\ref{appas}). It is convenient to set
$z_0=+\infty$, in which case
the correctly-normalised approximate solution can be written
for $M>1$ ($g<1/2$)
as~\cite{DTb}
\bea\fl ~~~
&& \phi^{\rm{WKB}}(z,E,\lambda) =\nn\\[3pt]
\fl &&\qquad \frac{1}{\sqrt{2i} \, R(z,E,\lambda)^{1/4}}
\exp \!\lf( \int_{z}^\infty \!\!\lf[ \sqrt{ R(u,E,\lambda)} -
e^{u/g} \,\right]du  -g\, e^{z/g} \ri).
\label{phiwkb}
\eea
Substituting into (\ref{newD}) gives the WKB approximation to the
spectral determinant:
\bea\fl \quad
 D^{\rm{WKB}}(E,l)
&=& \lim_{z \rightarrow -\infty}\left[2\lambda\,e^{\lambda
z}\,\phi^{\rm{WKB}}(z,E,l)\right] \nn\\[3pt]
\fl \quad&=&\sqrt{\frac{2\lambda}{i}}\,\lim_{z \rightarrow
-\infty}\left[
\exp\lf(\lambda z+ \int_{z}^\infty \!\lf[ \sqrt{ R(u,E,\lambda)} -
e^{u/g} \,\ri] du  \ri)\right].
\label{Dwkb}
\eea
To eliminate extraneous factors it is convenient to divide through by
$D^{\rm WKB}(0,l)$, giving
\eq
\frac{D^{\rm WKB}(E,l)}{D^{\rm WKB}(0,l)}=\exp(I(E,\lambda))
\label{Dwkbr}
\en
where
\eq
I(E,\lambda)=\int_{-\infty}^{\infty} \left[ \sqrt{ e^{2u/g}
-E e^{2u} + \lambda^2} -  \sqrt{ e^{2u/g} + \lambda^2}\,\right]du\,.
\label{Iwkb}
\en
The remaining step is to analyse this expression as $|E|\to \infty$
with $\arg(-E)<\pi$. Changing variables
$u\to u+\frac{g}{2-2g}\ln(-E)$ and setting
$\mu=1/(2{-}2g)=(M{+}1)/(2M)$\,,
\bea
\hspace{-1cm}
I(E,\lambda)
&=&(-E)^{\mu}\!
\int_{-\infty}^{\infty} \left[ \sqrt{ e^{2u/g}
+ e^{2u} + \lambda^2(-E)^{-2\mu}} -  \sqrt{ e^{2u/g} +
\lambda^2(-E)^{-2\mu}}\,\right]du\nn\\[3pt]
&\sim &(-E)^{\mu}\!
\int_{-\infty}^{\infty} \left[ \sqrt{ e^{2u/g}
{+} e^{2u} } -  e^{u/g} \,\right]du
=(-E)^{\mu} \!\int_0^\infty \lf[\sqrt{t^{2M}{+}1} -t^M\ri]dt \,.
\eea
Performing the integral then gives the result quoted in the main text:
\eq
\ln D(E,l) \sim \frac{a_0}{2} (-E)^\mu
 \quad,\quad |E| \to
\infty \quad, \quad |\arg(-E) | <\pi~,
\en
where
\bea
a_0
&=&
\frac{1}{\sqrt \pi} \,\Gamma {(-\fract{1}{2} {-}\fract{1}{2M})} \,
\Gamma
{(1{+}\fract{1}{2M})}  \nn \\
&=& \frac{-1}{\sqrt \pi} \, \Gamma{( -\mu) } \, \Gamma{( \mu+\fract{1}{2})}
\quad , \quad \mu=(M{+}1)/2M~.
\eea
Notice that this result is independent of $l$, and applies equally to
$D_-(E,l)\equiv D(E,l)$ and (after analytic continuation) to
$D_+(E,l)\equiv D(E,-1{-}l)$.

\subsection{The `small-$E$/large-$l$' asymptotic of  $\ln D(E,l)$.}
\label{largep}
The Langer-transformed equation can also be used to extract
the large-$l$ behaviour of the spectral determinant, a piece
of data which is relevant for matching solutions of the quantum
Wronskian equation. Once
the initial differential equation for   $y(x,E,l)$ has been
transformed into the equivalent Schr{\"o}dinger  problem (\ref{sr1}),
the  r\^ole of the energy parameter  is taken by $-\lambda^2$, and the
WKB approximation is thus good in the large-$l$ limit, in addition to
the large-$E$ limit investigated above. It is therefore
possible to study the large-$l$  behaviour of $D(E,l)$ with
standard semiclassical techniques.
We start with the WKB expression (\ref{Dwkbr}) for the ratio
$D^{\rm{WKB}}(E,l)/D^{\rm{WKB}}(0,l)$, and expand the integrand of
$I(E,\lambda)$ about $E=0$ as
\eq
\fl\quad
\sqrt{ e^{2u/g} -E e^{2 u} + \lambda^2}- \sqrt{ e^{2u/g}  +
\lambda^2}= - \sum_{n=1}^{\infty} { (2n-3)!! \over n! 2^n} {E^n e^{2
n u} \over \left( e^{2 u/g}+\lambda^2 \right)^{n-1/2} }~,
\en
so that
\eq
I(E,\lambda)= -\sum_{n=1}^{\infty}  b_n(\lambda)  E^n~,
\label{ser}
\en
where the coefficients $b_n$ can be written in terms of standard
Beta integrals:
\bea
b_n(\lambda) &=& {  (2n-3)!! \over n! \, 2^n} \int_{-\infty}^{\infty}
 {e^{2 n \te} \over \left( e^{2 \te/g}+\lambda^2 \right)^{n-1/2}
}\,d\te \nn \\
&=& \lambda^{1-2n +2ng }\left({ g  \over 4\,n! \sqrt{\pi}} \right)
\Gamma(n g) \Gamma(-1/2+n(1-g))~.
\eea
To ease comparison with  \ref{app:qwronk}, we also give
this result in integrable models notation, using the dictionary
entries
\eq
\lambda= 2 p/g~~,~~E= \Gamma^{2}(1{-}g) \left({ 2 \over
g} \right)^{\!2 -2g }s
\en
to find
\eq
\ln \frac{D^{\rm{WKB}}(E,l)}{D^{\rm{WKB}}(0,l) }
= - \sum_{n=1}^\infty  a_n(p)\,s^n
\en
where in the limit $p\gg 1$
\eq
a_n(p)\sim  b_n(2 p/g)\,\Gamma^{2n}(1{-}g) \left({ 2 \over g}
\right)^{\!2n-2ng } =\alpha_n \,p^{1-2n+2ng}~,
\en
\eq
\alpha_n={ \Gamma(ng) \Gamma(-1/2+n(1{-}g)) \over 2n! \sqrt{\pi}}
\Gamma^{2n}(1{-}g)~,
\en
matching the results quoted in  \cite{Bazhanov:1998za} and in
(\ref{prop3}).


%
\resection{Quantum Wronskians and the Weiner-Hopf method}
 \label{app:qwronk}
In this appendix we show how the quantum Wronskian allows the power
series expansions of the functions $\ln Q_{\pm}$ to
be pinned down uniquely, following Appendix A of
\cite{Bazhanov:1998za}. A useful background reference for the
Weiner-Hopf method is \cite{Roos}.
The proof relies on the following properties of $Q(s,p)$:
\begin{enumerate}
\item The function $\ln Q_+(s,p)$ has a formal power series
expansion
\eq
\ln Q_+(s,p)=-\sum_{n=1}^{\infty}a_n(p)\,s^{n}~.
\label{lnexp}
\en
\item
The coefficients $a_n(p)$ are meromorphic functions of $p$, analytic
in the half-plane $\Re e(2p)>-g$, where $g=\beta^2$ so $q=e^{i\pi g}$.
\item
As $p\to\infty$ in the right half-plane,
\eq
a_n(p)\sim \alpha_n\,p^{1-2n+2ng}
\label{prop3}
\en
with
\eq
\alpha_1=\frac{1}{2\sqrt{\pi}}\Gamma(g)\Gamma(\frac{1}{2}-g)\Gamma(1-g)^2.
\label{alphaone}
\en
\item
$Q_-$ is related to $Q_+$ as
\eq
Q_-(s,p)=Q_+(s,-p)=\exp\lf (-\sum_{n=1}^{\infty}a_n(-p)\,s^{n}\ri).
\label{lnexpm}
\en
\item
The quantum Wronskian condition holds:
\eq
\fl
e^{2\pi ip}Q_+(qs,p)\,Q_-(q^{-1}s,p)-
e^{-2\pi ip}Q_+(q^{-1}s,p)\,Q_-(qs,p)=
2i\sin(2\pi p)~.
\label{qwr}
\en
\end{enumerate}
\medskip
These properties lead to a sequence of problems
of Weiner-Hopf type which fix the
coefficients $a_n(p)$ uniquely, as follows.

First, substituting the power series (\ref{lnexp}) and (\ref{lnexpm})
into the quantum Wronskian (\ref{qwr}) gives
\bea
\fl
&&\qquad e^{2\pi
ip}\exp\left(-\sum_{n=1}^{\infty}(q^na_n(p)+q^{-n}a_n(-p))s^n\right)\nn\\
\fl
&&\qquad\qquad\qquad{}- e^{-2\pi
ip}\exp\left(-\sum_{n=1}^{\infty}(q^{-n}a_n(p)+q^{n}a_n(-p))s^n\right)
= 2i\sin(2\pi p)~.\qquad
\label{expqwr}
\eea
Expanding and equating
coefficients
leads to a set of relations of the form
\eq
\sin(\pi ng+2\pi p)a_n(p)-
\sin(\pi ng-2\pi p)a_n(-p)=R_n(p)\,,\quad n=1,2,\dots~
\label{qwronkone}
\en
where each function $R_n(p)$ can be expressed in terms of
$a_k(\pm p)$ with $k=1\dots n{-}1$. For example,
\bea
R_1&=&0\\
R_2&=&-(q \;a_1(p) +q^{-1} a_1(-p))^2 e^{4 \pi i p} \sin(2 \pi
p)/2~.
\eea

Let's consider a homogeneous case first, setting $R_n=0$. Then
(\ref{qwronkone})
can be rewritten as
\eq
\frac{\Gamma(1-ng+2p)}{\Gamma(ng+2p)}\,a_n(p)=
\frac{\Gamma(1-ng-2p)}{\Gamma(ng-2p)}\,a_n(-p)
\label{qwrr}
\en
The left-hand side of this equation is analytic in the right
half-plane $\Re e(2p)>-g$, and the right-hand side is analytic in the
left half-plane $\Re e(2p)<g$. The common strip of analyticity
$-g<\Re e(2p)<g$ ensures that both sides correspond to a single
function $f_n(p)$ which is therefore entire: it is analytic on the
whole complex plane. Furthermore, combining
$\Gamma(z+a)/\Gamma(z+b)\sim z^{a-b}$ with the asymptotic
(\ref{prop3}) shows that $f_n(p)=O(|p|^{2-2n})$ in the right
half-plane; and likewise in the left half-plane. For $n>1$ Liouville's
theorem then implies that
$f_n(p)$ is identically zero (i.e. there is no zero mode)
while for $n=1$, $f_1(p)$ is a constant, which is then fixed by
(\ref{alphaone}).

\begin{figure}[ht]
\begin{center}
\includegraphics[width=0.54\linewidth]{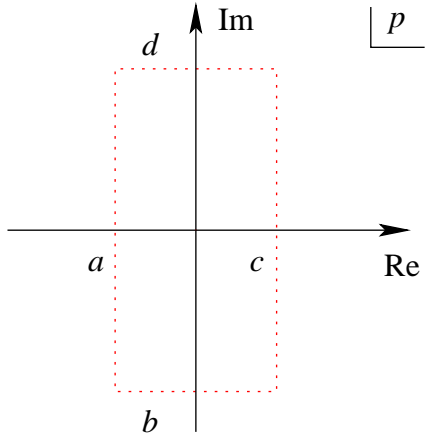}
\end{center}
\caption{The contour for the analytic decomposition of $S_n(p)$.%
\label{contourfig}}
\end{figure}

However, for $n>1$ the RHS of (\ref{qwronkone}) is not zero. In
addition the faster rate of decay of $a_n(p)$ allows for a more
effective rearrangement of (\ref{qwronkone}). Start by rewriting
this equation as
\eq
\fl\qquad
\sin(\pi ng+2\pi p-\pi K)a_n(p)-
\sin(\pi ng-2\pi p-\pi K)a_n(-p)=(-1)^KR_n(p)
\label{qwronktwo}
\en
for $K\in\ZZ$,
and then as
\eq
\fl\qquad
\frac{\Gamma(1-ng+2p+K)}{\Gamma(ng+2p-K)} a_n(p)-
\frac{\Gamma(1-ng-2p+K)}{\Gamma(ng-2p-K)} a_n(-p)
=S_n(p)
\label{qwronkthree}
\en
where
\eq
\fl\qquad
S_n(p)=
\frac{(-1)^K}{\pi}\,
\Gamma(1-ng+2p+K)\,\Gamma(1-ng-2p+K)\,
R_n(p)\,.
\label{sndef}
\en
This function has poles at the points
\eq
2p=\pm(K+1-ng+m)\,\qquad m=0,1,2,\dots
\en
coming from the gamma functions. Thus it has a strip of analyticity
$|\Re e\,p|<(K+1-ng)$.
To maximise the
width of this strip, $K$ should be as
large as possible. At the same time, we must
preserve the property that the first term on
the LHS of (\ref{qwronkthree}) decays in the right half-plane, and the
second term decays on the left half-plane. The asymptotic behaviour of
the first term in the right half-plane is
\eq
\frac{\Gamma(1-ng+2p+K)}{\Gamma(ng+2p-K)}\,a_n(p)\sim p^{2-2n+2K}
\en
so the largest value we can take for $K$ is $n{-}2$. The equation to be
solved becomes
\eq
\fl\qquad
\frac{\Gamma(n-1-ng+2p)}{\Gamma(2-n+ng+2p)}\,a_n(p)-
\frac{\Gamma(n-1-ng-2p)}{\Gamma(2-n+ng-2p)}\,a_n(-p)
=S_n(p)
\label{qwronkfour}
\en
where
\eq
\fl\qquad
S_n(p)=
\frac{(-1)^n}{\pi}\,
\Gamma(n-1-ng+2p)\,\Gamma(n-1-ng-2p)\,
R_n(p)\,.
\en
The final step is to find an `analytic decomposition' of the right-hand
side of (\ref{qwronkfour}) into the sum of functions analytic and
decaying in the
left and right half-planes:
\eq
S_n(p)=G_+(p)-G_-(p)~.
\label{ssplit}
\en
Assuming for the moment that this can be achieved, (\ref{qwronkfour})
can be rewritten as
\eq
\fl\qquad
\frac{\Gamma(n-1-ng+2p)}{\Gamma(2-n+ng+2p)}\,a_n(p)
-G_+(p)
=
\frac{\Gamma(n-1-ng-2p)}{\Gamma(2-n+ng-2p)}\,a_n(-p)
-G_-(p)~.
\label{qwronkfive}
\en
Now the same arguments as given  above
can be used to show
that the left- and right-hand sides of (\ref{qwronkfive}) are
identically zero. Hence
\eq
a_n(p)=
\frac{\Gamma(2-n+ng+2p)}{\Gamma(n-1-ng+2p)}\,
G_+(p)~.
\label{anexp}
\en
It only remains to find the decomposition (\ref{ssplit}). So long as
$S_n(p)$ has a non-empty strip of analyticity and remains bounded
there, this can be achieved
using Cauchy's theorem. Write
\eq
S_n(p)=\frac{1}{2\pi i}\oint_{\cal C}
\frac{S_n(p')}{(p'-p)}\,dp'
\label{cntrint}
\en
where ${\cal C}$ is the contour $a\cup b\cup c\cup d$ shown in
figure \ref{contourfig}, enclosing the point $p$ and lying entirely
within the strip of analyticity.

Moving the segments $b$ and $d$ down and up to $-i\infty$ and
$+i\infty$ respectively, their
contributions to $S_n(p)$ vanish, allowing us to write
\eq
S_n(p)=
G_+(p)-G_-(p)
\en
where $G_+(p)$ is the contribution to
(\ref{cntrint}) from $a$, and $G_-(p)$ is (minus) the contribution
from $c$, and these two functions do indeed decay in the right and
left half-planes, respectively. Changing integration variable to
$t\equiv p'/i$ and using (\ref{sndef}) for $S_n(p)$, (\ref{anexp})
becomes
\eq
\fl
a_n(p)=
\frac{(-1)^n}{2\pi^2}
\frac{\Gamma(2{-}n{+}ng{+}2p)}{\Gamma(n{-}1{-}ng{+}2p)}\,
\int_{-\infty}^{\infty}\!
\frac{\Gamma(n{-}1{-}ng{+}2it)\,
\Gamma(n{-}1{-}ng{-}2it)\,R_n(it)}{(p{-}it)}\,dt~,
\en
and $a_n(p)$ has  been determined uniquely in terms of the $a_m(p)$
with $m<n$. Since the first coefficient, $a_1(p)$, has already been
fixed via the homogeneous equation and the asymptotic
(\ref{alphaone}), this completes the proof that the power series
(\ref{lnexp}) is uniquely determined by properties (i) -- (v). Note
that the question of the convergence of the series is at this stage
moot. However, once the identification with the expansion of a
spectral determinant has been made, this follows from standard
properties of the solutions of ordinary differential equations.


%
\resection{Derivation of the TBA equations and of the NLIE}
\label{deriv}
In this appendix we describe how to turn certain functional
relations into nonlinear integral equations. We also indicate how
the equations encode the effective central charges of the
associated integrable models.

\subsection{TBA equations from truncated fusion hierarchies}
\label{TBAsection}
Finite sets of functional equations  such as the T-system
(\ref{twl}) -- or equivalently (\ref{ctba}) -- can be transformed
into sets of nonlinear integral equations of the form known in the
integrable model community as  thermodynamic Bethe ansatz (TBA)
equations~\cite{ZamTBA}. Recall that the T-system (\ref{twl}) is the
following set of $h{-}1$ coupled functional equations
\eq
\fl \quad t^{(m)}( \omega^{-1}E)t^{(m)}( \omega E)= 1+
\prod_{j=1/2}^{(h-1)/2}
\lf ( t^{(j)}(E) \ri)^{G_{2j,2m}},~~~m=1/2,1,\dots,(h{-}1)/2
\en
where $\omega=e^{2\pi i/h(M{+}1)}$
and $G$ is the incidence matrix of
the $A_{h-1}$ Dynkin diagram. In fact the following arguments apply
equally for any simply-laced Dynkin diagram, and so -- in the spirit
of \cite{Zamy,Ravanini:1992fi} --
we shall give a slightly more general treatment
here than is strictly needed for the differential equation
application. Also, because we consider the more general situation we
shall use the integrable model notation $T$ rather than $t$ and $s$
rather than $E$. We
further streamline by
setting  $r=h{-}1$ and
\eq
T_a(s)=t^{(a/2)}(E)~,\quad a=1\dots r
\en
so that the T-system is
\eq
T_a( \omega^{-1}s)T_a( \omega s)= 1+
\prod_{b=1}^{r}
\lf ( T_b(s) \ri)^{G_{ab}},\quad a=1\dots r.
\label{Tsys}
\en
In the more general cases, $a$ and $b$ label nodes on the Dynkin
diagram described by $G_{ab}$\,, and $r$ is its rank.

The  first step towards the TBA is to introduce the
Y-functions, defined by
\eq
Y_{a}(s)=\prod_{b=1}^r T_b(s)^{G_{ab}}~, \quad a=1\dots r.
\label{Ydef}
\en
Clearly, (\ref{Tsys}) implies
\eq
T_b( \omega^{-1}s)T_b( \omega s)= 1+Y_b(s)
\label{TYsys}
\en
and a suitable product over $b$
results in the set of equations
\eq
Y_a(\omega s)Y_a(\omega^{-1} s)=\prod_{b=1}^{r}
(1+Y_{b}(s))^{G_{ab}}\,,
\label{ysys}
\en
called a Y-system.
Equivalently, on setting
$s=e^{\theta/\mu}$ with
$\mu=(M+1)/hM$, we
have
\eq
Y_a(\theta+i \fract{\pi}{h}) Y_a(\theta-i \fract{\pi}{h} )=
 \prod_{b=1}^{h-1} \lf(1+Y_{b}(\theta) \ri)^{G_{ab}}\,.
\label{yte}
\en
These coincide  with the
Y-systems of Al.B.\ Zamolodchikov~\cite{Zamy}, which for the $A_{h-1}$
cases encode the finite
size effects  of
certain integrable quantum field theories with $\ZZ_h$ symmetry. (We
shall comment on the  relationship between these models and the
continuum limit of the twisted six-vertex model at the relevant
values of  $\phi$ and $\eta$ at the end of this subsection.)

Since in our cases
$T$ and $Y$ are entire functions of $s$, they are $2\pi i\mu
=2\pi i (h{+}2)/h$ -periodic functions of $\theta$. In fact, given the
symmetry $Y_a=Y_{h-a}$, this periodicity also follows from the
Y-system (\ref{yte}); see \cite{Zamy,Gliozzi:1995wq,Frenkel:1995vx}
for more details.

Next, the Y-system is transformed
into a set of integral equations by
introducing  the `pseudoenergies'
\eq
\ep_a(\te)=\ln Y_a(\te)~,
\en
which as a result of (\ref{yte}) satisfy
\eq
\ep_a(\te+i \fract{\pi}{h} )+\ep_a(\te-i \fract{\pi}{h}) -
\sum_{b=1}^{r} G_{ab} \ep_b(\te) =
 \sum_{b=1}^{r}  G_{ab} \,L_b(\te)
\label{ft}
\en
with
\eq
L_a(\te)=\ln(1{+}e^{-\ep_a(\te)})~.
\en
Equation (\ref{ft}) has many solutions, and to pin things down some
more information about the properties of the functions involved is
required. We shall need the following. Since the functions
$t^{(a/2)}(s)$ are entire and hence regular at $E=0$,
the $Y_a(\theta)$  approach constant values
as $\theta\rightarrow -\infty$\,.  These constants solve
the stationary ($\theta$-independent)
version of (\ref{yte}); as will be justified
shortly, the particular
solution relevant here has all its components positive:
\eq
{\cal Y}_a=\lim_{\te \rightarrow -\infty} e^{\ep_a(\te)}= {\sin(
\fract{a \pi}{h+2}) \sin( \fract{(a+2)\pi}{h+2}) \over
\sin^2(\fract{\pi}{h+2})}~.
\label{negasy}
\en
On the other hand, as $|s|\to\infty$ with
$-\pi+\delta<\arg(s)<\pi-\delta$ (where $\delta$ is an
arbitrarily-small positive number), the leading asymptotics of the
$\ln(t^{(m)}(s))$, and hence also of the
$\ln(Y_a(s))$, are proportional to
$s^{\mu}$\,\footnote{Asymptotics of this nature
can be proved in the ODE world using well-established techniques -- see
for example \cite{Sha,Shin:2002ay}; given the ODE/IM correspondence
this is also now the most efficient approach for the integrable
models.}. Thus, as $\Re e\,\theta\to +\infty$ in any strip
$|\Im m\,\theta|<\pi(h{+}2)/h-\delta$, the pseudoenergies behave as
\eq
\ep_a(\theta)\sim m_aL\,e^{\theta}\,,
\label{epasy}
\en
with, in fact, exponentially-small corrections.
A consideration of (\ref{ft}) at large real values of $\theta$, where
the right-hand side vanishes,
shows that the constants $m_aL$ are proportional to the components
of the Perron-Frobenius eigenvector of $G$. For the $A_{h-1}$ case of
prime interest here, we can write these components as
\eq
m_a L= \fract{b_0}{2} \sin (\fract{ \pi a}{h}) ~, \quad b_0 \in
\RR\,.
\en
Finally, for the solution of (\ref{ft}) that we require here --
the ground state eigenvalue of the integrable model, or
equivalently the quantum-mechanical spectral determinant --
the zeros of the $t^{(a/2)}(s)$ all lie on the negative $s$
axis. This means that the Y-functions do not vanish on the
real $\theta$
axis, which when combined with (\ref{epasy}) justifies the choice
of the solution
(\ref{negasy}) of the stationary Y-system.
It also implies that the functions
\eq
f_a(\te)=\ep_a(\te)- m_a L \, e^\te
\en
are bounded in the `analyticity strip'
$|\Im m (\te)|<\pi/h$\footnote{The existence
of such a strip surrounding the real axis is
usually deduced numerically in finite-lattice models and
extrapolated to the continuum limit. Via the ODE/IM
correspondence this property has now been proven  directly in the
continuum: it follows from the entirety property
of the  $T$'s and the fact that their zeros are exactly on the
negative $s$-axis~\cite{DDTb, Shin:2002ay}.}
 and satisfy the
relation
\eq
f_a(\te+i \fract{\pi}{h} )+f_a(\te-i \fract{\pi}{h}) -
\sum_{b=1}^{r} G_{ab} f_b(\te) =
 \sum_{b=1}^{r}  G_{ab} \,L_b(\te)\,.
\label{ft1}
\en
Taking an ($\epsilon$-regularised)  Fourier transform
\eq
\tilde f(k)={\cal F}[f(\te)] = \lim_{\epsilon \rightarrow 0^+}
\int_{-\infty}^\infty d\te \, f(\te) e^{-ik\te+ \epsilon \te}
\en
 of both
sides of (\ref{ft1}),
\eq
\sum_{b=1}^{r} (2 \, \delta_{ab}  \cosh(\fract{\pi k}{h})-G_{ab})
\, \tilde{f}_b(k)=
 \sum_{b=1}^{r}  G_{ab} \, \tilde{L}_{b}(k)\,.
\en
Solving for $\tilde f_a(k)$ and transforming back to
$\te$-space yields the equations
\eq
\ep_a(\te)= m_a L e^\te  - {1  \over 2 \pi}\sum_{b=1}^{r}
\int_{-\infty}^{\infty} \, \phi_{ab}(\te-\te') L_b(\te')\,d \te'
\label{ztba}
\en
with
\eq
\tilde \phi_{ab}(k)= -2\pi\sum_{c=1}^{r} \lf (
2 \, \delta_{ac} \cosh( \fract{\pi k}{h}) - G_{ac}
  \ri)^{-1} G_{cb}~.
\label{ftk}
\en
The inverse Fourier transforms of the kernels
$\tilde\phi_{ab}(k)$ can be found exactly in terms of elementary
functions,
and the results written as
\eq
\phi_{ab}(\theta)= - i \frac{d}{d \theta}\ln S_{ab}(\theta)\,,
\en
where
\eq
S_{ab}(\te)=\prod_{x\in A_{ab}}
\{x\}~,\qquad a,b=1\dots r\,.
\label{Sform}
\en
The integer-valued index set $A_{ab}$ can be characterised via
the Weyl group of the related Lie algebra, and
\eq
\{x\}=(x{-}1)(x{+}1)\,~~~,~~~ (x)={\sinh(\fract{\te}{2}{+}
i\fract{\pi x}{2h})
 \over \sinh(\fract{\te}{2}{-} i\fract{\pi x}{2h})}~.
\en
For $A_{h-1}$ the explicit formula is
\eq
S_{ab}(\te)=\prod_{|a-b|+1 \atop {\rm step~2}}^{{}\atop a+b-1}
\{x\}~,\qquad a,b=1\dots r\,.
\en
For the other cases see, for example,
\cite{Braden:1989bu,Dorey:1990xa}; for their relation with the Fourier
transforms (\ref{ftk}), see
\cite{Zamy,Ravanini:1992fi}.

Nonlinear integral equations of the form (\ref{ztba}) also arise in
the framework of  the thermodynamic Bethe ansatz~\cite{ZamTBA,
Klassen:1990dx} method for finding the ground-state energies of
massive relativistic integrable field theories in $1+1$ dimensions,
and hence they are often called TBA equations. In this picture $L$
should be interpreted as the circumference of the infinite cylinder
on which the theory is defined, and $S_{ab}(\te)$ are the
two-particle S-matrix elements describing the factorised scattering
of $r$ particles of relative masses $m_a$.  Strictly speaking, the
equations just derived correspond to an ultraviolet limit of these
TBA equations, in which the overall mass scale is taken to zero.

Once the $\ep_a(\theta)$ and hence the $Y_a(\theta)$ have been found
via (\ref{ztba}),
it remains to recover the T-functions. This can be
done using much the same reasoning as led to the TBA equation,
starting this time from (\ref{TYsys}).
Dividing through by $Y_a(s)$,
using (\ref{Ydef}) and taking logs,
\eq
\ln T_a(\te+i \fract{\pi}{h} )+\ln T_a(\te-i \fract{\pi}{h}) -
\sum_{b=1}^{r} G_{ab} \ln T_b(\te) = L_a(\te)\,.
\label{lntform}
\en
From (\ref{negasy}) and (\ref{TYsys}), the functions
$\ln T_a(\te)-m_aL\,e^{\te}/(2\cos(\pi/h))$ are bounded in the
same analyticity strip as the
$f_a(\te)$ above, and can likewise be found by
a Fourier transformation. The final result is
\eq
\ln T_a(\te)= \frac{m_a L}{2\cos(\pi/h)}\,
 e^\te  - {1  \over 2 \pi}\sum_{b=1}^{r}
\int_{-\infty}^{\infty} \, \psi_{ab}(\te-\te') L_b(\te')\,d \te'
\label{intform}
\en
where the Fourier transform of the kernel $\psi_{ab}(\te)$ is
\eq
\tilde \psi_{ab}(k)=  -2\pi\lf (
2 \, \delta_{ab} \cosh( \fract{\pi k}{h}) - G_{ab}
  \ri)^{-1}.
\en
Again, the inverse transform can be performed, to find that
$\psi_{ab}(\te)$ is defined as $\phi_{ab}(\te)$, save for the replacement
of the blocks $\{x\}$ by $(x)$ in (\ref{Sform}). However,
as explained in section \ref{stba} below,
for spectral applications one can often get away without calculating
the $T_a$ explicitly \cite{DTa}.

It is also interesting to extract the ground state energy of the spin
chain, or of the corresponding quantum field theory. It might be
tempting to use (\ref{Hdef}) for this purpose, but one has to be
careful as the continuum limit involves an infinite shift of the
$\theta$-like variable $\nu$ in (\ref{Hdef}). This provides some
intuition for the fact that the universal part of the
continuum ground-state energy is to be found in the large-$\theta$
asymptotic of the fundamental transfer matrix, $T_1(\theta)$ in current
notations.

The relevant asymptotic expansion has the form \cite{BLZ1,BLZexc}
\eq
\ln T_1(\te)-\frac{m_1 L}{2\cos(\pi/h)}\,e^{\theta}\sim
-\sum_{n=1}^{\infty}C_nI^{\rm vac}_{2n{-}1}\,e^{(1{-}2n)\te}
\label{toneexp}
\en
where the numbers $I^{\rm vac}_{2n-1}$ are the eigenvalues of the
local conserved charges of the system on a cylinder of circumference
$L$. The first of these, $I_1$, gives the ground state energy;
and from the results of \cite{BLZ1,BLZexc},
\eq
\ln T_1(\te)-\frac{m_1 L}{2\cos(\pi/h)}\,e^{\theta}\sim
-\frac{4\sin(\pi/h)}{m_1}\,I^{\rm vac}_1\,e^{-\theta}-\dots~.
\en
This asymptotic also follows from (\ref{lntform}). Let $q^{(\sigma)}_a$ be
an orthonormal set of eigenvectors of $G_{ab}$, where $\sigma$ runs over
the exponents of $G$, so that
\eq
\sum_{b=1}^rG_{ab}\,q^{(\sigma)}_b=2\cos(\pi
\sigma/h)\,q^{(\sigma)}_a\quad ,\quad
\sum_\sigma q_a^{(\sigma)}q_b^{(\sigma)}=\delta_{ab}\,.
\en
Then
\eq
\tilde\psi_{ab}(k)=-\pi\sum_\sigma\frac{q_a^{(\sigma)}q_b^{(\sigma)}}{\cosh(\pi
k/h)-\cos(\pi \sigma/h)}
\en
and is easily seen to have poles at $k=(\pm \sigma+2hn)i$, $n\in
\ZZ$\,. For large $\theta$
the inverse Fourier transform yielding $\psi_{ab}(\theta)$ can be
expanded over the residues of these poles, the leading term coming
from the pole nearest to the real axis:
\bea
\psi_{ab}(\te)&=&\frac{1}{2\pi}\int_{-\infty}^{\infty}dk\,
\tilde\psi(k)e^{ik\theta}\nn\\
&=&\frac{-h\,}{\sin(\pi/h)}\,q^{(1)}_aq^{(1)}_b\,e^{-\theta}+\dots~.
\label{kerexp}
\eea
For the $A_r$ case relevant for comparison with the expansion
(\ref{toneexp}), the normalised eigenvector has components
$q^{(1)}_a=\sqrt{\frac{2}{h}}\,\sin(\pi a/h)$. Substituting
(\ref{kerexp}) into the integral formula (\ref{intform}) and swapping
sums with integrals gives an asymptotic expansion which matches
(\ref{toneexp}), and begins
\eq
\fl\qquad
\ln T_1(\te)-\frac{m_1 L}{2\cos(\pi/h)}\,e^{\theta}=
\frac{1}{\pi}\sum_{b=1}^r\int_{-\infty}^{\infty}\sin(\pi b/h)\,
e^{-\theta+\theta'}L_b(\theta')\,d\theta' +\dots~.
\en
Comparing leading terms,
\bea
I^{\rm vac}_1\equiv E_0(L)
&=&-\frac{1}{4\pi}\sum_{a=1}^r\int_{-\infty}^{\infty}m_1
\frac{\sin(\pi a/h)}{\sin(\pi/h)} e^{\theta}L_a(\theta)\,d\theta\nn\\
&=&-\frac{1}{4\pi}\sum_{a=1}^r\int_{-\infty}^{\infty}
m_a e^{\theta}L_a(\theta)\,d\theta
\label{Eform}
\eea
where $L_a(\theta)=\ln(1+e^{-\ep(\theta)})$ and $\ep(\theta)$ solves
(\ref{ztba}). Recalling that $c_{\rm eff}$ was defined via
\eq
E_0(L)=F(L)=-\frac{\pi c_{\rm eff}}{6L}\,,
\en
we see that (\ref{Eform}) is equivalent to
\eq
c_{\rm eff}=\frac{3}{2\pi^2}\sum_{a=1}^r\int_{-\infty}^{\infty}
d\theta\,m_aL\,e^{\theta}L_a(\theta)\,.
\label{cform}
\en
Even though $\ep(\theta)$ cannot be found in closed form, it turns out
that the integral in (\ref{cform}) can be calculated exactly as a sum
of Rogers dilogarithm  functions
\eq
{\cal L}(x)= - { 1 \over 2} \int_0^x dy \left[ {\ln y\over 1-y} +
{\ln(1-y) \over y} \right]~.
\en
(For more about
sum rules for Rogers dilogarithm
see   \cite{Kirillov:1994en}.)
For our model, the final result is
\bea
c_{\rm eff} &=& {3 \over \pi^2} \sum_{a=1}^{r} {\cal L} \left( {1
\over 1+{\cal Y}_a } \right) =  {3 \over \pi^2} \sum_{a=1}^{r}
{\cal L} \left( { \sin^2(\fract{\pi}{h+2})  \over \sin^2(\pi
\fract{a+1}{h+2})
  } \right) \nn \\
            &=& {h-1 \over h+2}={2M-1 \over 2M+2} \quad ,\quad
            (h=2M)~,
\label{ceff1}
\eea
which  is exactly half the central charge for the $\ZZ_h$ parafermion
models~\cite{Fateev:1985mm}.
 How does this compare with the continuum limit of  the
twisted six-vertex model? As mentioned before the untwisted
six-vertex model is
equivalent,  in its continuum limit,  to the theory of a free
compactified boson with central charge $c_{\rm eff}^{6V}=1$.  For
the twisted  variant of the six-vertex model we
have  instead
\eq
c_{\rm eff}^{T6V}=1-\frac{ 6 \phi^2}{\pi ( \pi -2 \eta)}.
\en
Inserting the relations
\eq
 \eta = {\pi \over 2} {M \over M+1}~,~~~~ \phi= { \pi \over 2M+2}
\en
we find
\eq
c_{\rm eff}^{T6V}={2M-1 \over 2 M+2 }~,
\label{ce6v}
\en
 which is the result
quoted in~(\ref{ceff1}). The factor of two discrepancy with the
parafermionic central charge is,
for $h$ odd, simply due to a double counting
that  can be avoided via a folding procedure that identifies
conjugate nodes $j \leftrightarrow h-j$ in the related $A_{h-1}$
Dynkin diagram. A slightly more subtle folding phenomena is at work
in the $h$ even case too~\cite{DTa}.

\subsection{NLIEs from  Bethe ansatz systems}
\label{snlie}
The TBA method relies on the truncation of the fusion hierarchy, and
so can only work at rational values of $\eta/\pi$ or $M$. In
addition the resulting equations depend in a complicated way on the
arithmetic properties of these numbers. However, there is another
approach which works more generally. Employing ideas developed in
\cite{DDV,KP,KBP,BLZ2,DTc}, the
Bethe ansatz equations associated with a  TQ relation of the type
discussed here can be transformed
 into a single nonlinear integral equation.  Starting with the TQ relation
(\ref{tq}), set $E=E_k$ and invoke the
 entirety of
  $C(E,l)$ and $D(E,l)$ to obtain
 \eq
 \omega^{2l+1}  \frac{D(\omega^2 E_k ,l) } {D(\omega^{-2} E_k ,l) }
 +1=0 \quad , \quad k=1,2,\dots~.
 \en
Replacing $D(E,l)$ with its Hadamard factorisation (\ref{had}),
valid for $M>1$, the Bethe ansatz equations for the eigenvalues of
$D(E,l)$ are recovered:
\eq
\prod_{j=1}^{\infty} \left(\frac{E_j-\omega^2 E_k }{E_j-\omega^{-2} E_k
} \right) = - \omega^{2l+1} \quad , \quad k=1\dots~.
\en
Setting
\eq
a (E)=
\omega^{2l{+}1}
\frac{D(\omega^2E,l)}{D(\omega^{-2}E,l)}
=\omega^{2l{+}1} \prod_{j=1}^{\infty}
\left( \frac{E_j-\omega^{2}E}
 {E_j-\omega^{-2}E} \right)~,
\label{dall}
\en
the Bethe ansatz equations are rephrased as conditions on the
function $a(E)$:
\eq
a (E_k)+1=0 \quad ,\quad k=1,2,\dots~.
\label{abae}
\en
In fact, it follows from the TQ relation that the zeros of $a(E)+1$
are precisely the zeros of $D(E)$, the Bethe roots $E_k$, together
with the zeros of $T(E)$. For simplicity, we consider the ground
state situation, for which all the $E_k$ are real and positive, and
all the zeros of $T(E)$ are real and negative. We begin by taking
the logarithm of
 (\ref{dall})
\eq
\ln a(E)=\frac{i\pi(2l{+}1)}{M{+}1} + \sum_{k=1}^{\infty} F(E/E_k)~,
\label{lna}
\en
where  $F(E)=\ln [ ( 1-\omega^{2}E)/(1-\omega^{-2} E)]$\,.
The logarithmic derivative $\partial_{E}\ln (1+a(E))$ has a simple
pole at each eigenvalue $E_k$. Applying Cauchy's theorem,
the infinite sum in (\ref{lna}) can be replaced by a contour
 integral, the contour
 encircling the points $E_k$ in an anticlockwise direction, while
 avoiding  all other singularities of $\ln (1+a(E))$.
Given our assumptions about the locations of the zeros of $D(E)$ and
$T(E)$, a suitable  contour   $\cal C$
  runs from $+\infty$
to $0$ above the real axis,  encircles the origin then returns to
$\infty$  below the real axis. With this in mind (\ref{lna}) becomes
\eq
\ln a(E)= \frac{i\pi(2l{+}1)}{M{+}1} + \int_{\cal C} \frac{dE'}{2\pi
i} \, F(E/E')\,
\partial_{E'}\ln(1+a(E'))\,.
\label{lnaE}
\en
We change  variables via
$E=\exp(2M\theta/(M{+}1))$ and
 define (with a mild abuse of notation)
$\ln a(\theta) \equiv \ln a(e^{2M\theta/(M+1)})\,$.
Integrating by parts we have
\bea
\ln a(\theta)&=& \frac{i\pi(2l{+}1)}{M{+}1}-\int_{{\cal C}_1} d\te'
\,\partial_{\te'}R(\te-\te') \ln(1+a(\te'))
\nn \\
&&\qquad \qquad  \qquad {}+\int_{{\cal C}_2} d\te'
\,\partial_{\te'}R(\te-\te') \ln(1+a(\te'))~,
\eea
 where
\eq
R(\theta)~=~
\frac{i}{2\pi}\,\partial_{\theta}F(e^{2M\theta/(M+1)})~.
\label{rdefn}
\en
The new integration contours ${\cal C}_1$ and ${\cal C}_2$ run from
$-\infty$ to $\infty$ just below and just above the real axis
respectively. Using the property
  $[a(\theta)]^*=a(\theta^*)^{-1}$, which follows from the Bethe
  ansatz equations, we  rewrite
 the integrals in terms of integrations along the real
 axis:
\bea
\ln a(\theta) - \int_{-\infty}^{\infty} d\theta' R(\theta-\theta')
\ln a(\theta')&=& \frac{i\pi(2l{+}1)}{M{+}1}
\nn  \\
&&\hskip-2.3cm {}-2i \int_{-\infty}^{\infty} d\theta'
R(\theta{-}\theta') \Im m \ln (1+a(\theta'{-}i 0))~.
\label{lnaeq2}
\eea
It is then a simple matter to solve this equation using Fourier
transforms.  Using the notation $\tilde f(k)$ and ${\cal
F}[f(\te)](k))$ introduced earlier,
 the Fourier-transformed equation is
\eq
\hspace{-1cm}(1-\tilde R(k))\,\CF [\ln a](k) = -\frac{2\pi^2
(2l{+}1)}{M{+}1}\delta(k) -2i \tilde R(k) \, \Im m \,\CF [\ln
(1+a)](k)~.
\en
We apply $(1-\tilde R(k))^{-1}$ to both sides, then take the inverse
Fourier transform to find
\bea
\ln a(\te) &= &\frac{i\pi(2l{+}1)}{M{+}1}
\CF^{-1}[(1-\tilde{R}(k))^{-1}](0) + i m L \, e^{\te}
\nn \\
&& \hskip 1.cm {}- 2i\int_{-\infty}^\infty
d\te'\,\varphi(\te-(\te'-i0))\Im m \,\ln(1+a(\te'-i0))\,,
\label{eqn3}
\eea
where
\eq
\varphi(\te)=\CF^{-1}\,[\, (1-{\tilde R}(k))^{-1} {\tilde
  R}(k)\,](\te)~,
\en
and $ m L$ is a real constant which
 arises from a zero mode and
can be traced to the pole in $(1-{\tilde R}(k))^{-1}$ at $k=i$.
It should be fixed by a consideration of the large-$\te$ asymptotic
of $\ln a(\te)$.
{}From the definition of $R(\te)$ and the relations
\bea
i\partial_{\te} \ln \frac{\sinh \sigma \te+i\pi\tau}{\sinh
\sigma\te-i\pi\tau} &=& \frac{2\sigma \sin 2\pi \tau}
{\cosh 2\sigma\te-\cos 2\pi\tau} \\
\int_{-\infty}^\infty \frac{d\te}{2\pi}\,  e^{-ik\te} \frac{2\sigma
  \sin 2\pi \tau}{\cosh 2\sigma\te -\cos 2\pi \tau} &=& \frac{ \sinh
  (1-2\tau) \frac{\pi k}{2\sigma} }{ \sinh \frac{\pi k}{2\sigma} }
\eea
we obtain a compact expression for the
 kernel $\varphi(\te)$:
\eq
\varphi(\te) = \int \frac{dk}{2\pi}\,e^{i k\te} \frac{\sinh
\frac{\pi k(M-1) }{2M}} {2\sinh \frac{\pi\, k}{2M} \cosh \frac{\pi
k}{2} }~.
\label{sgker}
\en
Finally we  rewrite the equation in terms of the integration
contours ${\cal C}_1$ and ${\cal C}_2$:
\bea \hspace {-1.5cm}
\ln a(\te) &=& i \pi(l+\fract{1}{2} )
- i m L \,e^{\te} \nn \\
&& \hspace {-.2cm}
 + \int_{{\cal C}_1} d\te' \, \varphi(\te{-}\te') \ln
(1+a(\te'))
 - \int_{{ \cal C}_2} d\te' \, \varphi(\te{-}\te') \ln
(1+a(\te')^{-1})~.
\label{nlie}
\eea
This equation holds for all
$\te$ within the strip $|\Im m \,\te |<
\min(\pi,\pi/M)$.   A little extra care is required for larger $\Im
m \, \te$
as the kernel
$\varphi(\te)$ has poles at $\te=
\pm i \pi$ and $ \pm i\pi/M$. If $M>1$,  the correct analytic
continuation  for positive values of $\Im m \,\te $
is
given by the second determination  \cite{DDV2}
\bea \hspace {-1.5cm}
\ln a(\te) &=&
-i  (1-e^{-i \pi /M})
m L \,e^\te  \nn \\
&& \hspace {-.8cm}
 + \int_{{\cal C}_1} d\te' \, \varphi_{II}(\te{-}\te') \ln
(1+a(\te'))
 - \int_{{ \cal C}_2} d\te' \, \varphi_{II}(\te{-}\te') \ln
(1+a(\te')^{-1})~,
\label{nlie2}
\eea
where
\eq
\varphi_{II}(\te) = \frac{2 i \cos(\frac{\pi }{2M})  \sinh {(\te-
\frac{i \pi }{2M}})} { \pi \lf ( \cosh (2 \te{-} \fract{i \pi}{M}) -
\cos(\frac{ \pi}{M})\ri
  )}~.
\en
This equation is valid for $\pi /M < \Im m \,\te < \pi$, when $M>1$.
A similar continuation can be performed for $M<1$. For a detailed
discussion of the effect of this continuation on eigenvalue
asymptotics, see \cite{DMST}.

The NLIE (\ref{nlie}) first arose in \cite{KP} as the continuum limit
of an equation describing the finite size effects of the six-vertex
model. It also appears
in relativistic
integrable scattering field theory in $1+1$ dimensions, where it
describes the finite size effects of the
ultraviolet limit of the massive sine-Gordon model \cite{DDV}.
In the latter situation the kernel $\varphi(\te)$ is
related to the scalar factor of the sine-Gordon soliton-soliton
scattering amplitude.
In the field theory context,
the effective central charge is given in terms of $\ln a(\te)$
according to
\eq
c_{\rm eff} = \frac{3imL}{\pi^2} \lf [
\int_{{\cal C}_1} \,d \te  \,   e^{\te} \log(1+a(\te'))-
\int_{{\cal C}_2} \,d \te  \, e^{\te} \log(1+a^{-1}(\te'))\ri].
\label{ceffNLIE}
\en
This integral can be evaluated exactly, with the result
\eq
c_{\rm eff} = 1- \frac{ 6(l+\hf)^2}{M+1}~,
\en
which, given the relations $\beta^2 =1/(M+1)$ and $p=(2l+1) /(4M+4)$,
 agrees perfectly with
\eq
c_{\rm eff}  = c-24 \Delta_p  = 1- 24\, \frac{p^2}{\beta^2}~.
\en

The above NLIE has been derived for the ground state, for which the
Bethe roots $\{E_k\}$ are all real and positive. However,  nonlinear
integral equations of both TBA and NLIE types can be found even when
some of the Bethe roots lie in the complex plane. See, for example,
 \cite{BLZ2,BLZexc,DDV2,Dorey:1996re,Fioravanti:1996rz,Dorey:1997rb}.

\newpage

\resection{Calculating the spectrum of an ODE}
 \label{num}
We briefly describe how to calculate the spectrum
of
\eq
 \left[
 -\frac{d^2}{dx^2}+ x^{2M}+\frac{l(l{+}1)}{x^2}-E\right]\,\psi(x)=0~.
 \en
using the nonlinear integral equations derived in
appendix~\ref{deriv}.

\subsection{TBA approach}
\label{stba}
When $M$ is an integer and $l(l+1)=0$
 the spectrum encoded in the spectral determinant
$D(E,l)$ can be determined from the fusion relations (\ref{ctba}) using TBA
equations. Recall the following  relation obtained in
section~\ref{fusion}:
\eq
\frac{i}{2} T_{M/2}(E) =\frac{i}{2} C^{(M)}(v^{-1} E) =D(-E)
\equiv  D_{+}(-E) D_{-}(-E)~.
\label{tc}
\en
{}From this we see that the real positive zeros
of $D(E)$ are
precisely the real negative zeros of $T_{M/2}(E)$ or
$C^{(M)}(E)$.  We set
\eq
t^{(m)}(\te+i \pi \fract{h+2}{2h}) \equiv T_{m}(-E) \quad , \quad h=2M~.
\en
Then from
\eq
1+Y_{2m}(\te+i \fract{\pi}{2})=t^{(m)}(\te+i \pi \fract{h+2}{2h})
t^{(m)}(\te+i\pi \fract{h-2}{2h})~,
\en
we see that the eigenvalues $\{E_k\}$ are those
  zeros of $1+Y_{M}(\theta)$ that lie
along   the line $\Im  m\, \te =  \pi/2$.

The
constants  $m_a L = \frac{b_0}{2} \sin \frac{ \pi a}{h}$
appearing in the TBA equations (\ref{ztba}) must be tuned
to match the
asymptotic  properties of the spectral determinants.
The correct result is found by setting  $b_0 =2 \cos (\pi/ 2M )
\,a_0$ where $a_0$ was defined in (\ref{das}).

A very simple iteration scheme can be used to solve the TBA equations
(\ref{ztba}) numerically.  Starting from
\eq
\fl
\quad  \ep_a^{(n+1)}(\te)=  \alpha \ep_a^{(n)}(\te) + (1-\alpha)\left[m_a L
e^\te - {1 \over 2 \pi}\sum_{b=1}^{h-1} \int_{-\infty}^{\infty} \,
\phi_{ab}(\te-\te') L_b^{(n)}(\te') {d \te' \over 2 \pi} \right]
\label{itba}
\en
with  $\ep^{(0)}_a(\theta) = m_a L \, e^\te$ and $0 < \alpha \le 1$,
(\ref{itba}) can be iterated until the desired accuracy
($\sim 10^{-14}$) is reached.  Empirically the value $\alpha=0.5$
 gives a very good rate of convergence usually in less than one
 hundred iterations.   Once the functions
 $\ep_a(\theta)$  are known
numerically along the real axis, the TBA equations~(\ref{ztba}) provide an
integral representation which can be used to reconstruct the
functions $\ep_a(\theta)$ and $Y_a(\te)=\exp (\ep_a(\theta))$
everywhere in the complex plane. The only point to watch is that the
singularities of the kernels $\phi_{ab}$'s at $\pm \imath \pi/h$ and
beyond necessitate the introduction of extra terms when the
$\ep_a$'s  are continued beyond the strip $-\pi/h<\Im m
\,\te<\pi/h$.

\subsection{NLIE approach}
\label{snlie2}
As mentioned above, the  single nonlinear integral
equation derived  in appendix \ref{snlie} encodes
 the zeros of both the spectral determinants $C$ and $D$.
The zeros
 of $1+a(E)$ on the positive real axis
are the zeros of $D(E,l)$, while
those on the negative real axis are the zeros of $C(E,l)$.
Under the variable change $E=\exp(2M\theta/(M{+}1))$ the positive real
axis of the
complex $E$-plane becomes the real-$\te$ axis. Therefore
the eigenvalues encoded in $D(E,l)$ can be found by searching
along the real axis for the zeros of $1+a(\te)$.
Similarly the eigenvalues of $C(E,l)$ can be found as
zeros of $1+a(\te)$ on the line $\Im m\,\te = \pi (M{+1})/2M$.
Before we can solve the NLIE (\ref{nlie}) we  must fix the
constants $mL$.
The large-$E$ asymptotics of $a(E)$ are
\eq
\fl \   \log a(E)\sim \left\{
\begin{array}{ll}
-\fract{1}{2}ib_0(1{-}e^{-i\pi/M}) (E)^{\mu} \quad
 & ~\frac{2\pi}{M{+}1}<\arg (E)<2\pi-\frac{2\pi}{M{+}1}\\[7pt]
-\fract{1}{2}ib_0 (E)^{\mu}
 & -\frac{2\pi}{M{+}1}<\arg (E)<\frac{2\pi}{M{+}1}\\[7pt]
-\fract{1}{2}ib_0(1{-}e^{i\pi/M}) (E)^{\mu}
 & -2\pi+\frac{2\pi}{M{+}1}<\arg (E)<-\frac{2\pi}{M{+}1}
\end{array}\right. ~,
\en
where $b_0=2\cos(\frac{\pi}{2M})a_0$ as before. Above, by
$(E)^{\mu}$ we imply $e^{i\mu\arg(E)}|E|^{\mu}$. Thus the first and
third asymptotics coincide, as indeed they must since $a$ is a
single-valued function of $E$. Given (\ref{ad}), we set $m L =
\fract{1}{2} b_0 v^{-\mu}$, though the factor of $v^{-\mu}$ is
arbitrary and is purely to match the conventions of
\cite{DTb,BLZ2}.

The NLIE can also be solved by iteration
and the function $a(\te)$ constructed
for all complex $\te$ using
the integral representations (\ref{nlie}) or (\ref{nlie2})
as appropriate.  Consequently the eigenvalues of the spectral
determinants $D(E)$ and $C(E)$ can be calculated for all $M>0$ and
$l$.
This method is more generally
applicable than the TBA approach simply because the
NLIE  depends on the parameters $M$ and $l$ in a continuous manner.


%




\begin{thebibliography}{99}
\bibitem{fk1}
F.~Kleefeld, `Non-Hermitian quantum theory and its holomorphic
representation: introduction and some applications',
\hepth{0408028}.
\end{thebibliography}
\end{document}